\documentclass[useAMS, usenatbib, fleqn]{mn2e}
\pdfoutput=1
\usepackage{aecompl}
\usepackage{amsmath}
\usepackage{amssymb}
\usepackage{bbm}
\usepackage{color}
\usepackage{graphicx}
\usepackage{mathtools}
\usepackage{multirow}
\usepackage{times}
\usepackage{url}
\usepackage{hyperref}
\hypersetup{colorlinks=true,citecolor=blue,linkcolor=blue,filecolor=blue,urlcolor=blue}

\newcommand*\diff{\mathop{}\!\mathrm{d}}
\renewcommand\vec{\mathbf}
\newcommand\vechat[1]{\mathbf{\hat #1}}
\newcommand\vectilde[1]{\mathbf{\tilde #1}}
\DeclareMathOperator{\sgn}{sgn}
\DeclareMathOperator{\Wker}{W}

\title[Simulating galactic dust grain evolution]
    {Simulating galactic dust grain evolution on a moving mesh}
\author[R.~McKinnon et al.]
    {\parbox{18cm}{Ryan McKinnon,$^{1}$\thanks{E-mail: ryanmck@mit.edu}
     Mark Vogelsberger,$^{1}$\thanks{Alfred P.~Sloan Fellow}
     Paul Torrey,$^{1}$\thanks{Hubble Fellow}
     Federico Marinacci$^{1}$ \\
     and Rahul Kannan$^{2,1}$\thanks{Einstein Fellow}
     }\vspace{0.3cm}\\
     $^{1}$Department of Physics and Kavli Institute for Astrophysics and Space Research,
           Massachusetts Institute of Technology,
           Cambridge, MA 02139, USA \\
     $^{2}$Harvard-Smithsonian Center for Astrophysics,
           Cambridge, MA 02138, USA
    }
\begin{document}

\date{Accepted ???. Received ???; in original form ???}

\pagerange{\pageref{firstpage}--\pageref{lastpage}}
\pubyear{2018}

\maketitle

\label{firstpage}

\begin{abstract}
Interstellar dust is an important component of the galactic ecosystem, playing
a key role in multiple galaxy formation processes.  We present a novel
numerical framework for the dynamics and size evolution of dust grains
implemented in the moving-mesh hydrodynamics code \textsc{arepo} suited for
cosmological galaxy formation simulations.  We employ a particle-based method
for dust subject to dynamical forces including drag and gravity.  The drag
force is implemented using a second-order semi-implicit integrator and
validated using several dust-hydrodynamical test problems.  Each dust particle
has a grain size distribution, describing the local abundance of grains of
different sizes.  The grain size distribution is discretised with a
second-order piecewise linear method and evolves in time according to various
dust physical processes, including accretion, sputtering, shattering,
and coagulation.  We present a novel scheme for stochastically forming dust
during stellar evolution and new methods for sub-cycling of dust
physics time-steps.  Using this model, we simulate an isolated disc galaxy to
study the impact of dust physical processes that shape the interstellar grain
size distribution.  We demonstrate, for example, how dust shattering shifts the
grain size distribution to smaller sizes resulting in a significant rise of
radiation extinction from optical to near-ultraviolet wavelengths.  Our
framework for simulating dust and gas mixtures can readily be extended to
account for other dynamical processes relevant in galaxy formation, like
magnetohydrodynamics, radiation pressure, and thermo-chemical processes.
\end{abstract}


\section{Introduction}

Interstellar dust is a crucial ingredient for the formation and evolution of
galaxies, which is produced through condensation of metals expelled into the
interstellar medium (ISM) by supernovae (SNe) and stellar winds.  About
$30-50$ per cent of the metals condense into the dust component~\citep{Draine2007}.
Within the ISM dust plays an important role for multiple physical processes.
For example, dust grains provide a source of opacity to radiation from sources
like active galactic nuclei~(AGN) and massive stars.  Radiation pressure acting
on dust grains can inject momentum in the ISM and help drive galactic winds
\citep{Murray2005, Novak2012, Zahid2013, Ishibashi2015, Thompson2015}.  Dust
grain surfaces also aid the formation of molecular hydrogen
\citep{Hollenbach1971, Cazaux2004} and contribute to photoelectric heating of
gas \citep{Bakes1994, Weingartner2001b}, which both affect star formation in
galaxies.  Dust grains can also develop electric charge \citep{Feuerbacher1973,
Burke1974, Draine1987, Weingartner2001b} and are therefore affected by
magnetic fields, which alters the dynamics of dust in a turbulent ISM
\citep{Lazarian2002, Yan2004}.

Besides influencing interstellar chemistry and galaxy physics, importantly dust
also affects the detectability and observed properties of galaxies.  Dust
grains absorb ultraviolet (UV) light and re-emit the radiation at infrared (IR)
wavelengths~\citep{Draine1984, Mathis1990, Tielens2005}.  Especially at high
redshifts, where many surveys are executed in the UV rest frame, the measured
properties of galaxies critically depend on dust extinction.  Dust has such a
strong effect on galaxy properties despite the dust-to-gas ratio in galaxies
being a few per cent at most \citep{Draine2007, RemyRuyer2014}.  Emission from
dust is an important foreground not only for observation of galaxies but also
for the cosmic microwave background \citep{Planck2014}.

Over the last decade, observations from \textit{Herschel} \citep{Pilbratt2010}
have yielded several dust scaling relations tying dust to fundamental ISM
properties.  For example, there are observed relations between dust mass and
gas mass \citep{Corbelli2012}, dust-to-stellar mass ratio and gas fraction
\citep{Cortese2012}, and dust-to-stellar flux and mass ratios
\citep{Skibba2011}.  Data at high redshift is less abundant, but dust has
recently been detected in reionisation-era galaxies using the Very Large
Telescope and the Atacama Large Millimetre Array \citep{Watson2015,
Laporte2017}, challenging models to explain the production of dust at such
early times.  The importance of addressing high-redshift dust is likely to
increase, given the upcoming \textit{James Webb Space Telescope} mission and
its capability to witness the formation of galaxies.

Studying the abundance, distribution, and impact of dust in galaxies requires
detailed models that are capable of evolving the dust population of a galaxy
along with a plethora of other galaxy formation processes.  The specific
impact of dust can only be quantified by understanding its spatial and grain
size distribution.  This grain size distribution evolves over time within a
galaxy.  Dust is produced as stars return metals to the ISM \citep{Todini2001,
Nozawa2003, Ferrarotti2006, Bianchi2007, Zhukovska2008, Schneider2014}, setting
the initial size distribution for a population of dust grains.  The grain size
distribution is then subject to processes that conserve grain number but grow
or destroy dust mass.  For example, grain sizes grow through accretion of
gas-phase metals \citep{Liffman1989, Draine1990, Dwek1998, Michalowski2010,
Asano2013a} but shrink through sputtering \citep{Ostriker1973, Burke1974,
Barlow1978, Draine1979b, Dwek1992, Tielens1994} and SN shocks
\citep{Nozawa2006, Bianchi2007, Nozawa2007}.  Other physical processes conserve
total dust mass but shape the interstellar size distribution by increasing or
decreasing the number of grains: these include dust-dust collisional processes
like shattering \citep{ODonnell1997, Hirashita2009, Asano2013b, Mattsson2016}
and coagulation \citep{Chokshi1993, Jones1996, Dominik1997, Hirashita2009,
Mattsson2016}.

Without a detailed knowledge of the grain size distribution and the overall
arrangement of dust in galaxies, the modelling of dust physical processes
remains uncertain.  Studying, for example, the interplay of radiation and dust
as a feedback mechanism within galaxies requires very detailed knowledge about
both the radiation fields and dust content.  Simplified feedback prescriptions
motivated by radiation pressure coupling to dust grains have been included in
some cosmological simulations \citep[e.g.][]{Hopkins2014, Roskar2014,
Agertz2015}.  However, none of these studies self-consistently model either the
radiation field or the dust content.  Other studies based on
radiation-hydrodynamics simulations improve on those by coupling
self-consistent radiation fields to dust but without evolving the dust
component self-consistently \citep[e.g.][]{Rosdahl2015, Costa2018}.
Overall, most modern cosmological simulations of large scale structure
\citep{Vogelsberger2014a, Vogelsberger2014b, Schaye2015, Khandai2015} do not
directly treat dust within galaxies, despite analysing statistics like the
mass-metallicity relation \citep{Torrey2017, Torrey2017b, DeRossi2017} and
cluster metal distribution \citep{Vogelsberger2018} that could be affected by
depletion of metals onto dust. It is therefore highly desirable to have a
self-consistent dust model coupled to a comprehensive galaxy formation model in
combination with radiation-hydrodynamics to capture the impact of dust on
galaxy formation more reliably.

Various numerical models have been developed to evolve the grain size
distribution of galaxies in time \citep[e.g.][]{Liffman1989, ODonnell1997,
Asano2013b, Hirashita2015}.  These models suggest, for example, that changes in
the grain size distribution can strongly affect the overall dust mass.  For
instance, the process of shattering may temporarily conserve dust mass but, by
shifting grains to smaller sizes and increasing the total grain surface area,
subsequently leads to rapid increases in dust mass through accretion
\citep{Asano2013b}.  However, these models are often ``one zone'' in nature and
focus only on the total size distribution, ignoring dust and gas dynamics
because they lack spatial resolution within a galaxy.  While many of these
previous models are idealised in nature, recent galaxy formation simulations
are beginning to evolve dust physics in more detail.  These simulations attempt
to predict the distribution of dust mass within and around galaxies, include
the dynamical forces that impact dust motion, and model the processes that
shape the grain size distribution.  Initial attempts have been made to track
dust in non-cosmological smoothed particle hydrodynamics~(SPH) simulations
using ``live'' dust particles that are subject to different dynamics (e.g.~drag
and radiation pressure) than gas particles \citep{Bekki2015b}.  However, these
simulations assume grains to be of fixed size and thus do not make predictions
about the interstellar grain size distribution.  Recent simulations using the
moving-mesh code \textsc{arepo} \citep{Springel2010} have modelled the
formation of dust in a fully cosmological context \citep{McKinnon2016,
McKinnon2017}, albeit assuming perfect coupling between dust and gas and not
tracking the grain size distribution either.  Cosmological simulations by
\citet{Aoyama2018} model a simplified grain size distribution, dividing grains
into ``small'' and ''large'' sizes, but do not account for dynamical forces
like drag or radiation pressure.  Such cosmological results make predictions
for the dust content of a diverse sample of galaxies and the distribution of
dust on large scales.  So far, no simulation has been able to perform
cosmological galaxy formation simulations with a state-of-the-art galaxy
formation model combined with a dust model that traces both the spatial
distribution and full range of sizes of dust grains.

This paper aims to close this gap by presenting a novel dust framework,
modelling aspects of grain dynamics and size evolution and implemented
alongside the galaxy formation physics in the moving-mesh hydrodynamics code
\textsc{arepo}.  Section~\ref{SEC:drag_force} describes our implementation of
the drag force that couples dust grains to hydrodynamical motion and a series
of test problems.  In Section~\ref{SEC:grain_size_evolution}, we discuss the
modelling of the size distribution and evolution of dust grains.
Section~\ref{SEC:dust_production} details our implementation for
stochastically producing dust during stellar evolution.  Using the dust model,
in Section~\ref{SEC:hernquist_spheres} we perform simulations of isolated disc
galaxies.  Our conclusions are presented in Section~\ref{SEC:conclusions}.

\section{Dust dynamics and drag}\label{SEC:drag_force}

In this Section we first discuss the dynamics of dust particles as they
interact with surrounding gas.  Solid dust grains travelling through a gaseous
medium experience a drag force that alters their dynamical
behaviour~\citep[e.g.][]{Baines1965, Draine1979b}, and which effectively
couples dust dynamics to gas dynamics.  The strength of this drag force depends
on both gas and grain properties and affects the distribution of dust within
the ISM.  For example, a grain size dependent drag force impacts the grain size
distribution that results from SN shocks \citep[e.g.][]{Nozawa2006}.

Various numerical works have studied two-fluid dust and gas mixtures using a
particle-based SPH framework~\citep{Monaghan1995, Monaghan1997, Laibe2011,
Laibe2012, Laibe2012b, Bekki2015b, Booth2015, Booth2016, Price2017}.  In the
limit of a strong drag force, it can be advantageous to adopt a one-fluid
approach and solve for the mixture's barycentric motion and dust-to-gas
ratio~\citep{Barranco2009, Laibe2014, Laibe2014b, Laibe2014c, Price2015,
Tricco2017}.  Drag dynamics have also been studied using grid-based
methods~\citep{Cuzzi1993, Balsara2009} and hybrid methods that combine grid
techniques and particle approaches~\citep{Johansen2006, Balsara2009,
Miniati2010, Hopkins2016}.  Also the influence of grain size on drag forces has
been explored in many ways.  For example, \citet{Goodson2016} evolve
dust particles of different grain sizes in an expanding Sedov-Taylor blast wave
using a drag force and study the loss of grain mass due to sputtering.  Other
simulations treating drag adopt one fixed grain size~\citep{Saito2002,
Saito2003, Miniati2010, Laibe2012, Hopkins2016}.  Newer work accounts for drag
acting on multiple dust phases simultaneously when following barycentric motion
in the one-fluid approach \citep{Hutchison2018}.  Other models couple drag
force strength to an evolving grain size distribution in idealised SN
studies~\citep{Bocchio2016}.

In our work, we model dust with a particle-based framework that exists
alongside moving-mesh hydrodynamical calculations.  In
\textsc{arepo}~\citep{Springel2010}, a finite-volume scheme is used to solve
hydrodynamics on a mesh generated by a Voronoi tessellation of space and
allowed to move with the local fluid velocity.  The mesh can consist of
irregularly-shaped gas cells and is (de-)refined so that gas cells have roughly
equal mass~\citep{Vogelsberger2012}.

We could treat dust as a property of each gas cell and model dust
dynamics by transferring dust across cell interfaces.  However, while gas
exists throughout the computational domain, dust might only exist in more
localised regions.  Thus, it is advantageous to model dust using particles
representing ensembles of individual dust grains, with particle motion
unconstrained by mesh geometry.  This parallels the treatment of collisionless
star or black hole particles in
\textsc{arepo}~\citep[e.g.][]{Vogelsberger2013}.  The formulation for drag
below assumes dust is treated in this particle-based manner.

\subsection{Drag force calculation}

Our drag implementation follows the standard approach taken by~\citet{Booth2015}
and~\citet{Hopkins2016}. The acceleration of a dust particle of mass
$m_\text{d}$ is given by
\begin{equation}
\frac{\diff \vec{v}_\text{d}}{\diff t} = -\frac{K_\text{s} (\vec{v}_\text{d} - \vec{v}_\text{g})}{m_\text{d}} + \vec{a}_\text{d,ext},
\label{EQN:dvdt_dust}
\end{equation}
where $K_\text{s}$ is a drag coefficient determined below, $\vec{v}_\text{d}$
and $\vec{v}_\text{g}$ are the dust and gas velocity, respectively, and
$\vec{a}_\text{d,ext}$ denotes external sources of acceleration (e.g.~gravity,
radiation pressure, or magnetic fields), while the backreaction on the gas is
given by
\begin{equation}
\frac{\diff \vec{v}_\text{g}}{\diff t} = -\frac{\nabla P}{\rho_\text{g}} + \frac{\rho_\text{d} K_\text{s} (\vec{v}_\text{d} - \vec{v}_\text{g})}{\rho_\text{g} m_\text{d}} + \vec{a}_\text{g,ext},
\label{EQN:dvdt_gas}
\end{equation}
for gas pressure $P$, dust and gas densities $\rho_\text{d}$ and
$\rho_\text{g}$, respectively, and external gas acceleration
$\vec{a}_\text{g,ext}$. We assume the dust is pressureless.

The drag force can be written in terms of relative velocity as
\begin{equation}
\frac{\diff (\vec{v}_\text{d} - \vec{v}_\text{g})}{\diff t} = - \frac{\vec{v}_\text{d} -
\vec{v}_\text{g}}{ t_\text{s}},
\label{EQN:relative_velocity}
\end{equation}
using the stopping time-scale
\begin{equation}
t_\text{s} = \frac{m_\text{d} \rho_\text{g}}{K_\text{s} (\rho_\text{g} + \rho_\text{d})}.
\end{equation}
Shorter stopping time-scales correspond to the high-drag regime in which
relative velocities quickly decay.  In this work, we focus on
collisional drag and neglect Coulomb drag resulting from grain charge.

To lowest order, the aerodynamic drag force has magnitude
\begin{equation}
F_\text{D} = \frac{1}{2} C_\text{D} \pi a^2 \rho_\text{g} |\vec{v}_\text{d} - \vec{v}_\text{g}|^2,
\end{equation}
the product of a drag parameter $C_\text{D}$, grain cross-section, and ram
pressure.  A typical interstellar grain of radius $a$ satisfies $a < 9 \lambda /
4$, where $\lambda$ is the gas mean free path.  This corresponds to the Epstein
drag regime \citep{Epstein1924, Weidenschilling1977, Stepinski1996}, in which
drag effects build up through collisions with individual gas atoms.  This is in
contrast to the Stokes limit, $a > 9 \lambda / 4$, in which the gas
behaves as a fluid and the drag force depends on the Reynolds number of the
flow.  In the Epstein limit, the drag parameter is given by
\begin{equation}
C_\text{D} = \frac{16 \sqrt{2} c_\text{s}}{3 \sqrt{\pi \gamma} |\vec{v}_\text{d} - \vec{v}_\text{g}|},
\end{equation}
where $c_\text{s}$ is the local sound speed and $\gamma$ is the
adiabatic index. In this regime the drag force is therefore linear in the
relative velocity.  The drag coefficient entering into the acceleration
equations is
\begin{equation}
K_\text{s} = \frac{1}{2} C_\text{D} \pi a^2 \rho_\text{g} |\vec{v}_\text{d} - \vec{v}_\text{g}| = \frac{8 \sqrt{2 \pi} c_\text{s} a^2 \rho_\text{g}}{3 \sqrt{\gamma}}.
\end{equation}
Furthermore, for ISM studies with $\rho_\text{d} / \rho_\text{g} \ll 1$, we can
ignore the drag force in the gas equation of motion; i.e.~we solve
equation~(\ref{EQN:dvdt_dust}) for dust motion including the drag force but
solve gas motion using equation~(\ref{EQN:dvdt_gas}) neglecting the
backreaction of dust dynamics on the gas.  Inclusion of the backreaction of
drag on gas will be necessary in future studies of radiation-driven outflows.

Assuming spherical grains with mass $m_\text{d} = 4\pi a^3 \rho_\text{gr} /
3$, this implies a stopping time-scale of
\begin{equation}
t_\text{s} = \frac{m_\text{d}}{K_\text{s}} = \frac{\sqrt{\pi \gamma} a \rho_\text{gr}}{2 \sqrt{2} \rho_\text{g} c_\text{s}},
\label{EQN:t_s_zeroth}
\end{equation}
where $\rho_\text{gr}$ is the internal density of a dust grain.
The derivation of equation~(\ref{EQN:t_s_zeroth}) implicitly assumed
subsonic relative dust-gas velocities.  Supersonic relative motion requires a
further correction factor for the stopping time-scale~\citep{Kwok1975,
Draine1979b, Paardekooper2006, Price2017}, which is approximated by the
following fit
\begin{equation}
t_\text{s} = \frac{\sqrt{\pi \gamma} a \rho_\text{gr}}{2 \sqrt{2} \rho_\text{g} c_\text{s}} \left( 1 + \frac{9 \pi}{128} \left|\frac{\vec{v}_\text{d} - \vec{v}_\text{g}}{c_\text{s}} \right|^2 \right)^{-1/2}.
\label{EQN:t_s_full}
\end{equation}
To remain consistent with previous works, we calculate all stopping time-scales
using equation~(\ref{EQN:t_s_full}) and take the internal density to be
$\rho_\text{gr} \approx 2.4 \, \text{g} \, \text{cm}^{-3}$~\citep{Draine2003}.
A stopping time-scale of this form is valid for supersonic dust-gas
relative velocity and has been used in turbulent giant molecular cloud
simulations reaching Mach numbers $\mathcal{M} > 10$ \citep{Hopkins2016,
Lee2017}. We note that in the subsonic limit, this reduces to the form of
stopping time-scale from equation~(\ref{EQN:t_s_zeroth}) above, similar to that
used in~\citet{Booth2015}.

Since we apply drag acceleration only to dust in the $\rho_\text{d} /
\rho_\text{g} \ll 1$ limit, we only need to interpolate $\rho_\text{g}$,
$\vec{v}_\text{g}$, and $c_\text{s}$ to the position of a dust particle in
order to calculate its stopping time-scale. To this end, we perform a
kernel-smoothing around a given dust particle at position $\vec{r}_\text{d}$.
We first iteratively solve for its smoothing length $h_\text{d}$ using
\begin{equation}
N_\text{ngb} = \frac{4\pi h_\text{d}^3}{3} \sum_{i}^\text{gas} \Wker(|\vec{r}_{i} - \vec{r}_\text{d}|, h_\text{d}),
\label{EQN:N_ngb}
\end{equation}
where $N_\text{ngb}$ is the desired number of gas neighbours and the cubic
spline kernel is given by
\begin{equation}
\Wker(r, h) = \frac{8}{\pi h^3}
\begin{dcases}
1 - 6 \left(\frac{r}{h}\right)^2 + 6 \left( \frac{r}{h} \right)^3, & 0 \leq \frac{r}{h} \leq \frac{1}{2}, \\
2 \left(1 - \frac{r}{h} \right)^3, & \frac{1}{2} < \frac{r}{h} \leq 1, \\
0, & \frac{r}{h} > 1.
\end{dcases}
\label{EQN:cubic_spline}
\end{equation}
Then we can estimate
\begin{equation}
\rho_\text{g}(\vec{r}_\text{d}) = \sum_{i=1}^{N_\text{ngb}} m_{i} \Wker(|\vec{r}_{i} - \vec{r}_\text{d}|, h_\text{d}),
\label{EQN:rho_kernel}
\end{equation}
and
\begin{equation}
\vec{v}_\text{g}(\vec{r}_\text{d}) = \frac{\sum_{i=1}^{N_\text{ngb}} m_{i} \vec{v}_{i} \Wker(|\vec{r}_{i} - \vec{r}_\text{d}|, h_\text{d})}{\sum_{i=1}^{N_\text{ngb}} m_{i} \Wker(|\vec{r}_{i} - \vec{r}_\text{d}|, h_\text{d})},
\label{EQN:kernel_weighting}
\end{equation}
which amounts to a mass-weighted gas velocity calculation.  Throughout our
work, we perform all kernel smoothings in a similar manner.  The kernel
framework above is written for three spatial dimensions but can also be
generalised to one or two dimensions.

\subsection{Time integration}

An explicit drag integrator requires us to resolve $\Delta t < t_\text{s}$ for
dust particles, meaning drag time-steps may be more restrictive than
hydrodynamical or gravitational time-steps.  To get an idea of typical stopping
time-scales, we can use equation~(\ref{EQN:t_s_full}) to write
\begin{equation}
\begin{split}
t_\text{s} &\approx 6.2 \, \text{Myr} \, \left( \frac{a}{0.1 \, \mu\text{m}} \right) \left( \frac{\rho_\text{gr}}{2.4 \, \text{g} \, \text{cm}^{-3}} \right) \\
&\quad \times \left( \frac{\rho_\text{g}}{10^{-24} \, \text{g} \, \text{cm}^{-3}} \right)^{-1} \left( \frac{c_\text{s}}{1 \, \text{km} \, \text{s}^{-1}} \right)^{-1},
\label{EQN:t_s_Myr}
\end{split}
\end{equation}
where we assume $\gamma = 5/3$ and neglect the higher-order stopping time-scale
correction for supersonic relative gas-dust velocity.  As noted by
\citet{Laibe2012}, resolving stopping time-scales is most prohibitive when gas
and dust are highly coupled and thus show little relative motion.  In essence,
we require high temporal resolution only to find that dust and gas move as one.
Worse yet, if dust is not treated in the test-particle limit (where we assumed
$\rho_\text{d} / \rho_\text{g} \ll 1$) and backreaction on the gas is included,
a high spatial resolution is also needed to avoid artificial overdissipation of
kinetic energy when dust-gas dephasing is not resolved \citep{Laibe2012}.

One alternative approach eschews the two-fluid formalism in favor of a
one-fluid method following the gas-dust barycentre~\citep{Laibe2014,
Laibe2014b}.  In the limit of small dust-to-gas ratio, this simply treats dust
as a passive scalar perfectly coupled to gas motion.  Another alternative
approach, valid in the test-particle limit, maintains the two-fluid formalism
from Section~\ref{SEC:drag_force} and employs semi-implicit integrators to
avoid the need for prohibitively small drag time-steps.  Here we therefore
follow the semi-implicit time-stepping approaches detailed in previous works
\citep{Monaghan1997, Laibe2012b, LorenAguilar2014, Booth2015, LorenAguilar2015}
that make use of the analytical solution of
equation~(\ref{EQN:relative_velocity}) in the case of constant stopping
time-scale.

\begin{figure}
\centering
\includegraphics{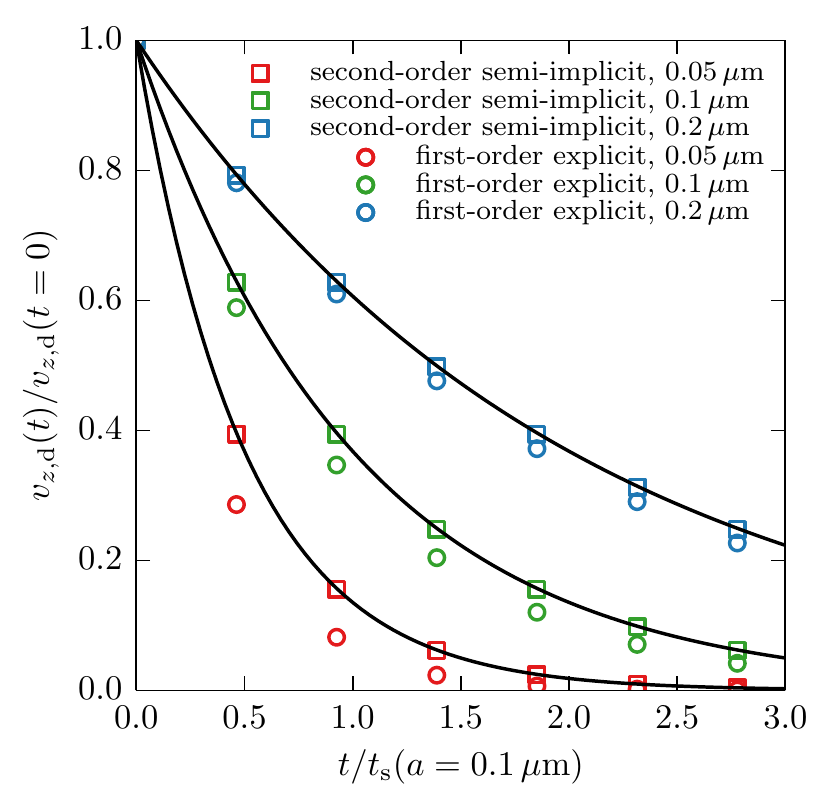}
\caption{Mean velocity evolution for dust particles travelling with initial
velocity $v_{z,\text{d}}(t = 0) = 1 \, \text{km} \, \text{s}^{-1}$ in a box
where gas is at rest.  Circles denote velocities calculated using the explicit
first-order integrator for fixed grain sizes $a = 0.05$, $0.1$, and $0.2 \,
\mu\text{m}$, while squares indicate velocities from the semi-implicit method
of equation~(\ref{EQN:semi_implicit_kick}).  Black lines show analytic results.
Times are normalised to the stopping time-scale for the run with $a = 0.1 \,
\mu\text{m}$.  For the explicit method, we require $\Delta t < t_\text{s}$,
while the semi-implicit method can adopt time-steps independent of stopping
time-scale.  The accuracy of the explicit method could be improved by adopting
smaller time-steps at the expense of computational cost.}
\label{FIG:dustybox_velocity}
\end{figure}

In practice, we employ the $\rho_\text{d} / \rho_\text{g} \ll 1$
limit of \citet{LorenAguilar2015}, whose semi-implicit, split-update method is
well-suited to the time integration routine in \textsc{arepo}.  The method
presented in \citet{LorenAguilar2015} fixes the limitations of the method in
\citet{LorenAguilar2014} pointed out by \citet{Booth2015}, namely the incorrect
behaviour of relative velocity in cases of a net dust-gas relative acceleration
from external sources.  We note that equation~(17) in \citet{LorenAguilar2015},
the basis for our drag kicks, can be recast in the form of equation~(16) in
\citet{Booth2015}, which is shown to be a second-order scheme.  However,
\citet{Booth2015} suggest that a simpler first-order scheme may be acceptable
for general use.  We refer the reader to \citet{LorenAguilar2014} for
discussion on the stability of semi-implicit drag integrators.

Our semi-implicit second-order time integration is implemented in the following
way. Suppose the system is at time $t$ and a dust particle's velocity is being
updated over time-step $\Delta t$.  Let $\vectilde{v}_\text{d}(t + \Delta t)$
and $\vectilde{v}_\text{g}(t + \Delta t)$ denote the dust particle's velocity
and SPH-averaged gas velocity at time $t + \Delta t$ after non-drag
(e.g.~gravity) kicks are applied, but before drag acts on velocities.  Then, we
update the dust particle's velocity to
\begin{equation}
\begin{split}
\vec{v}_\text{d}(t + \Delta t) &= \vectilde{v}_\text{d}(t + \Delta t) - \xi \left[\vectilde{v}_\text{d}(t + \Delta t) - \vectilde{v}_\text{g}(t + \Delta t)\right] \\
&\;\; + \left[ (\Delta t + t_\text{s}) \xi - \Delta t \right] \left[\vec{a}_\text{d,ext}(t) - \vec{a}_\text{g,ext}(t) + \frac{\nabla P}{\rho_\text{g} }\right],
\end{split}
\label{EQN:semi_implicit_kick}
\end{equation}
where we define $\xi \equiv 1 - \exp(-\Delta t / t_\text{s})$.  To maintain
consistency with equation~(\ref{EQN:dvdt_gas}), our notation differs slightly
from that used in \citet{LorenAguilar2015}, where $-\nabla P / \rho_\text{g}$
is folded into $\vec{a}_\text{g,ext}$.  While we adopt this semi-implicit
approach and use it in most cases, we also implement an explicit first-order
time-stepping framework for comparison purposes.

Dust particles are dynamically assigned individual time-steps in the following
manner.  For each dust particle, we first calculate the minimum hydrodynamical
time-step for gas cells within the smoothing kernel radius $h_\text{d}$, which
we denote $\Delta t_\text{g,ngb}$.  Next, we determine a
Courant-Friedrichs-Lewy~(CFL) type time-step using
\begin{equation}
\Delta t_\text{CFL} \equiv \frac{C_\text{CFL} h_\text{d}}{\sqrt{c_\text{s}^2 + |\vec{v_\text{g}} - \vec{v_\text{d}}|^2}},
\end{equation}
where $C_\text{CFL} \sim 0.3$ and $c_\text{s}$ is the kernel-averaged gas sound
speed.  In the case of the explicit integrator, we also calculate a time-step
using $\Delta t_\text{stop} \equiv \beta_\text{stop} t_\text{s}$ where
$\beta_\text{stop}$ controls what fraction of the stopping time-scale must be
resolved.  Typically $\beta_\text{stop}$ is a factor of order $0.1$, although
in practice we do not use the explicit drag integrator beyond a simple test
problem.  Using these time-step values, the dust particle time-step is chosen
to satisfy all of these constraints via
\begin{equation}
\Delta t_\text{d} = \min(\Delta t_\text{g,ngb}, \Delta t_\text{CFL}, \Delta t_\text{stop}),
\label{EQN:timestep_constraint}
\end{equation}
where the final term involving $\Delta t_\text{stop}$ only applies if using an
explicit integrator (i.e.~the term is not included when using
equation~\ref{EQN:semi_implicit_kick}).  In addition, gravitational time-steps
for dust particles are calculated in the same manner as for dark matter, stars,
and other collisionless particles in \textsc{arepo}.

In the following we will present test problems to demonstrate the performance
of our dust integrator.

\begin{figure}
\centering
\includegraphics{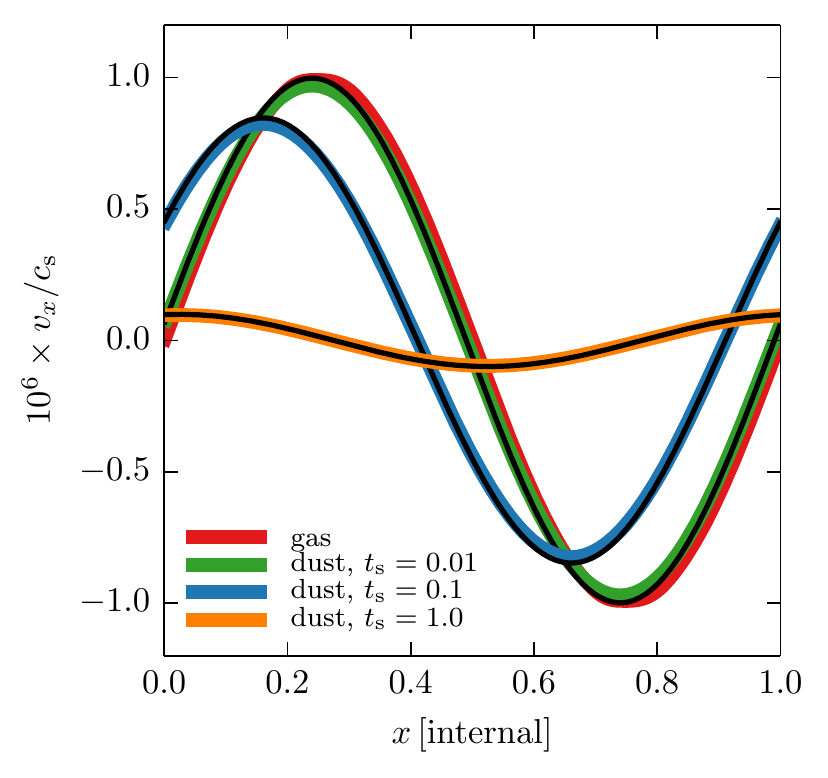}
\caption{Velocity profiles in the dusty travelling wave test at $t = 1$
after one wave crossing.  Coloured points show velocities for gas (red) and
dust using various stopping time-scales: $t_\text{s} = 0.01$ (green),
$t_\text{s} = 0.1$ (blue), and $t_\text{s} = 1.0$ (orange).  Black lines show
numerically-integrated dust velocity profiles.  Dust most closely follows the
gas when the stopping time-scale is short, corresponding to high drag.}
\label{FIG:dustywave_plots}
\end{figure}

\subsection{Drag in uniform gas flow}

We start with a first simple test by simulating a periodic, three-dimensional
box of volume $(1 \, \text{kpc})^3$ using $16^3$ gas cells and $16^3$ dust
particles, arranged in a body-centred configuration.  Dust is given an initial
velocity $\vec{v}_\text{d} = 1 \, \text{km} \, \text{s}^{-1} \, \vechat{z}$,
and gas has uniform density $\rho_\text{g} = 2 \times 10^{7} \, \text{M}_\odot
\, \text{kpc}^{-3}$, corresponding to an ISM-like number density $n \sim 1 \,
\text{cm}^{-3}$.  The uniform dust density is taken to be $\rho_\text{d} =
\rho_\text{g} / 100$, and grains are assumed to have a fixed radius $a$.  We
perform runs with $a = 0.05$, $0.1$, and $0.2 \, \mu\text{m}$.

\begin{figure}
\centering
\includegraphics{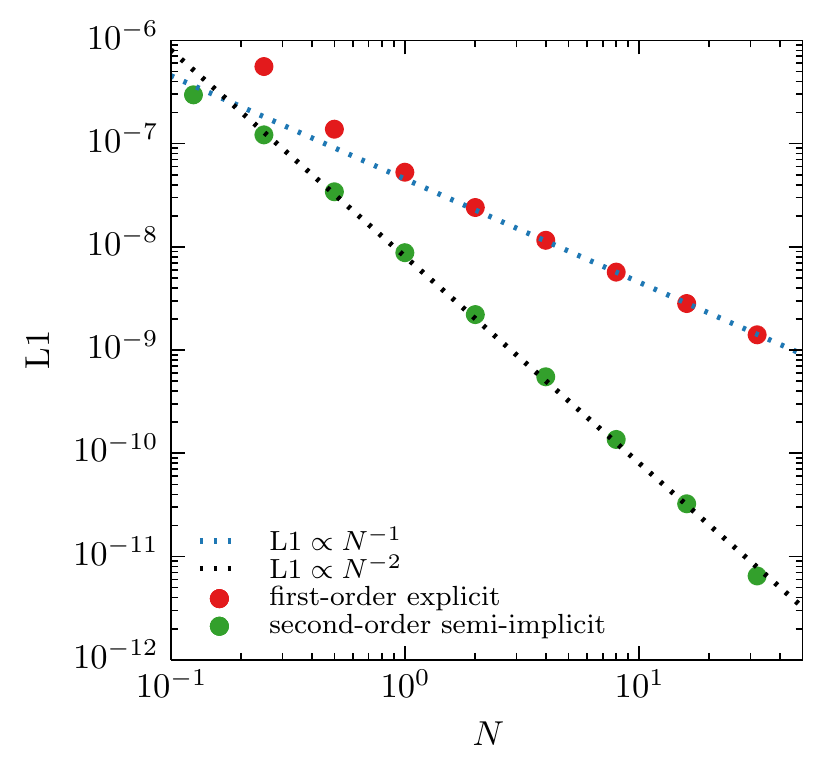}
\caption{Convergence results for the dusty travelling wave test, plotting L1 error
in dust velocity at $t = 1$ after one gas wave-crossing for a constant stopping
time-scale of $t_\text{s} = 0.1$.  Here, $N$ controls time-step resolution via
the constraint $\Delta t < t_\text{s} / N$, and we show results for explicit,
forward Euler (red) and semi-implicit (green) drag updates.  We hard-code the
analytic gas velocity when computing dust drag forces to avoid interpolation
noise.  Dotted lines show first-order (blue) and second-order (black) scalings.
The semi-implicit velocity update given by
equation~(\ref{EQN:semi_implicit_kick}) produces a second-order drag solver.}
\label{FIG:dustywave_convergence}
\end{figure}

\begin{figure*}
\centering
\includegraphics{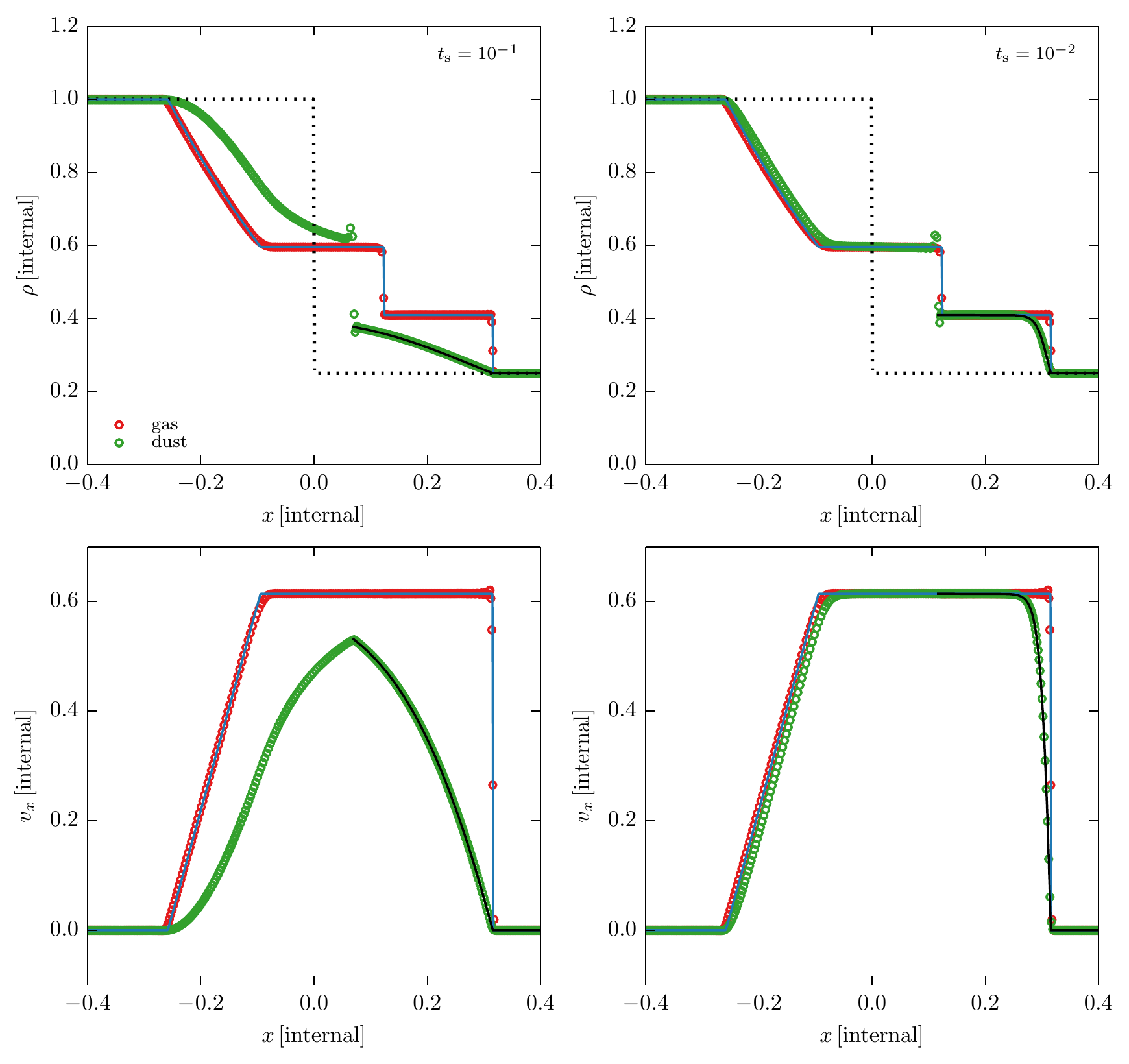}
\caption{Density (top) and velocity (bottom) profiles for gas (red) and dust
(green) in the shock test at $t = 0.2$, using fixed stopping
time-scales $t_\text{s} = 0.1$ (left) and $t_\text{s} = 0.01$ (right).
Solid lines denote analytic profiles for gas (blue) and dust (black), while the
dotted line in the top panels marks the initial density jump.  The dust density
profile is constructed using kernel interpolation at dust particle positions
and is multiplied by the overall gas-to-dust ratio to enable comparison with
gas density.  The hydrodynamics methods from \citet{Springel2010} prevent the
spurious gas velocity ringing present in Figure~5 of \citet{Booth2015}, leading
to reduced noise in the dust velocity profiles.  The dust profiles more closely
follow the gas profiles in the high-drag case with shorter stopping
time-scale.}
\label{FIG:dustyshock_plots}
\end{figure*}

We turn off self-gravity, so that only hydrodynamic forces act.  Since there is
no drag backreaction on gas cells, their velocities remain unchanged as the
system evolves.  We integrate these dust particles over several stopping times,
using the two integrators: the explicit first-order method (requiring $\Delta t
< t_\text{s}$) and the semi-implicit second-order method given by
equation~\ref{EQN:semi_implicit_kick}.  Figure~\ref{FIG:dustybox_velocity}
shows the evolution of dust velocity as a function of time.  We note again that
in general the semi-implicit integrator chooses dust time-steps independent of
stopping time-scale, but for this test we force it to use the same time-steps
as the explicit first-order integrator.  Both integrators yield exponential
velocity decay, but the explicit first-order method overdamps the dust velocity
when resolving the stopping time-scale.  By contrast, the second-order
semi-implicit drag integrator offers much better agreement with the
analytically calculated expected velocity evolution tracks.

Our initial analysis of the benefits of semi-implicit drag integrators agrees
with findings from earlier two-fluid studies \citep{Monaghan1997, Laibe2012,
Laibe2012b, Booth2015}.  The conclusion of the test in
Figure~\ref{FIG:dustybox_velocity} is \emph{not} that an explicit integrator is
unsuitable for gas-dust drag in theory, but rather that high-accuracy solutions
may require prohibitively small time-steps.  This is especially the case in
highly-coupled flows, where the stopping time-scale can be much smaller than
the hydrodynamical time-scale.  We investigate the convergence properties of
these integrators in more detail in the following section.

\subsection{Dusty travelling wave}

The propagation of linear sound waves that transport dust is a well-studied test
problem~\citep{Laibe2011, Laibe2012, Booth2015} that we explore next.  We
perform the travelling wave test in one dimension, where in internal units the
periodic domain has length $1$ and sound speed $c_\text{s} = 1$.  At
equilibrium, gas and dust are at rest, with the gas having density
$\rho_\text{g} = 1$ and adiabatic index $\gamma = 5/3$.  To produce a linear
wave, we add sinusoidal perturbations to the gas density and velocity with
amplitudes $\Delta \rho_\text{g} / \rho_\text{g} = \Delta v_\text{g} /
c_\text{s} = 10^{-6}$.  As this wave propagates, it accelerates the dust via
the drag force.  We use various fixed stopping time-scales to test our
implementation.

\begin{figure*}
\centering
\includegraphics{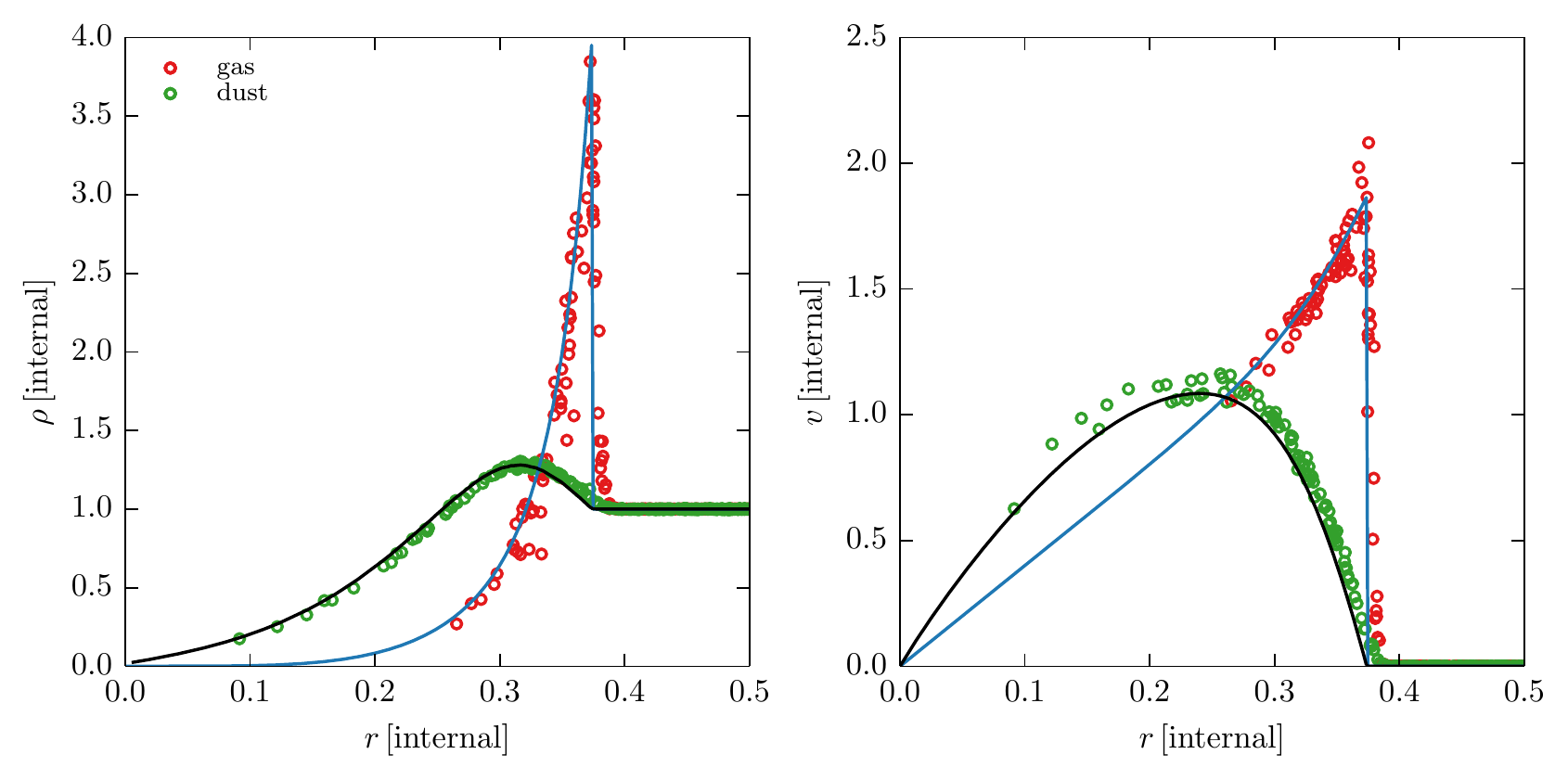}
\caption{Density (left) and velocity (right) profiles for gas (red) and dust
(green) in the Sedov dust test at $t = 0.06$.  We multiply the dust
density by the gas-to-dust ratio to compare densities on the same axes.  The
drag stopping time-scale is fixed to $t_\text{s} = 0.04$.  Solid blue lines
denote analytic gas profiles, and solid black lines indicate dust profiles
predicted by numerically integrating the dust drag acceleration using the
self-similar Sedov-Taylor solution.  We randomly subsample gas cells and dust
particles to improve readability.  Dust lags behind the gas and does not
display sharp peaks in density and velocity.}
\label{FIG:dustysedov_plots}
\end{figure*}

Figure~\ref{FIG:dustywave_plots} shows the velocity structure of the wave at $t
= 1$, after one full period.  While the gas wave returns to its original state,
the behaviour of the dust is more complex.  When the stopping time-scale is
small ($t_\text{s} = 0.01$), the drag force acts quickly and produces a dust
wave closely mirroring the gas wave.  However, when the stopping time-scale is
large ($t_\text{s} = 1.0$), dust is not strongly coupled to the gas and
experiences velocity amplitudes roughly one-tenth of the gas velocity.
Furthermore, as the drag strength decreases, there is a clear phase offset
between the gas and dust waves.  A run with medium stopping time-scale
($t_\text{s} = 0.1$) displays a hybrid of these two limiting cases.

We next study how the test results are affected by changes in
time-step.  We use the parameter $N$ to indicate how many time-steps fit into
one stopping time-scale: that is, we enforce $\Delta t < t_\text{s} / N$.  In
this test, we fix $t_\text{s} = 0.1$.  To focus strictly on the accuracy of the
drag integrator, we do not use kernel smoothing to estimate the local gas
velocity in performing drag updates but instead use the known analytic gas
solution.  As in Figure~\ref{FIG:dustywave_plots}, we let the wave propagate
for one full period.  We estimate the error after one period using the L1 norm
\begin{equation}
\text{L1} = \frac{1}{N_\text{d}} \sum_{i} |v_i - v_\text{d}(x_i)|,
\end{equation}
where $N_\text{d} = 256$ is the number of dust particles, $x_i$ and $v_i$ are
the position and velocity of the dust particle $i$ and, following
\citet{Booth2015}, $v_\text{d}(x_i)$ is the dust velocity at $t = 1$ computed
using a high resolution numerical integrator.

Figure~\ref{FIG:dustywave_convergence} shows the L1 error for dust after one
wave-crossing as the fineness of the time-steps (given by the parameter $N$) is
increased.  As expected we find that the first-order explicit scheme has an
error scaling as $N^{-1}$, while the second-order semi-implicit method
converges faster with an error proportional to $N^{-2}$.  In all subsequent
tests and simulations we only use the second-order semi-implicit scheme.

\begin{figure*}
\centering
\includegraphics{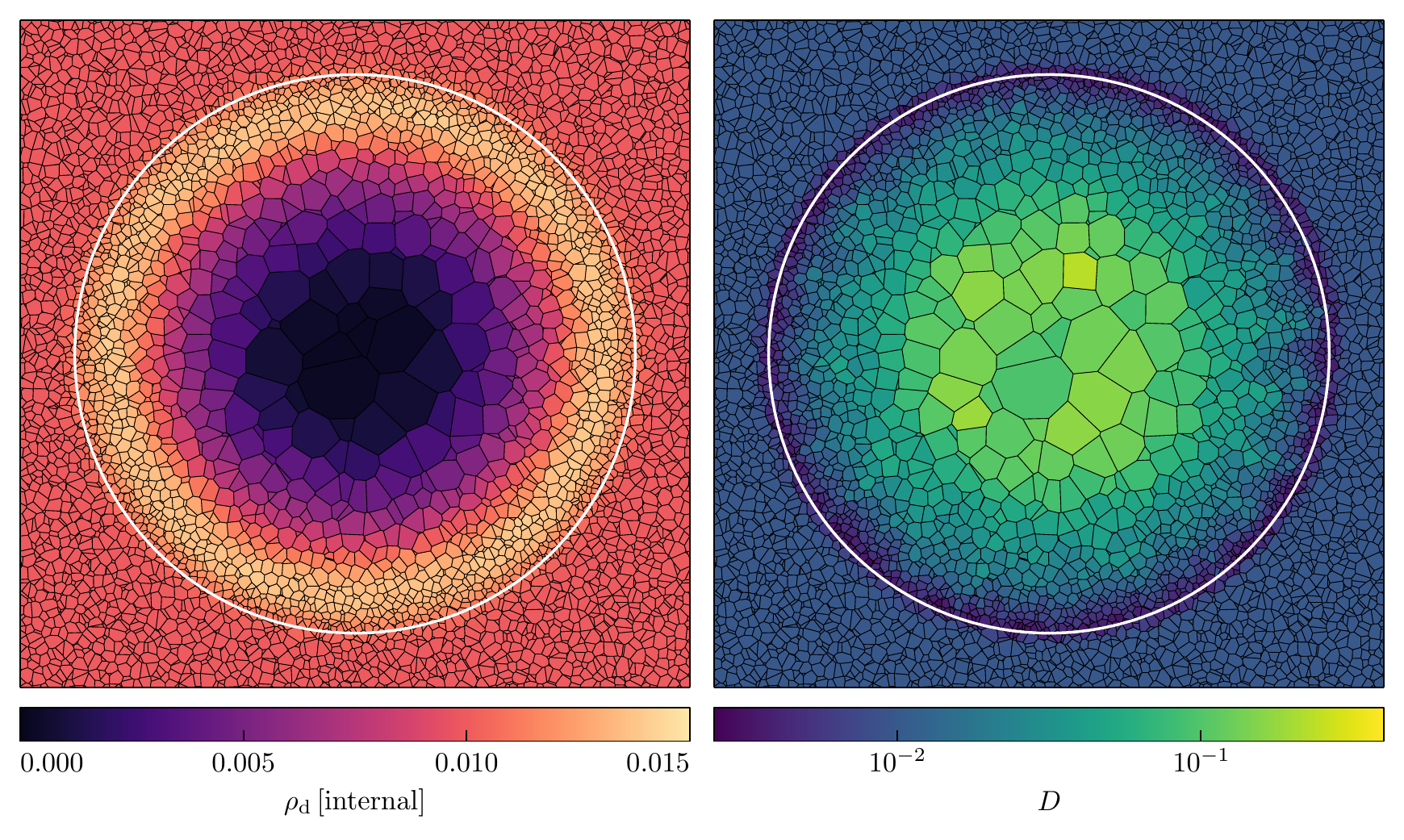}
\caption{Two-dimensional slices of the mesh structure in the
Sedov dust test at $t = 0.08$, where gas cells are coloured by the
local dust density (left) and dust-to-gas ratio (right).  These quantities are
computed in a kernel-smoothed manner about the centroid of each
two-dimensional cell.  White circles denote the radius where gas density peaks,
computed using the analytic self-similar solution.  As in
Figure~\ref{FIG:dustysedov_plots}, the stopping time-scale is set to
$t_\text{s} = 0.04$.  Because dust is not perfectly coupled to the
hydrodynamical motion, the dust density peaks at a smaller radius than the gas
density.  Thus, there is a drop in dust-to-gas ratio near the blast radius.}
\label{FIG:dustysedov_mesh}
\end{figure*}

\subsection{Hydrodynamical shock in a dusty medium}

Next we test the dynamics of dust particles in a \citet{Sod1978} shock tube,
which has been studied both for high dust-to-gas ratio \citep{Paardekooper2006,
Laibe2012, Laibe2012b} and in the test-particle limit \citep{Booth2015}.

We use an elongated box of dimensions $1.25 \times 0.15625 \times 0.15625$ in
code units, with $512 \times 64 \times 64$ equally-spaced gas cells and dust
particles initially at rest and arranged in a body-centred lattice with
reflective boundary conditions.  Following \citet{Booth2015}, gas cells have
$\rho_\text{g} = 1$ and $P = 1$ for $x < 0$ and $\rho_\text{g} = 0.25$ and $P =
0.1795$ for $x \geq 0$.  The adiabatic index is $\gamma = 5/3$.  The
dust-to-gas ratio is set to $\rho_\text{d}/\rho_\text{g} = 0.01$.  As a
result of this configuration, gas cells and dust particles across the jump have
unequal mass.

Figure~\ref{FIG:dustyshock_plots} shows the density and velocity profiles
obtained in this shock tube at $t = 0.2$ for two choices of fixed stopping
time-scale, $t_\text{s} = 0.1$ (low drag) and $t_\text{s} = 0.01$ (high
drag).  Fixing the stopping time-scale enables comparison with analytic dust
profiles for particles satisfying $x > 0$ at $t = 0$ \citep[see equations 20
and 21 in][]{Booth2015}.  The density of a gas cell is obtained directly from
the hydrodynamics solver in \textsc{arepo}, while we calculate the dust density
via kernel smoothing using an equivalent version of
equation~(\ref{EQN:rho_kernel}).  Smoothing lengths are calculated to ensure
dust particles have $N_\text{ngb} = 64 \pm 8$ neighbours.

Qualitatively, the dust profiles show good agreement with the analytic
predictions and are more similar to those of the gas for shorter stopping
time-scale.  However, while the gas density profiles show two discontinuities,
corresponding to the contact discontinuity and shock, the dust density has only
one discontinuity.  We note that \textsc{arepo} robustly captures the expected
gas dynamics, and this in turn improves the accuracy of our drag calculations.
In contrast, the shock test presented in Figure~5 of \citet{Booth2015} displays
gas velocity ringing near the contact discontinuity (i.e.~gas velocity
dispersions of roughly $5-10$ per cent of the sound speed).  This leads to
numerical noise when integrating dust particles, although to some degree this
problem is ameliorated by smoothing over the velocities of many gas neighbors.
This test demonstrates that accurate dust dynamics in part requires accurate
gas dynamics.

\subsection{Drag acceleration in an expanding Sedov blast wave}

The Sedov blast wave test studies the dynamics of dust in a standard
three-dimensional \citet{Sedov1959} blast wave.  There exist analytical
solutions for the gas dynamics in the purely hydrodynamical case
\citep[e.g.][]{Landau1959}, and these are still valid in the dust test-particle
limit.

This dust test has been introduced in \citet{Laibe2012}, and we
largely parallel that setup.  We simulate a periodic, cubic volume of unit side
length with $128^3$ gas cells and dust particles.  The initial gas cells are
determined by choosing random mesh-generating points and relaxing the mesh
using Lloyd's algorithm \citep{Lloyd1982}, while dust particles are
superimposed using a Cartesian lattice.  In code units, the initially uniform
gas and dust densities are $\rho_\text{g} = 1$ and $\rho_\text{d} = 0.01$,
respectively.  We inject a total energy $E = 1$ into the gas cell at the volume
centre.  For comparison, \citet{Laibe2012} spreads this blast energy over
multiple gas particles using kernel-smoothing.  Outside of this blast
cell, we assign the gas pressure such that the sound speed $c_\text{s} = 2
\times 10^{-5}$.  The gas has adiabatic index $\gamma = 5/3$.  For this test,
we fix the stopping time-scale at $t_\text{s} = 0.04$.

We note that our test focuses strictly on grain dynamics and ignores
high-temperature sputtering \citep{Ostriker1973, Burke1974, Barlow1978,
Draine1979b, Dwek1992, Tielens1994}, although hot blast waves are expected to
modify the grain size distribution \citep{Nozawa2006, Bianchi2007, Nath2008,
Kozasa2009, Silvia2010, Silvia2012, Goodson2016}.  Evolution in the grain size
distribution would in turn affect the strength of dust-gas drag.  The purpose
of this test is not to realistically model a SN remnant but to assess
grain motion in a well-known hydrodynamical problem.

Figure~\ref{FIG:dustysedov_plots} shows the resulting density and velocity
profiles at $t = 0.06$ for both gas and dust.  We compare against analytic gas
profiles predicted by the Sedov solution and dust profiles predicted by
numerically integrating the dust drag equation of motion. Here,
we see that dust shows qualitatively different features: the density and
velocity profiles peak before the radius of the blast wave and do not show
discontinuities.  Because gas and dust are decoupled and interact only through
the drag force, dust lags behind the gas and experiences smaller-amplitude
increases in density and velocity.  The simulated dust profiles show good
agreement with the numerical predictions, although we note that the dust
velocity near the blast wave tends to lie above its predicted value, exceeding
the peak velocity by about ten per cent.

Two-dimensional slices of the mesh are shown in
Figure~\ref{FIG:dustysedov_mesh}.  To improve the visibility of the mesh, this
figure has been generated from a run using only $64^3$ gas cells and dust
particles and at $t = 0.08$, when the blast has expanded to fill more of the
volume than in Figure~\ref{FIG:dustysedov_plots}.  For each gas cell in this
two-dimensional slice, we compute the local dust density by kernel
interpolation in three dimensions over nearby dust particles, centring the
interpolation about the cell centroid.  The dust-to-gas ratio is then estimated
by dividing the local dust density by the cell's known gas density.

Figure~\ref{FIG:dustysedov_plots} shows that the dust density increases
radially outward from the blast but peaks before reaching the radius of
the blast wave.  Because the drag force coupling dust to the hydrodynamical
motion takes some time to act, dust appears to chase the expanding blast.  This
results in a clear negative radial gradient for the dust-to-gas ratio: the
dust-to-gas ratio is highest near the centre of the blast, since dust is
delayed in expanding outward, and lowest at the blast radius, since gas
compresses to higher density more rapidly than the lagging dust.  Simulations
treating dust as perfectly coupled to the hydrodynamical motion would not
resolve these dust-to-gas ratio variations.

\begin{figure}
\centering
\includegraphics{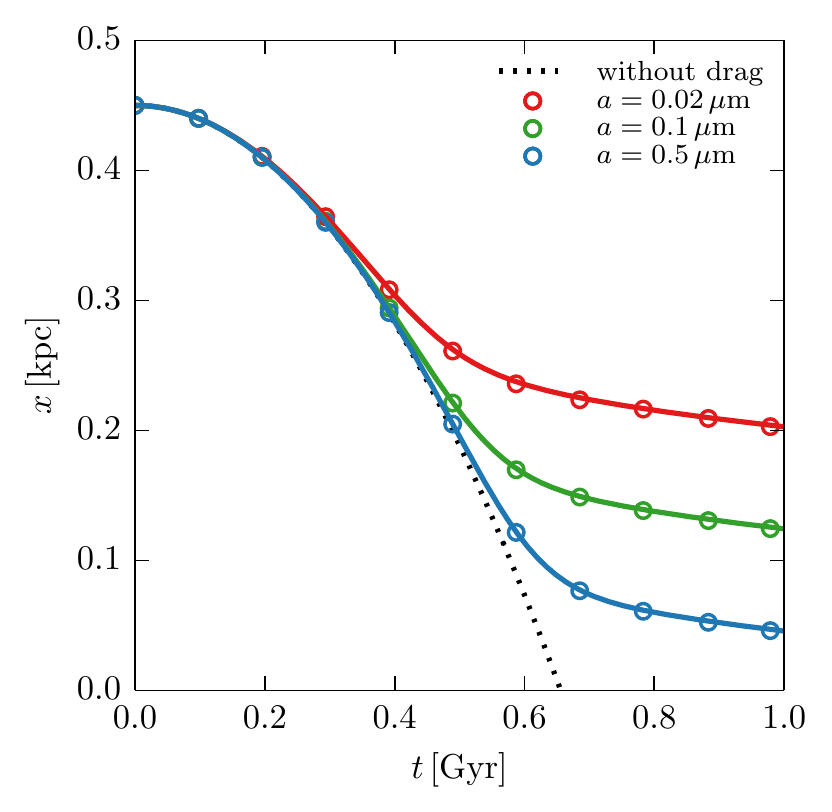}
\caption{Position as a function of time for dust grains starting at
rest in isothermal, hydrostatic gas with constant external gravity $\vec{g} =
-2 \, \vechat{x} \, \text{km}^{2} \, \text{kpc}^{-1} \, \text{s}^{-2}$.
Coloured circles show simulation results for grains of different radii, while
coloured lines show analytic predictions.  Grains initially follow the
parabolic profiles expected in the drag-free case (dotted black line), but
eventually the magnitude of the drag force is sufficient to slow grain motion.
Drag acts more quickly for small grains with shorter stopping time-scale.}
\label{FIG:dustysettle_evolution}
\end{figure}

\subsection{Dust falling through gas under gravitational acceleration}

Next we study the dynamics of dust grains subject to an external gravitational
acceleration in a gaseous medium in hydrostatic equilibrium
\citep[e.g.][]{Monaghan1997}.  We generate an equispaced lattice of $128^3$ gas
cells in a box of length $1 \, \text{kpc}$ centred on the origin and
apply an external gravitational acceleration pointing to the box midplane,
$\vec{g} = -2 \, \sgn(x) \, \vechat{x} \, \text{km}^{2} \,
\text{kpc}^{-1} \, \text{s}^{-2}$, where $\sgn$ is the sign function.  The gas
has adiabatic index $\gamma = 5/3$ and initial density profile $\rho(x)
= 10^{8} \exp(-|x| / h) \, \text{M}_\odot \, \text{kpc}^{-3}$, where $h = 0.05
\, \text{kpc}$ is a scale height.  We assume an isothermal gas, and the
choices for $\vec{g}$ and $\rho(x)$ above determine the gas temperature needed
for hydrostatic equilibrium.  Thus, the gas has a pressure distribution that is
also exponential and a uniform sound speed $c_\text{s} = \sqrt{\gamma
|g| h}$.

We place a dust particle at position $\vec{r} = 0.45 \, \vechat{x} \,
\text{kpc}$, such that gravity pushes the dust particle towards the box
centre.  The dust particle starts with zero initial velocity.  We assume a
fixed grain radius $a$, as described below, and an internal grain density
$\rho_\text{gr} = 2.4 \, \text{g} \, \text{cm}^{-3}$. Note that because the gas
density is not uniform, the stopping time-scale given by
equation~(\ref{EQN:t_s_full}) varies with position and is smallest near the box
centre, where the gas is most dense.  We include the velocity-dependent
correction factor in equation~(\ref{EQN:t_s_full}) in our test, although it
does not qualitatively impact our results.  Finally, we neglect self-gravity.

While the gas maintains its pressure gradient to counteract the external
gravity and remain at rest, the dust particle is accelerated by gravity and
begins to move.  However, as the dust velocity increases, so too does the
strength of the drag force opposing gravity.
Figure~\ref{FIG:dustysettle_evolution} shows the dust particle's position
versus time, for three different choices of grain radius $a$: $0.5$,
$1.0$, and $2.0 \, \mu\text{m}$.  Initially, the dust particle follows the
parabolic trajectory expected for drag-free motion in a uniform gravitational
field.  However, as the dust particle moves towards $x = 0 \,
\text{kpc}$, both its velocity and the local gas density increase.  This
results in a shorter stopping time-scale and thus a stronger drag acceleration.
Around $t \approx 0.5 \, \text{Gyr}$, the dust particle deviates from
the drag-free motion.  As expected, a smaller grain feels the effects of drag
more quickly, since stopping time-scale varies linearly with grain
radius.

In Figure~\ref{FIG:dustysettle_evolution}, we compare our simulations results
with predictions obtained by numerically integrating the dust particle's
position and velocity using a high-accuracy differential equations solver.  The
gravitational acceleration is constant, while the drag acceleration depends on
the dust velocity and stopping time-scale.  We compute the stopping time-scale
as a function of position using the analytic gas density profile.  Our
simulations agree well with these expected profiles.

\section{Grain size evolution}\label{SEC:grain_size_evolution}

\begin{figure*}
\centering
\includegraphics[width=0.95\textwidth]{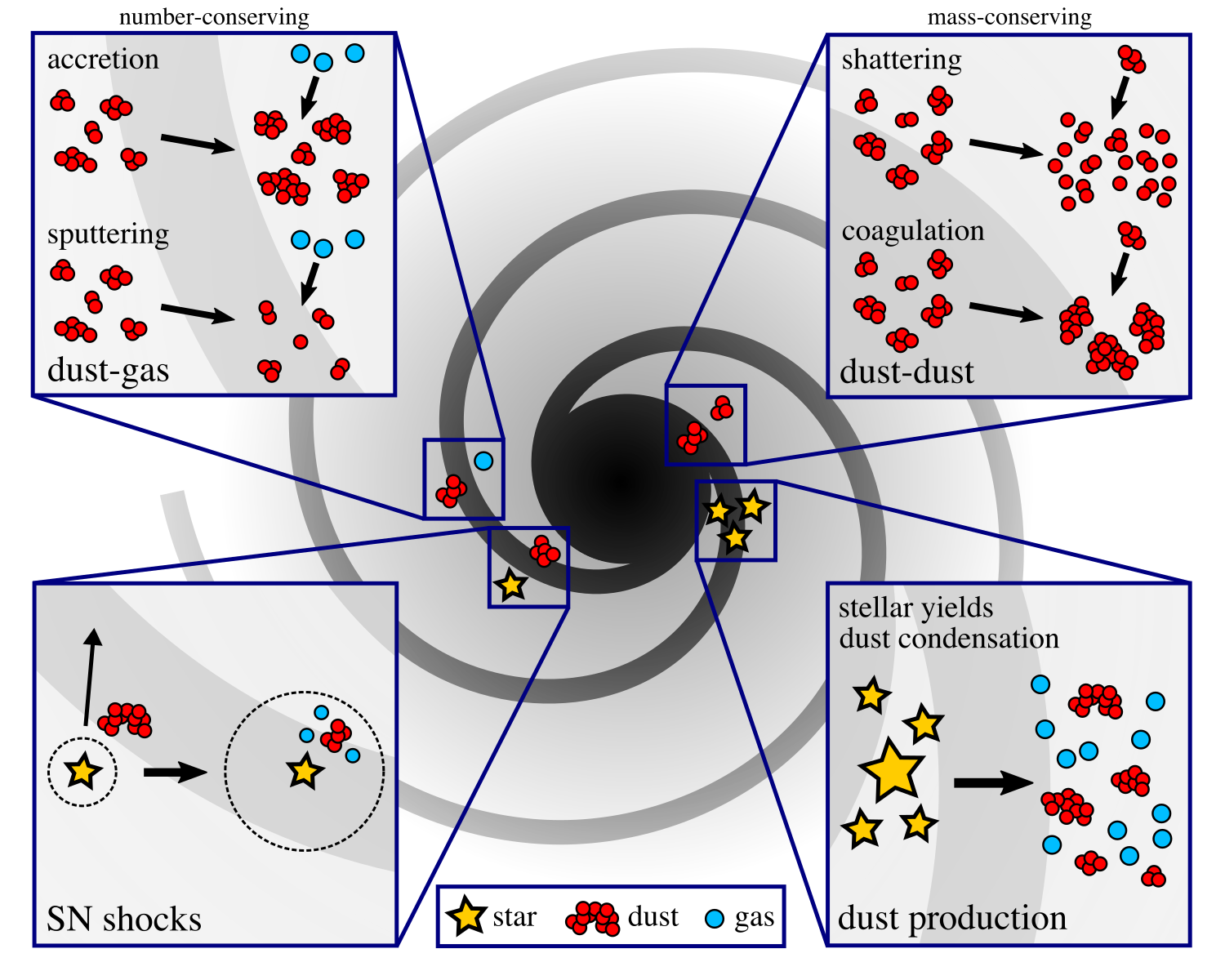}
\caption{Schematic illustration of the lifecycle of dust within a galaxy.
Graphics depict stars (yellow), dust grains (red), and gas-phase metals (blue)
in the ISM.  Dust grains are produced through stellar evolution, interact with
other dust grains and gas-phase metals through collisional processes, and can
be destroyed near SNe.  Collisional processes are divided into those that
conserve grain number (top left) and those that conserve grain mass (top
right).  Accretion and sputtering change total dust mass by growing or
shrinking individual grains, while shattering and coagulation preserve overall
mass but affect the number of grains.}
\label{FIG:galaxy_schematic}
\end{figure*}

Dust grains injected into the ISM by stars experience a range
of physical processes -- accretion, sputtering, shattering, and coagulation,
among others -- that affect their size distribution, as illustrated in
Figure~\ref{FIG:galaxy_schematic}.  In turn, the grain size distribution
affects the strength of dust-gas drag (e.g.~see Section~\ref{SEC:drag_force}),
interstellar extinction~\citep[e.g.][]{Mathis1977,
Weingartner2001}, and other processes like radiation pressure.  Thus, it is
important to properly model the evolution of the grain size distribution when
using a two-fluid approach.

Many theoretical and computational dust studies either evolve a grain size
distribution but track only total dust mass \citep{Liffman1989, ODonnell1997,
Hirashita2015}, or evolve dust masses for various chemical elements but assume
fixed grain radii \citep{Zhukovska2008, Bekki2015b, Popping2017, McKinnon2017}.
Here, we wish to do both.  However, to combine chemical element and grain size
distribution tracking without unwieldy complexity, we make several assumptions.

First, we distinguish between a dust grain (a single, physical object) and a dust
particle (an element of our simulation consisting of an ensemble of dust
grains).  In this work, we always assume grains are spherical, so that a grain
with radius $a > 0$ has mass
\begin{equation}
m(a) \equiv \frac{4 \pi \rho_\text{gr} a^3}{3}.
\label{EQN:m_a}
\end{equation}
To simplify notation later, define $m(a) \equiv 0$ for the unphysical case $a
\leq 0$.  In reality, dust grains have some degree of nonsphericity and
internal voids \citep{Mathis1998, Draine2003, Draine2009b}, but this spherical,
compact approximation is sufficient for our applications.

Second, we assume that a dust particle's grain size distribution is agnostic as
to the chemical composition of the grains.  That is, we do not have separate
grain size distributions for grains of different composition (e.g. SiO$_2$,
MgSiO$_3$, etc.).  This reduces computational complexity and also acknowledges
the limitations of our galaxy formation model~\citep{Vogelsberger2013}, which
tracks mass for nine chemical elements: H, He, C, N, O, Ne, Mg, Si, and Fe.
Because we track mass only for chemical elements as a whole and not
individual chemical compounds, it would not be feasible to assign different
grain size distributions to different grain types.  As in previous works
\citep{McKinnon2016, McKinnon2017}, only C, O, Mg, Si, and Fe are allowed to
condense into dust.

Although we do not track complex grain compositions, we do follow the mass of
individual chemical elements locked in dust.  When a dust particle is created,
we store what fraction of the total dust mass came from each chemical element.
The total dust mass can be calculated just from the grain size distribution.
When we add or subtract dust mass (e.g.~grain growth or sputtering), we keep
track of what masses of each element are being added from or returned to gas
cells, and update the dust mass fractions accordingly.  In this manner, the
total masses of individual chemical elements in gas and dust are conserved
during a time-step.  Thus, dust particles have one array of dust mass fractions
describing chemical composition and one array describing the overall grain size
distribution.

In what follows, we begin with a generic, analytical description of grain size
evolution.  Then, we describe the discretization used in our simulations and
the various physical processes that modify our grain size distribution.  Our
framework builds off of \citet{Dwek2008} and \citet{Hirashita2009}.
Conceptually, our methods handle two sorts of processes: those that conserve
grain number and those that conserve grain mass.

We first introduce methods to handle number-conserving processes that grow or
shrink the radii of individual grains.  A dust particle's grain size
distribution thus satisfies the continuity equation
\begin{equation}
\frac{\partial}{\partial t} \left[ \frac{\partial n(a, t)}{\partial a} \right] + \frac{\partial}{\partial a} \left( \frac{\partial n(a, t)}{\partial a} \times \frac{\diff a}{\diff t} \right) = 0,
\label{EQN:continuity_equation}
\end{equation}
where $\partial n(a, t) / \partial a \times \diff a$ is the number of
grains with radii in the interval $[a, a + \diff a]$ at time $t$ for a given
dust particle. This differs from the hydrodynamical continuity equation
because the ``velocity'' term $\mathrm{d}a / \mathrm{d}t$ for the grain size
distribution may be independent of $a$ and only a function of gas quantities
(see discussion of grain growth and thermal sputtering in
Sections~\ref{SEC:grain_growth} and~\ref{SEC:thermal_sputtering},
respectively).  Thus, unlike the hydrodynamical case where changes in density
lead to changes in velocity, shifting the grain size distribution to smaller or
larger radii does not directly affect $\mathrm{d}a / \mathrm{d}t$.  In the
limit where $\mathrm{d}a / \mathrm{d}t$ is constant (e.g.~small dust-to-gas
ratios where the accretion of dust does not materially affect gas
metallicities), the grain size distribution would simply obey the solution
$\partial n(a, t + \Delta t) / \partial a = \partial n(a - \dot{a} \Delta t, t)
/ \partial a$.  In practice, although $\mathrm{d}a / \mathrm{d}t$ may not
explicitly depend on grain size, shifts in the grain size distribution lead to
changes in dust and metal mass, which in turn can affect gas properties like
metallicity and temperature.  Thus, $\mathrm{d}a / \mathrm{d}t$ evolves as the
gas evolves, and we develop methods to discretise this problem.

Second, we address mass-conserving processes like shattering and coagulation in
a framework that accounts for grain-grain collisions.  These processes do not
conserve grain number (i.e.~shattering one large grain produces many smaller
grains) and do not involve mass transfer to or from gas cells.  The underlying
physics shares similarities to a wide class of population balance equations
\citep{Smoluchowski1916, Vigil1989, Dubovskii1992}.

Our methods below discretise the grain size distribution into $N$ bins in a
general way.  The $N = 1$ case models a fixed grain size, where
changes in dust mass result only from changes in number of grains, not changes
in grain radii.  The $N = 2$ case is similar to the simplified two-size grain
distribution used in recent works \citep{Hirashita2015b, Hou2016, Hou2017,
Chen2018}.

\subsection{Analytic formulation}\label{SEC:analytic_formulation}

We assume that grains can have radii in the interval $I_\text{full} \equiv
[a_\text{min}, a_\text{max}]$.  Define a differential grain size distribution
$\partial n(a, t) / \partial a$ over $I_\text{full}$ such that $\partial n(a,
t) / \partial a \times \diff a$ denotes the number of grains with radii in the
range $[a, a + \diff a]$ at time $t$.

Because we will later discretise this formulation, partition $I_\text{full}$
into $N$ bins with edges $(a^\text{e}_{0}, a^\text{e}_{1}, \dots,
a^\text{e}_{N})$, where $a^\text{e}_{0} \equiv a_\text{min}$ and
$a^\text{e}_{N} \equiv a_\text{max}$.  At this point, we do not make any
assumptions about the spacing of these bins.  Bin $i$ covers the interval
$I_{i} \equiv [a^\text{e}_{i}, a^\text{e}_{i+1}]$ with midpoint
\begin{equation}
a^\text{c}_{i} \equiv \frac{a^\text{e}_{i} + a^\text{e}_{i+1}}{2}.
\end{equation}
We write the number of grains in bin $i$ at time $t$ as
\begin{equation}
N_{i}(t) \equiv \int_{I_i} \frac{\partial n(a, t)}{\partial a} \diff a,
\label{EQN:N_i_t}
\end{equation}
and their mass as
\begin{equation}
M_{i}(t) \equiv \int_{I_i} m(a) \frac{\partial n(a, t)}{\partial a} \diff a.
\label{EQN:M_i_t}
\end{equation}

We discuss in later sections how various physical processes affect the
time-evolution of grain radius.  For now we assume that we have a known form of
$\dot{a}(a, t) \equiv \mathrm{d}a / \mathrm{d}t$.  This may in principle be a
function of radius and time (the latter because, e.g., if grain radius is
changing through collisions with gas atoms, gas properties like density and
temperature may evolve in time).

We next consider the time evolution by a small time-step $\Delta t$.  We can
rewrite the number of grains in bin $j$ at time $t + \Delta t$ as the number of
grains in any bin at time $t$ that evolve over the time-step to lie in bin $j$,
using
\begin{equation}
\begin{split}
N_{j}(t + \Delta t) &= \int_{I_{j}} \frac{\partial n(a, t + \Delta t)}{\partial a} \diff a \\
&= \int_{I_\text{full}} \mathbbm{1}_{j}(a, \dot{a}, t) \frac{\partial n(a, t)}{\partial a} \diff a,
\end{split}
\end{equation}
where the indicator function is
\begin{equation}
\mathbbm{1}_{j}(a, \dot{a}, t) \equiv
\begin{cases}
1, \quad \text{if} \, a + \dot{a}(a, t) \Delta t \in I_{j}, \\
0, \quad \text{else}.
\end{cases}
\end{equation}
Using the partition of $I_\text{full}$,
\begin{equation}
N_{j}(t + \Delta t) = \sum_{i = 0}^{N-1} \int_{I_{i}} \mathbbm{1}_{j}(a, \dot{a}, t) \frac{\partial n(a, t)}{\partial a} \diff a.
\label{EQN:N_j_partition}
\end{equation}
In general, the form of $\dot{a}(a, t)$ determines where the integrands are
non-zero.  If $\dot{a}(a, t) = \dot{a}(t)$, suitable for collisional processes
like grain accretion \citep[e.g.][]{Hirashita2011} or thermal sputtering
\citep[e.g.][]{Draine1979b}, equation~(\ref{EQN:N_j_partition}) can be
simplified as
\begin{equation}
N_{j}(t + \Delta t) = \sum_{i = 0}^{N-1} \int_{I_{i} \cap (I_{j} - \dot{a} \Delta t)} \frac{\partial n(a, t)}{\partial a} \diff a,
\label{EQN:N_j_intersection}
\end{equation}
where we use the shorthand $I_{j} - \dot{a} \Delta t \equiv [a^\text{e}_j -
\dot{a}(t) \Delta t, a^\text{e}_{j+1} - \dot{a}(t) \Delta t]$ to indicate the
range of grain radii at time $t$ that later evolve to fall in bin $j$ at time
$t + \Delta t$.  This expresses the number of grains in each bin at time $t +
\Delta t$ as a summation of integrals of the time $t$ grain size distribution
over overlapping intervals.  In many cases, $I_{i} \cap (I_{j} - \dot{a} \Delta
t)$ may trivially be the empty set: for example, in handling grain growth with
$\dot{a} > 0$, this overlap is non-empty only for $i \leq j$ since grains in
bins $j+1$ and above will not shrink.

To this point, we have neglected boundary conditions that enforce $a_\text{min}
\leq a \leq a_\text{max}$ in the grain size distribution.  However, grains may
erode or grow such that $a + \dot{a} \Delta t < a_\text{min}$ or $a + \dot{a}
\Delta t > a_\text{max}$ and thus require rebinning.  For notational
convenience, we define ``bin $-1$'' and ``bin $N$'' as the intervals
$I_{-1} \equiv (-\infty, a_\text{min}]$ and $I_{N} \equiv [a_\text{max}, \infty)$,
respectively.  With these definitions, equation~(\ref{EQN:N_j_intersection})
can be extended to bins $-1$ and $N$, where $N_{-1}(t + \Delta t)$ and $N_{N}(t
+ \Delta t)$ represent the number of grains whose radius evolves below
$a_\text{min}$ or above $a_\text{max}$, respectively.  This formulation
conserves total grain number, i.e.~$N(t + \Delta t) = N(t)$.

While total grain number is conserved, total mass evolves.  Paralleling
equation~(\ref{EQN:N_j_intersection}), the mass in bin $j$ at time $t + \Delta
t$ is given by
\begin{equation}
M_{j}(t + \Delta t) \equiv \sum_{i = 0}^{N-1} \int_{I_{i} \cap (I_{j} - \dot{a} \Delta t)} m(a + \dot{a} \Delta t) \frac{\partial n(a, t)}{\partial a} \diff a,
\label{EQN:M_j_intersection}
\end{equation}
where integrals are over the time $t$ grain size distribution but use the mass
$m(a + \dot{a} \Delta t)$ to account for mass at time $t + \Delta t$.  Using
the definitions of $I_{-1}$ and $I_{N}$ above and $m(a + \dot{a} \Delta t)
\equiv 0$ for $a + \dot{a} \Delta t \leq 0$,
equation~(\ref{EQN:M_j_intersection}) is valid for $-1 \leq j \leq N$.  We note
that that if $\dot{a} < 0$ (e.g.~thermal sputtering) and $\Delta t \to \infty$,
$m(a + \dot{a} \Delta t) \to 0$, implying that all grain mass is destroyed.

An overall grain size distribution update from time $t$ to $t + \Delta t$ takes
place as follows.  First, the numbers of grains in bins $0, 1, \dots, N-1$ at
time $t + \Delta t$ are updated using equation~(\ref{EQN:N_j_intersection}) and
the time $t$ grain size distribution.  Then, $M_{j}(t + \Delta t)$ is
calculated using equation~(\ref{EQN:M_j_intersection}) for $-1 \leq j \leq N$.
The change in mass $\Delta m_\text{d} \equiv m_\text{d}(t + \Delta t) -
m_\text{d}(t)$ for the dust particle over this time-step is
\begin{equation}
\Delta m_\text{d} = \sum_{j=-1}^{N} M_{j}(t+\Delta t) - \sum_{j=0}^{N-1} M_{j}(t).
\end{equation}
Our rebinning procedure places mass $M_{-1}(t + \Delta t)$ back into bin $0$
and mass $M_{N}(t + \Delta t)$ into bin $N-1$.  This rebinning process
conserves the grain mass calculated at time $t + \Delta t$ (and thus $\Delta
m_\text{d}$) but does not conserve total grain number.  For example, if
$\dot{a} > 0$ and $M_{N}(t + \Delta t) > 0$, rebinning will cause the number of
grains to increase since grains in bin $N-1$ are less massive than those in bin
$N$.  In the case of a continuous grain size distribution, there are various
ways the grain size distributions in bins $0$ and $N-1$ can be modified to
increase the bin mass.  In the following section, we describe how to discretise
the grain size distribution using a piecewise linear approximation.

\subsection{Discrete formulation}\label{SEC:piecewise_linear}

\subsubsection{Evolution of dust mass between grain size bins}

\begin{figure*}
\centering
\includegraphics{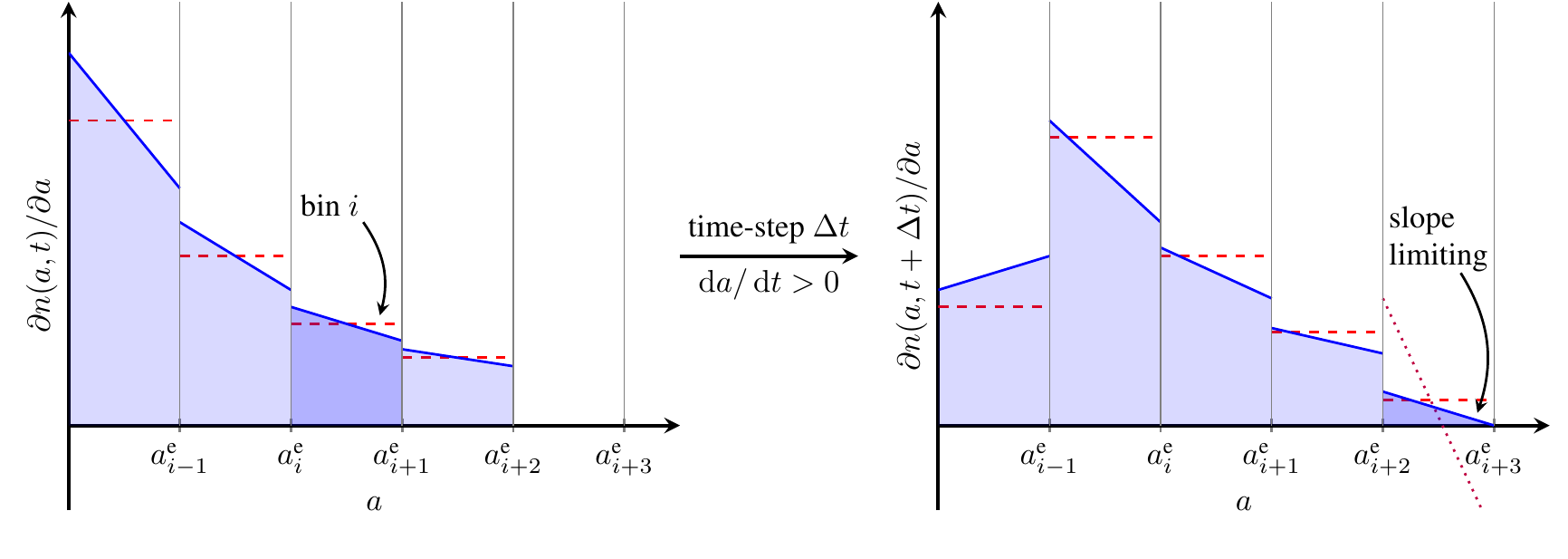}
\caption{Schematic illustration of a piecewise linear grain size distribution
evolving from time $t$ (left) to $t + \Delta t$ (right).  We assume that
$\mathrm{d}a / \mathrm{d}t > 0$ during the time-step, although the opposite
case is similar.  Solid blue lines mark the piecewise linear discretisation,
with the shaded region in each bin giving the number of grains.  Dashed red
lines show a possible piecewise constant discretisation.  At time $t$, we
assume the piecewise linear and constant methods yield the same number of
grains in a given bin (i.e.~dashed red lines pass through the midpoints of the
solid blue lines).  By the end of the time-step, this property is not
maintained (e.g.~here, the piecewise constant method overestimates the number
of grains leaving the leftmost bin).  At time $t + \Delta t$, the rightmost bin
is subject to slope limiting: if the mass and number of grains entering this
bin yield a grain size distribution that drops below zero (dotted purple line),
the slope is limited to remove this unphysical behaviour.  Slope limiting
preserves a bin's total grain mass.  To improve readability, this figure adopts
linearly-spaced bins.  In practice, the formulation outlined in
Section~\ref{SEC:piecewise_linear} uses log-spaced bins.}
\label{FIG:dnda_schematic}
\end{figure*}

Following \citet{Hirashita2009}, we discretise the grain size distribution into
$N$ log-spaced bins in the following manner.  Using the minimum and maximum
grain sizes $a_\text{min}$ and $a_\text{max}$, respectively, we define the
logarithmic bin width
\begin{equation}
\log \delta \equiv \frac{\log{a_\text{max}} - \log{a_\text{min}}}{N}.
\label{EQN:bin_delta}
\end{equation}
The edges of the $N$ bins are then $(a^\text{e}_{0}, a^\text{e}_{1}, \dots,
a^\text{e}_{N})$, where $a^\text{e}_{i} \equiv \delta^{i} a_\text{min}$.
This specifies the exact partition of $I_\text{full}$ that we use in the
formulation from Section~\ref{SEC:analytic_formulation}.

We then assume that the differential grain size distribution in bin $i$ at time
$t$ takes the linear form
\begin{equation}
\frac{\partial n(a, t)}{\partial a} = \frac{N_i(t)}{a^\text{e}_{i+1} - a^\text{e}_{i}} + s_i(t) (a - a^\text{c}_i),
\label{EQN:dnda_linear}
\end{equation}
where $a^\text{c}_i$ is the midpoint of the bin and $s_i(t)$ denotes the slope.
We note that the number of grains in bin $i$ is determined only by the first
term, since the second term integrates to zero over the bin's interval.  The
piecewise linear grain size distribution at time $t$ is fully determined by the
set of $N_i(t)$ and $s_i(t)$ values for all bins.
Figure~\ref{FIG:dnda_schematic} shows a schematic of this discretisation and
its evolution in time, which is described below in detail.

Discretising equation~(\ref{EQN:N_j_intersection}), the number of grains in bin
$j$ at time $t + \Delta t$ is
\begin{equation}
N_{j}(t + \Delta t) = \sum_{i = 0}^{N-1} \int_{I_{i,j}} \left( \frac{N_i(t)}{a^\text{e}_{i+1} - a^\text{e}_{i}} + s_i(t)(a - a^\text{c}_i) \right) \diff a,
\label{EQN:N_j_overlap_linear}
\end{equation}
where $I_{i,j} \equiv I_i \cap (I_j - \dot{a} \Delta t)$ denotes the portion of
bin $i$ that ends up in bin $j$ after the time-step.  To help determine whether
the intersection of these two intervals $I_i$ and $I_j - \dot{a} \Delta t$ is
non-empty, we first set $x_1(i, j) \equiv \max(a^\text{e}_{i}, a^\text{e}_{j} -
\dot{a} \Delta t)$, the maximum of the intervals' left edges, and $x_2(i, j)
\equiv \min(a^\text{e}_{i+1}, a^\text{e}_{j+1} - \dot{a} \Delta t)$, the
minimum of the intervals' right edges.  Then, $I_i \cap (I_j - \dot{a} \Delta
t) \neq \emptyset$ if and only if $x_2(i, j) \geq x_1(i, j)$, in which case the
intersection interval is $[x_1(i, j), x_2(i, j)]$.  We define the indicator function
\begin{equation}
\mathbbm{1}_{x_2 \geq x_1}(i, j) =
\begin{cases}
1, \quad \text{if} \, x_2(i, j) \geq x_1(i, j), \\
0, \quad \text{else},
\end{cases}
\end{equation}
which is unity when any portion of bin $i$ evolves into bin $j$ over the time-step.
To improve readability below, we will often label $x_1(i, j)$ and $x_2(i, j)$
without their implied arguments $i$ and $j$.  Simplifying
equation~(\ref{EQN:N_j_overlap_linear}) yields
\begin{equation}
\begin{split}
&N_{j}(t + \Delta t) \\
&\quad = \sum_{i = 0}^{N-1} \mathbbm{1}_{x_2 \geq x_1}(i, j) \int_{x_1}^{x_2} \left( \frac{N_i(t)}{a^\text{e}_{i+1} - a^\text{e}_{i}} + s_i(t)(a - a^\text{c}_i) \right) \diff a \\
&\quad = \sum_{i = 0}^{N-1} \mathbbm{1}_{x_2 \geq x_1}(i, j) \left[ \frac{N_i(t) a}{a^\text{e}_{i+1} - a^\text{e}_{i}} + s_i(t) \left( \frac{a^2}{2} - a^\text{c}_i a \right) \right]^{a = x_2}_{a = x_1}.
\end{split}
\label{EQN:N_j_update_linear}
\end{equation}
This reduces the calculation of the number of grains in bin $j$ at time $t +
\Delta t$ to a sum over factors involving the time $t$ grain size distribution.
Similarly, the mass in bin $j$ at time $t + \Delta t$ comes from discretising
equation~(\ref{EQN:M_j_intersection}) as
\begin{equation}
\begin{split}
M_{j}(t + \Delta t) &= \sum_{i = 0}^{N-1} \mathbbm{1}_{x_2 \geq x_1}(i, j) M_{i \to j}(t, \Delta t),
\end{split}
\end{equation}
where
\begin{equation}
M_{i \to j} \equiv \int_{x_1}^{x_2} m(a + \dot{a} \Delta t) \left( \frac{N_i(t)}{a^\text{e}_{i+1} - a^\text{e}_{i}} + s_i(t)(a - a^\text{c}_i) \right) \diff a,
\end{equation}
denoting mass transfer from bin $i$ to $j$.  Then,
\begin{equation}
\begin{split}
M_{j}(t + \Delta t) &= \frac{4 \pi \rho_\text{gr}}{3} \sum_{i = 0}^{N-1} \mathbbm{1}_{x_2 \geq x_1}(i, j) \bigg[ \frac{N_i(t) (a + \dot{a} \Delta t)^4}{4(a^\text{e}_{i+1} - a^\text{e}_{i})}  \\
& \qquad \qquad \qquad \quad + s_i(t) f^M_i(a, \dot{a}, \Delta t) \bigg]^{a = x_2}_{a = x_1},
\end{split}
\label{EQN:M_j_update_linear}
\end{equation}
where we use equation~(\ref{EQN:m_a}) to evaluate $m(a + \dot{a} \Delta t)$ and
define
\begin{equation}
\begin{split}
f^M_i(a, \dot{a}, \Delta t) &\equiv \frac{a^5}{5} + (3 \dot{a} \Delta t - a^\text{c}_i) \frac{a^4}{4} + \dot{a} \Delta t (\dot{a} \Delta t - a^\text{c}_i) a^3 \\
&\quad + (\dot{a} \Delta t)^2 (\dot{a} \Delta t - 3 a^\text{c}_i) \frac{a^2}{2} - \dot{a}^3 \Delta t^3 a^\text{c}_i a.
\end{split}
\end{equation}
Equation~(\ref{EQN:M_j_update_linear}) also holds for the two boundary bins
with $j = -1$ and $j = N$, although the case $j = -1$ requires a small
modification.  Since bin $-1$ covers the interval $I_{-1} = (-\infty,
a_\text{min}]$ and $m(a) = 0$ for $a \leq 0$, we need to ensure we only
integrate over grain sizes $a$ with $a + \dot{a} \Delta t > 0$.  To do this,
define $a^* \equiv -\dot{a} \Delta t$ so that $a > a^*$ implies $a + \dot{a}
\Delta t > 0$.  Then, for the boundary bin $j = -1$ only, modify the integrals
in equation~(\ref{EQN:M_j_update_linear}) to be over the intervals $[x_1, x_2]
\cap [a^*, \infty)$.

Alternatively, if the number of grains $N_j(t +
\Delta t)$ and slope $s_j(t + \Delta t)$ are known, the mass in bin $j$ at time
$t + \Delta t$ can be expressed as
\begin{equation}
\begin{split}
&M_j(t + \Delta t) \\
&\enskip = \int^{a^\text{e}_{j+1}}_{a^\text{e}_j} \frac{4 \pi \rho_\text{gr} a^3}{3} \left( \frac{N_j(t+\Delta t)}{a^\text{e}_{j+1} - a^\text{e}_{j}} + s_j(t+\Delta t)(a - a^\text{c}_j) \right) \diff a \\
&\enskip = \frac{4 \pi \rho_\text{gr}}{3} \left[ \frac{N_j(t+\Delta t) a^4}{4(a^\text{e}_{j+1} - a^\text{e}_{j})} + s_j(t+\Delta t) \left(\frac{a^5}{5} - \frac{a^\text{c}_j a^4}{4} \right) \right]^{a^\text{e}_{j+1}}_{a^\text{e}_j}.
\end{split}
\label{EQN:M_j_slope_linear}
\end{equation}
One can think of $M_j(t + \Delta t)$ not as an explicit function of time but as
a function of $N_j(t + \Delta t)$ and $s_j(t + \Delta t)$.  We summarise how to
update the grain size distribution in bin $j$ from $t$ to $t + \Delta t$.
First, apply equations~(\ref{EQN:N_j_update_linear})
and~(\ref{EQN:M_j_update_linear}) to the grain size distribution at time $t$ to
calculate the number and mass of grains at time $t + \Delta t$.  Then, use
equation~(\ref{EQN:M_j_slope_linear}) to solve for the slope in bin $j$, $s_j(t
+ \Delta t)$.  This choice of slope ensures bin $j$ has the expected mass of
grains.

However, this procedure may result in a slope $s_j(t + \Delta t)$ whose
magnitude is so large that the grain size distribution becomes negative at one
of the edges of bin $j$.  Since this is unphysical, we introduce the following
slope limiting step.  We therefore calculate
\begin{equation}
\left. \frac{\partial n(a, t+\Delta t)}{\partial a}\right|_{a^\text{e}_j} \equiv \frac{N_j(t + \Delta t)}{a^\text{e}_{j+1} - a^\text{e}_j} + s_j(t + \Delta t) (a^\text{e}_j - a^\text{c}_j),
\end{equation}
and
\begin{equation}
\left. \frac{\partial n(a, t+\Delta t)}{\partial a}\right|_{a^\text{e}_{j+1}} \equiv \frac{N_j(t + \Delta t)}{a^\text{e}_{j+1} - a^\text{e}_j} + s_j(t + \Delta t) (a^\text{e}_{j+1} - a^\text{c}_j).
\end{equation}
If both of these values are non-negative, no slope limiting is necessary.
Furthermore, since the grain size distribution is piecewise linear and the
number of grains $N_j(t + \Delta t) > 0$, at most one of these values could be
negative.  Without loss of generality, we assume $\partial n(a, t+\Delta t) /
\partial a |_{a=a^\text{e}_{j+1}} < 0$, so that $s_j(t + \Delta t) < 0$.  Let
$M_j(t + \Delta t)$ be the mass in bin $j$ computed using
equation~(\ref{EQN:M_j_update_linear}).
We will find a new number of grains $\tilde{N}_j(t + \Delta t)$ and slope
$\tilde{s}_j(t + \Delta t)$ so that the grain size distribution at edge
$a^\text{e}_{j+1}$ is zero (thus ensuring the grain size distribution is
non-negative everywhere in bin $j$), while keeping the mass in bin $j$ is
unchanged.  To do this, we use equation~(\ref{EQN:M_j_slope_linear}) and the
unlimited $N_j(t + \Delta t)$ and $s_j(t + \Delta t)$ values to simultaneously
solve the linear system
\begin{equation}
M_j(\tilde{N}_j(t + \Delta t), \tilde{s}_j(t + \Delta t)) = M_j(N_j(t + \Delta t), s_j(t + \Delta t)),
\end{equation}
and
\begin{equation}
\frac{\tilde{N}_j(t + \Delta t)}{a^\text{e}_{j+1} - a^\text{e}_j} + \tilde{s}_j(t + \Delta t)(a^\text{e}_{j+1} - a^\text{c}_j) = 0,
\end{equation}
where the unknowns are $\tilde{N}_j(t + \Delta t)$ and $\tilde{s}_j(t + \Delta
t)$.  This procedure keeps the slope negative but
limits its magnitude.  Flattening the bin's slope causes the number of grains
in the bin to drop, since the average grain mass increases and mass is
conserved.  We employ a similar procedure when $\partial n(a, t+\Delta t) /
\partial a |_{a=a^\text{e}_{j}} < 0$, an alternative case that causes the
number of grains to increase as the positive slope is flattened.  In both
cases, this slope limiting preserves the mass in the bin, at the cost of
changing the number of grains away from the value predicted by
equation~(\ref{EQN:N_j_update_linear}).  Afterwards, we omit the tildes and
assume that $N_j(t + \Delta t)$ and $s_j(t + \Delta t)$ refer to the possibly
slope limited values in bin $j$.

\subsubsection{Rebinning dust mass to obey grain size limits}\label{SEC:rebinning}

In order to complete the time-step update, we need to address grains
whose radii grow above $a_\text{max}$ or shrink below $a_\text{min}$.
There are several approaches one could take.  In this work, we move
grains that evolve beyond the allowed size limits back into the closest grain
size bin in a mass-conserving manner.  Alternatively, we could assume that
grains whose radii evolve below $a_\text{min}$ are destroyed and set their mass
to zero.  However, for the galaxy simulations presented in
Section~\ref{SEC:hernquist_spheres}, we have found that these two approaches
yield similar results.

Below, we describe our procedure for rebinning grains that become too
large or too small.  Our steps are given for bin $N-1$, which contains the
largest grains.  The steps for bin $0$ are similar.  As in the case of slope
limiting, tildes indicate quantities after rebinning.

Before any rebinning, the average grain size in bin $N-1$ is
\begin{equation}
\begin{split}
&\langle a \rangle_{N-1}(t + \Delta t) \\
&\quad = \frac{1}{N_{N-1}} \int_{a^\text{e}_{N-1}}^{a^\text{e}_{N}} a \left( \frac{N_{N-1}}{a^\text{e}_{N} - a^\text{e}_{N-1}} + s_{N-1}(a - a^\text{c}_{N-1}) \right) \diff a \\
&\quad = \left[ \frac{a^2 / 2}{a^\text{e}_N - a^\text{e}_{N-1}} + \frac{s_{N-1}}{N_{N-1}} \left( \frac{a^3}{3} - \frac{a^\text{c}_{N-1} a^2}{2} \right) \right]^{a = a^\text{e}_N}_{a = a^\text{e}_{N-1}},
\end{split}
\label{EQN:avg_a_slope_linear}
\end{equation}
where on the right we drop the arguments of $N_{N-1}(t + \Delta t)$ and
$s_{N-1}(t + \Delta t)$ for brevity.  The mass $M_N(t + \Delta t)$ to be added
to bin $N-1$ consists of grains with radii larger than $a^\text{e}_N$, the
maximum radius allowed in bin $N-1$.  During rebinning, let us suppose we
shrink these grains to have radius $a^\text{e}_N$, so that
$N^\text{rebin}_{N-1}(t + \Delta t) = M_N(t + \Delta t) / (4 \pi \rho_\text{gr}
{a^\text{e}_N}^3 / 3)$ denotes the equivalent number of grains.  Then, by
rebinning this excess mass at the maximum possible radius, the average grain
size in bin $N-1$ increases to
\begin{equation}
\langle \tilde{a} \rangle_{N-1}(t + \Delta t) = \frac{N_{N-1} \langle a \rangle_{N-1} + N^\text{rebin}_{N-1} a^\text{e}_N}{N_{N-1} + N^\text{rebin}_{N-1}},
\label{EQN:avg_a_after_rebin}
\end{equation}
where for readability we omit the argument $t + \Delta t$ in quantities on the
right.  We note that we can also rewrite equation~(\ref{EQN:avg_a_slope_linear})
to express the average grain size after rebinning in terms of unknowns
$\tilde{N}_{N-1}(t + \Delta t)$ and $\tilde{s}_{N-1}(t + \Delta t)$ that
characterise the grain size distribution in bin $N-1$ after rebinning.  As in
the case of slope limiting, we enforce mass conservation, so that
\begin{equation}
\begin{split}
&M_{N-1}(\tilde{N}_{N-1}(t + \Delta t), \tilde{s}_{N-1}(t + \Delta t)) \\
&\quad = M_{N-1}(N_{N-1}(t + \Delta t), s_{N-1}(t + \Delta t)) + M_N(t + \Delta t),
\end{split}
\label{EQN:mass_after_rebin}
\end{equation}
where $M_{N-1}$ is computed using equation~(\ref{EQN:M_j_slope_linear})
and $M_{N}$ using equation~(\ref{EQN:M_j_update_linear}).  We perform the
rebinning step by simultaneously solving for $\tilde{N}_{N-1}(t + \Delta t)$
and $\tilde{s}_{N-1}(t + \Delta t)$ from
equations~(\ref{EQN:avg_a_after_rebin}) and~(\ref{EQN:mass_after_rebin}), which
can be expressed as a linear system.  This ensures that rebinning conserves
mass and places rebinned grains at the largest possible grain radius.  If
necessary, we slope limit bin $N-1$ after rebinning.  The procedure for bin
$0$ is essentially identical, with grains that evolve below the minimum grain
radius $a^\text{e}_0$ shifted back to this edge.

This converts the continuous grain size distribution framework from
Section~\ref{SEC:analytic_formulation} into a piecewise linear framework.  In
some of the tests below, we also simulate a piecewise constant grain size
distribution by forcing the slope in every bin to be zero.  This considerably
simplifies the number and mass updates in
equations~(\ref{EQN:N_j_update_linear}) and~(\ref{EQN:M_j_slope_linear}) and
alleviates the need for slope limiting.  We rebin boundary mass during a
time-step by adding $M_N(t + \Delta t) / \langle m \rangle_{N-1}$ grains to bin
$N-1$ and $M_{-1}(t + \Delta t) / \langle m \rangle_{0}$ grains to bin $0$,
where $\langle m \rangle_j$ is the average mass of a grain in bin $j$ and is
completely specified only by the edges of bins.

\subsubsection{Transfer of mass between gas and dust}

One additional complexity to discuss is the transfer of mass between gas and
dust.  Let us assume that, during a time-step, changes in the grain size
distribution cause a dust particle to change in mass by $\Delta m_\text{d}$.
We carry out this mass transfer over $N_\text{ngb}$ neighbouring gas cells in a
kernel-weighted fashion, similar to equation~(\ref{EQN:kernel_weighting}).
If $\Delta m_\text{d} > 0$, the dust particle expects to gain mass from gas
cells, and there is a risk that those cells do not contain enough metals.  We
discuss this complication later.

As discussed at the start of Section~\ref{SEC:grain_size_evolution}, each dust
particle and gas cell tracks what fraction $f_k$ of its mass comes from each
chemical element $k$.  If $\Delta m_\text{d} < 0$ and dust mass is being
returned to gas cells, we choose to keep these dust mass fractions constant.
For example, if a dust particle with mass fractions $f_k$ is set to transfer
mass $w_i \Delta m_\text{d}$ to gas cell $i$ for some weight $w_i$, the gas
cell gains mass $f_k w_i \Delta m_\text{d}$ in species $k$.  Similarly, if
$\Delta m_\text{d} > 0$ and dust mass is being accreted from gas cells, we
choose to keep constant the relative ratios of gas cell mass fractions
corresponding to \emph{those chemical elements which can condense onto dust
grains}.  We reiterate that only C, O, Mg, Si, and Fe can contribute to dust
grains in our model.  Using the notation above, suppose gas cell $i$ has mass
fractions $f_k$, and define $f_\text{sum} \equiv f_\text{C} + f_\text{O} +
f_\text{Mg} + f_\text{Si} + f_\text{Fe} \leq 1$.  Let $\hat{f}_k \equiv f_k /
f_\text{sum}$ for $k \in \{\text{C}, \text{O}, \text{Mg}, \text{Si},
\text{Fe}\}$.  Then, the gas cell loses mass $\hat{f}_k w_i \Delta m_\text{d}$
in each of these five elements.  Using these accreted masses, the dust
particle's normalised mass fractions for each chemical element are
recalculated.  Regardless of the sign of $\Delta m_\text{d}$, the mass
fractions in affected gas cells are also recomputed.

This procedure assumes that gas cells always have enough metals for dust
particles to accrete in a time-step.  However, this may not be the case,
particularly if a dust particle has already accreted many nearby metals and
surrounding gas cells have low or zero metal mass.  To account for this, we
break the dust particle update into two steps.  First, we perform the grain
size distribution calculations above to determine the new number of grains
$N_j(t + \Delta t)$ and slope $s_j(t + \Delta t)$ in each bin $j$, assuming
surrounding gas cells have enough metals to accrete the expected mass $\Delta
m_\text{d}^\text{exp}$ over the time-step.  When performing the mass transfer
from gas cells to the dust particle, we keep track of the actual metal mass
$\Delta m_\text{d}^\text{act}$ that gas cells are able to transfer.  A gas
cell $i$ with kernel weight $w_i$ transfers the minimum of $w_i \Delta
m_\text{d}^\text{exp}$ and its available metal mass, so that summing over
nearby gas cells gives $\Delta m_\text{d}^\text{act} \leq \Delta
m_\text{d}^\text{exp}$.  As the second step, once mass transfer is complete,
we perform the grain size distribution's time-step update by setting the number
of grains in bin $j$ at time $t + \Delta t$ to be $N_j(t) + \Delta
m_\text{d}^\text{act} / \Delta m_\text{d}^\text{exp} \times (N_j(t + \Delta t)
- N_j(t))$ and the slope to be $s_j(t) + \Delta m_\text{d}^\text{act} / \Delta
m_\text{d}^\text{exp} \times (s_j(t + \Delta t) - s_j(t))$.  This approach
ensures that the change in dust particle mass equals $\Delta
m_\text{d}^\text{act}$, the actual amount of accreted metals.  The case of dust
mass loss is much simpler: we are always able to transfer all desired mass back
to nearby gas cells (i.e.~$\Delta m_\text{d}^\text{act} = \Delta
m_\text{d}^\text{exp}$), and so no special handling is needed.

When transferring mass between gas cells and dust particles, we also
update other conserved quantities like momentum.  When a dust particle of mass
$m_\text{d}$ and velocity $\vec{v}_\text{d}$ transfers mass $\Delta m_\text{d}$
to a surrounding gas cell of mass $m_\text{g}$ and velocity $\vec{v}_\text{g}$,
the dust particle and gas cell's momenta are updated to $m_\text{d}
\vec{v}_\text{d} - \Delta m_\text{d} \vec{v}_\text{d}$ and $m_\text{g}
\vec{v}_\text{g} + \Delta m_\text{d} \vec{v}_\text{d}$, respectively.  We
employ this exchange not only for $\Delta m_\text{d} > 0$ but also for $\Delta
m_\text{d} < 0$, when dust accretes from surrounding gas.  In general galaxy
applications, the stopping time-scale (equation~\ref{EQN:t_s_Myr}) is short
enough that local gas and dust velocities are similar.

During mass transfer, we keep a gas cell's internal energy per unit
mass constant.  Using its updated mass and momentum, the gas cell's energy is
then recomputed as the sum of thermal and kinetic components.  More complicated
momentum and energy exchanges based on detailed fluid-solid interactions are
beyond the scope of this work.

\begin{figure}
\centering
\includegraphics{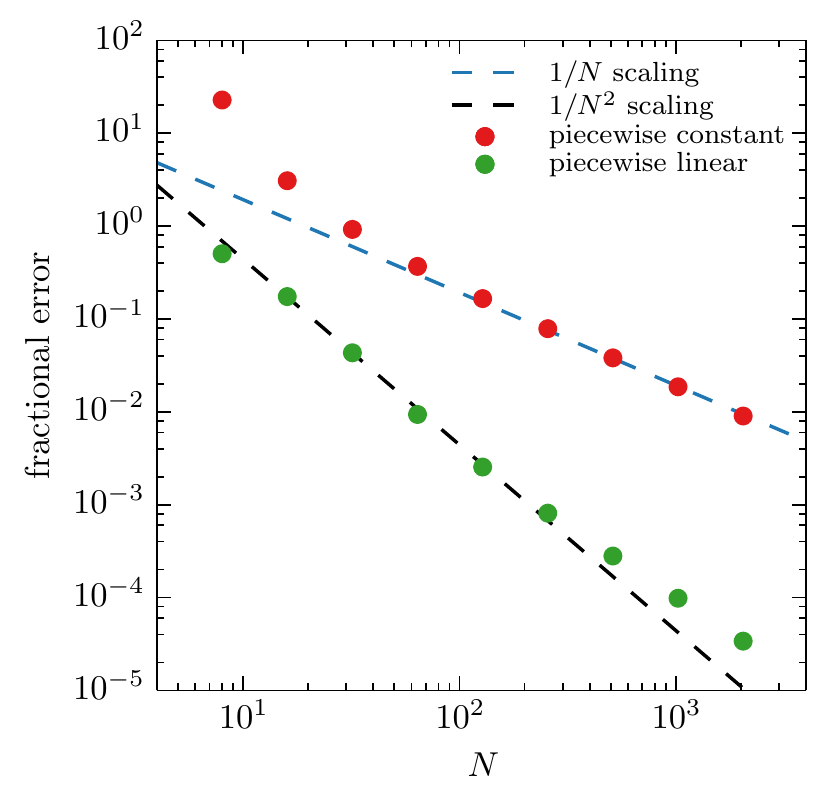}
\caption{Convergence results for the grain growth test, plotting
fractional dust mass error as a function of number of grain bins $N$.  We
evolve a single dust particle whose grain size limits are $a_\text{min} = 0.001
\, \mu\text{m}$ and $a_\text{max} = 1 \, \mu\text{m}$ and whose initial grain
size distribution $\partial n(a, t=0 \, \text{Gyr}) / \partial a \propto
\Pi^{a_\text{cut}}_{a_\text{min}}(a)$ is non-zero and uniform over the interval
from $a_\text{min}$ to $a_\text{cut} \equiv a_\text{min} (a_\text{max} /
a_\text{min})^{1/4}$.  As a result, only a quarter of bins have a non-zero
number of grains at $t = 0 \, \text{Gyr}$.  We fix the grain radius growth rate
$\dot{a} = 0.005 \, \mu\text{m} \, \text{Gyr}^{-1}$.  Results are shown at $t =
5 \, \text{Gyr}$ for the piecewise constant (red) and piecewise linear (green)
discretisations.  Dashed lines show $1/N$ (blue) and $1/N^2$ scalings (black).
The piecewise constant discretisation provides a first-order method, while the
piecewise linear discretisation deviates slightly from a second-order scaling
only at large $N$.}
\label{FIG:growthtest_convergence}
\end{figure}

\subsubsection{Grain size evolution test problems}

Figure~\ref{FIG:growthtest_convergence} shows a test of the convergence
properties of the piecewise linear and piecewise constant methods.  Using
various choices for number of bins $N$, we evolve the same initial grain size
distribution and compare with the expected analytic result.  The limits of the
grain size distribution are $a_\text{min} = 0.001 \, \mu\text{m}$ and
$a_\text{max} = 1 \, \mu\text{m}$, and the initial grain size distribution
$\partial n(a, t=0 \, \text{Gyr}) / \partial a \propto
\Pi^{a_\text{cut}}_{a_\text{min}}(a)$ is given in terms of the ``boxcar''
function
\begin{equation}
\Pi^b_a(x) \equiv
\begin{cases}
1, \quad \text{if} \, a \leq x \leq b, \\
0, \quad \text{else}.
\end{cases}
\end{equation}
Here, $a_\text{cut} \equiv a_\text{min} (a_\text{max} / a_\text{min})^{1/4}$
lies one-quarter of the way between $a_\text{min}$ and $a_\text{max}$ on a
logarithmic scale.  We note that $\Pi^b_a(x) = H(x-a) - H(x-b)$, where $H$ is the
Heaviside step function.  Thus, the initial grain size distribution takes a
constant, non-zero value over the interval $a_\text{min}$ to $a_\text{cut}$.
The grain growth rate is fixed at $\dot{a} = 0.005 \, \mu\text{m} \,
\text{Gyr}^{-1}$, and we calculate the fractional error in dust mass at
$t = 5 \, \text{Gyr}$, after grains grow by $0.025 \, \mu\text{m}$.
The analytic grain size distribution is simply $\partial n(a, t) / \partial a
\propto \Pi^{a_\text{cut} + \dot{a} t}_{a_\text{min} + \dot{a} t}(a)$.

The piecewise constant method yields first-order accuracy, while the piecewise
linear method largely displays second-order behaviour apart from a slight
softening of the convergence rate for $N \gtrsim 512$.  In this test, the
fractional mass error for $N = 64$ bins is roughly 40 per cent for the
piecewise constant discretisation and just 1 per cent for the piecewise linear
one.  In the tests and applications below, we use the piecewise linear method
for its improved accuracy and convergence properties.

\begin{figure}
\centering
\includegraphics{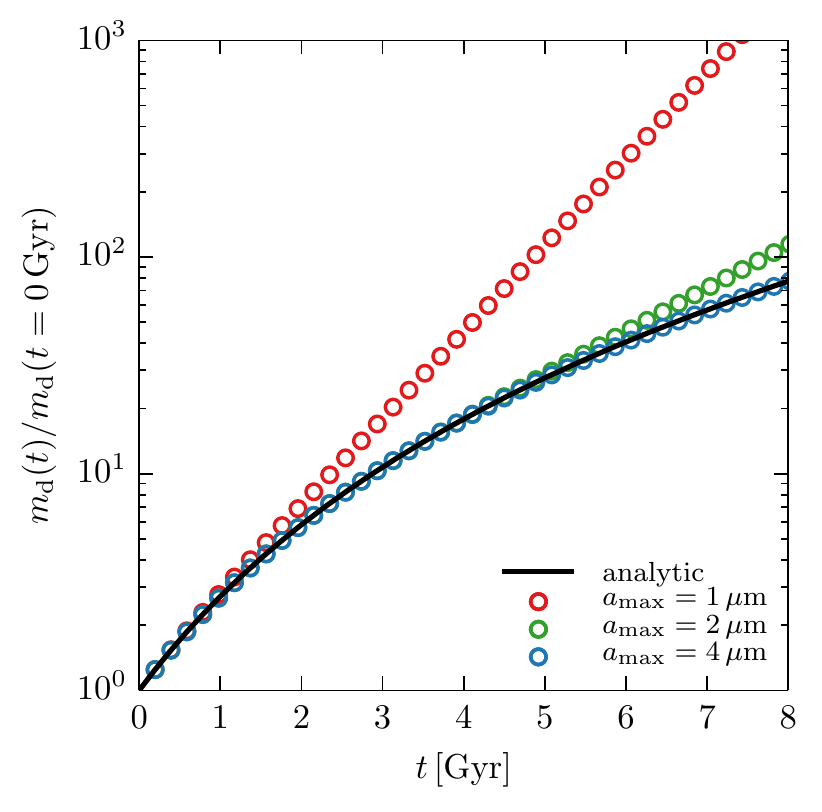}
\caption{Mass evolution of a dust particle with a grain size
distribution that is uniform over a fixed range at $t = 0 \, \text{Gyr}$ and
whose largest grains are $1 \, \mu\text{m}$ in size.  We perform three runs
that share the same minimum grain size.  These runs increase the maximum
allowable grain size $a_\text{max}$ but also increase the number of bins to
keep the same bin-spacing factor $\delta$ from equation~(\ref{EQN:bin_delta}).
We fix $\dot{a} = 0.3 \, \mu\text{m} \, \text{Gyr}^{-1}$ and evolve the dust
particle until $t = 8 \, \text{Gyr}$ so that grains grow by $2.4 \,
\mu\text{m}$.  Colored circles show the mass evolution of the dust particle,
normalised to its initial mass, for these runs using the piecewise linear
formulation with boundary mass rebinning.  The black line denotes the expected
analytic result.  By increasing the maximum allowable grain size, we reduce
inaccuracies from the rebinning procedure that artificially limits grain
size.}
\label{FIG:rebintest_plot}
\end{figure}

We next study the impact of the mass rebinning procedure given by
equation~(\ref{EQN:mass_after_rebin}) in order to highlight the fact that
rebinning may conserve mass during each time-step but not yield the expected
long-term behaviour.  Figure~\ref{FIG:rebintest_plot} shows the mass evolution
of a dust particle whose initial grain size distribution has minimum grain size
$a_\text{min} = 0.015625 \, \mu\text{m}$ and takes the form $\partial n(a, t=0)
/ \partial a \propto \Pi^{a_\text{r}}_{a_\text{l}}(a)$, where $a_\text{l} =
a_\text{min} \delta^{45}$, $a_\text{r} = 1 \, \mu\text{m}$, and
\begin{equation}
\log \delta = \frac{\log a_\text{r} - \log a_\text{min}}{60}.
\end{equation}
Thus, $a_\text{l}$ lies three-quarters of the logarithmic distance between
$a_\text{min}$ and $a_\text{r}$.  As a result, the initial grain size
distribution is non-zero and uniform over $[a_\text{l}, a_\text{r}]$.  We
perform three runs, all of which use the same minimum grain size $a_\text{min}$
and bin-spacing factor $\delta$ but vary the number of bins $N$.  The maximum
grain size $a_\text{max} = a_\text{min} \delta^N$ for these runs is $1 \,
\mu\text{m}$ (for $N = 60$ bins), $2 \, \mu\text{m}$ ($N = 70$), and $4 \,
\mu\text{m}$ ($N = 80$).  Since the initial grain size distribution and bin
spacing is the same across all three runs, this test allows us to determine the
impact of the maximum allowable grain size and rebinning procedure on mass
evolution while keeping resolution fixed.  The grain growth rate is fixed to
$\dot{a} = 0.3 \, \mu\text{m} \, \text{Gyr}^{-1}$.

\begin{figure}
\centering
\includegraphics{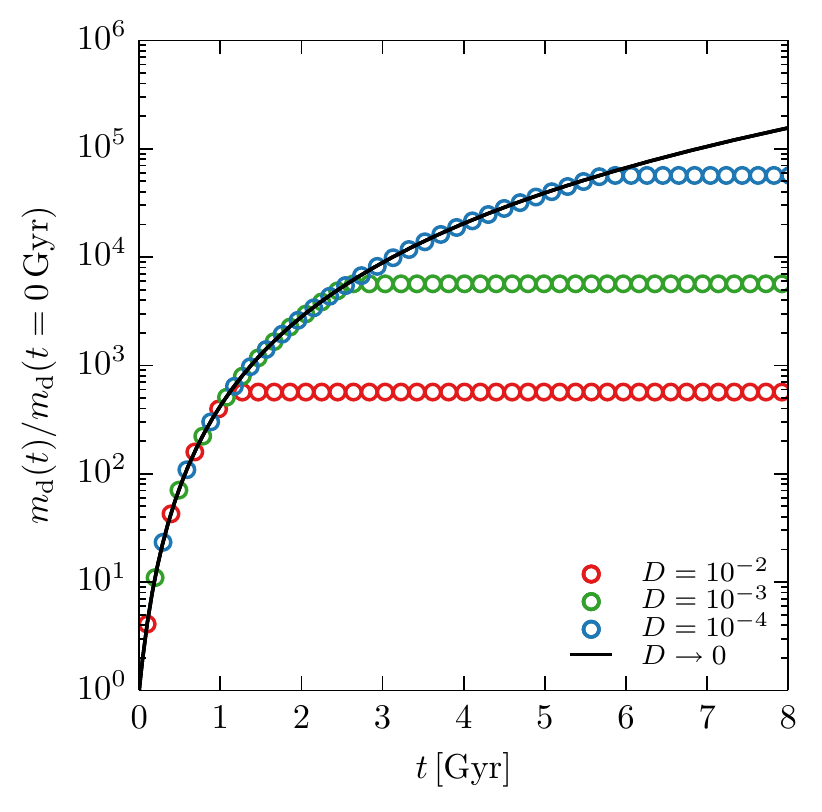}
\caption{Demonstration of grain growth in gas with a fixed amount of
metals.  The grain growth rate is fixed at $\dot{a} = 0.025 \, \mu\text{m} \,
\text{Gyr}^{-1}$, and the dust particle's smoothing length is chosen so that it
encloses $N_\text{ngb} = 64 \pm 8$ neighbouring gas cells.  Neighbouring gas
cells are located on a uniform lattice with equal mass and have metallicity $Z
= 0.1$.  The dust-to-gas ratio $D$ is the initial mass ratio between the dust
particle and a neighbouring gas cell.  The dust particle increases its mass
$m_\text{d}(t)$ by a factor of $N_\text{ngb} Z / D$ before the
surrounding gas runs out of metals.  The black line shows the expected analytic
mass growth if the gas is treated as an infinite reservoir of metals.}
\label{FIG:saturatetest_plot}
\end{figure}

The dust particle's mass most closely follows the analytic result when
$a_\text{max}$ is large and the effect of rebinning is small.  Because the
largest grains at $t = 0 \, \text{Gyr}$ are $1 \, \mu\text{m}$ in
size, when $a_\text{max} = 1 \, \mu\text{m}$ some grains are subject to
rebinning starting on the very first time-step.  In contrast, grains are
rebinned less often for $a_\text{max} = 2 \, \mu\text{m}$, whose profile
displays better accuracy.  Although the $a_\text{max} = 4 \, \mu\text{m}$ test
should not involve any rebinning in theory (even the largest grains that start
at $a = 1 \, \mu\text{m}$ will not grow larger than $a_\text{max}$), in
practice the slope limiting procedure will introduce some diffusion that
populates then largest grain size bins over time.  However, this effect is
sufficiently small that the test with $a_\text{max} = 4 \, \mu\text{m}$ yields
mass evolution visually indistinguishable from the analytic result.

This behaviour is easy to understand intuitively: consider a grain of radius $a
= 1 \, \mu\text{m}$ and time-steps such that $\dot{a} \Delta t = 1 \,
\mu\text{m}$.  Over two time-steps without rebinning, the grain will grow to
have radius $3 \, \mu\text{m}$.  Next, suppose we adopt rebinning so that
grains are not allowed to grow beyond $1 \, \mu\text{m}$: then, after the first
time-step, the grain grows to radius $2 \, \mu\text{m}$ and is replaced with
eight grains of radius $1 \, \mu\text{m}$.  After the second time-step, this
process repeats for each of these eight grains, so that at the end we have 64
grains of radius $1 \, \mu\text{m}$.  The ratio of final mass with rebinning to
final mass without rebinning is $64 / 3^3 > 1$: in this case, mass has
artificially grown too quickly.  Since mass scales nonlinearly as radius cubed,
artificially limiting grain radii can allow mass discrepancies to build up over
time.

\begin{figure*}
\centering
\includegraphics{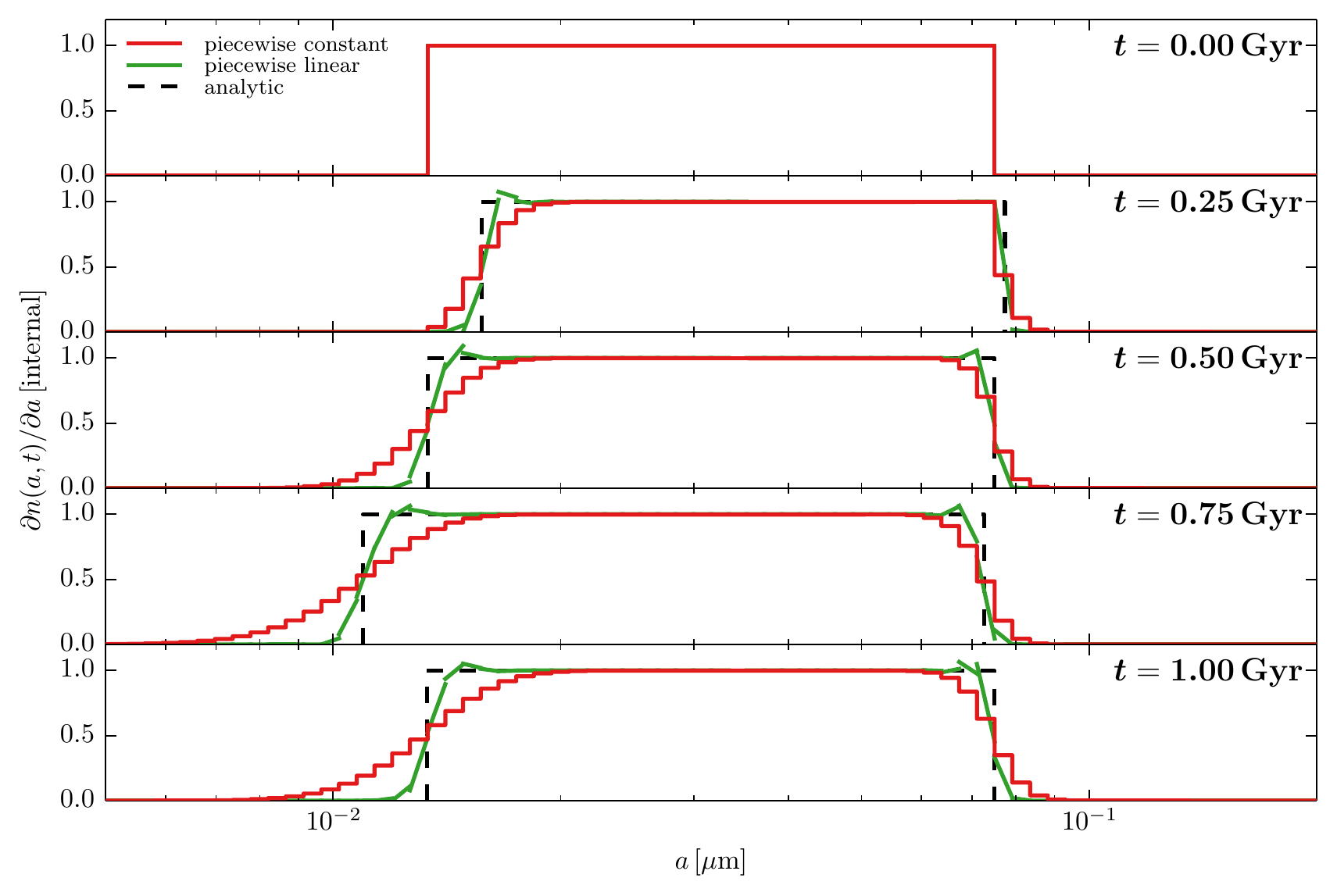}
\caption{Grain size evolution for dust where $\dot{a}$ varies
sinusoidally with amplitude $0.015 \, \mu\text{m} \, \text{Gyr}^{-1}$ and
period $1 \, \text{Gyr}$.  We compare the piecewise constant~(red) and
piecewise linear~(green) discretisations with the expected analytic
solution~(black).  From top to bottom, we show the grain size distribution as
it evolves over quarter-periods.  The piecewise linear method better captures
discontinuities where the grain size distribution jumps to zero.}
\label{FIG:sinetest_evolution}
\end{figure*}

These results suggest rebinning mass is acceptable when the fraction of dust
particle mass affected is small (as in the $a_\text{max} = 2 \, \mu\text{m}$
run), not large (as in the $a_\text{max} = 1 \, \mu\text{m}$ run).  Rebinning
is not guaranteed to provide the correct long-term behaviour, but it can
preserve mass from time-step to time-step.  In practice, we recommend using
knowledge of typical time-scales and grain growth rates (e.g.~in cosmological
contexts, $t \sim 14 \, \text{Gyr}$) to estimate a rough maximum grain size and
adopting this as $a_\text{max}$.  For example, interstellar grain size
distributions typically extend from $a_\text{min} = 0.001 \, \mu\text{m}$ to
$a_\text{max} = 1 \, \mu\text{m}$ \citep{Weingartner2001}, but the size
distribution for, say, protoplanetary applications would extend to much larger
radii.

Up to this point, we have considered cases where the gas surrounding a dust
particle always contains enough metals to deplete onto grains.  However, if the
gas has a limited supply of metals, the growth of dust may deviate from the
expected behaviour.  Figure~\ref{FIG:saturatetest_plot} shows the mass
evolution of a single dust particle surrounded by a uniform lattice of
equal-mass gas cells.  The ratio of initial dust particle mass to gas cell
mass, $D$, is chosen to be $10^{-4}$, $10^{-3}$, or $10^{-2}$, and the initial
gas metallicity is $Z = 0.1$.  We intentionally choose a large value of
metallicity to provide a reservoir of metals for dust to deplete.  The grain
radius rate of growth is fixed to $\dot{a} = 0.025 \, \mu\text{m} \,
\text{Gyr}^{-1}$.  The dust particle is able to accrete metals in a
kernel-weighted fashion from neighbouring gas cells within its smoothing
length, determined using equation~(\ref{EQN:N_ngb}) and $N_\text{ngb} = 64 \pm
8$.  The limits and initial condition of the grain size distribution are the
same as those used in Figure~\ref{FIG:growthtest_convergence}, although they do
not affect this test.

As expected, the dust particle is able to grow its mass by a factor of
$N_\text{ngb} Z / D$, at which point neighbouring gas cells within the smoothing
length run out of metals.  Afterwards, the dust particle's mass is constant.
Decreasing $D$ increases the relative abundance of metals to dust and prolongs
the point at which dust accretion stops.  Of course, in a realistic setting it
is possible for gas to be re-enriched with metals (e.g.~through stellar
evolution) and dust to resume its accretion.

Figure~\ref{FIG:sinetest_evolution} demonstrates how the piecewise constant and
piecewise linear methods reproduce a grain size distribution as it evolves
under mass growth and mass loss.  We adopt $a_\text{min} = 0.001 \,
\mu\text{m}$ and $a_\text{max} = 1 \, \mu\text{m}$ and use $N = 128$ bins.  The
initial grain size distribution is non-zero and constant over the middle
quarter of bins covering the interval $[a_\text{min} \delta^{3N/8},
a_\text{min} \delta^{5N/8}]$, where $\delta$ is the usual bin-spacing factor
from equation~(\ref{EQN:bin_delta}).  The grain growth rate $\dot{a}$ is a
sinusoid with amplitude $0.015 \, \mu\text{m} \, \text{Gyr}^{-1}$ and
period $1 \, \text{Gyr}$.  Grain size distribution boundary effects are
unimportant for this choice of amplitude.

As the grain size distribution evolves over one full period, the piecewise
constant and piecewise linear discretisations experience some numerical
diffusion in reproducing the jump discontinuities in the grain size
distribution.  However, the piecewise linear method is better able to preserve
the steepness of the discontinuity.  After one period, the piecewise linear
grain size distribution takes an extra three bins beyond the left-most analytic
discontinuity to become visually consistent with zero.  In contrast, the
piecewise constant method requires an extra nine bins.  In a test like this,
combining mass growth and mass loss, the piecewise linear method does a far
better job of reproducing the analytic result and reducing numerical diffusion.

In the above tests, we used arbitrary choices for $\dot{a}$ to enable
comparison with analytic results.  Below, we describe the form that $\dot{a}$
takes for various physical processes.

\subsection{Grain growth}\label{SEC:grain_growth}

Dust grains in the ISM can grow by accreting gas-phase metals
\citep{Draine1990, Dwek1998, Michalowski2010}, and a number of accretion
parameterisations have been used in models in recent years
\citep[e.g.][]{Zhukovska2008, Hirashita2011, Hirashita2012, Hirashita2014b,
Asano2013b, deBennassuti2014, Popping2017}.  In this work, we follow equation~5
from \citet{Hirashita2011} and equations~19 and~20 from \citet{Hirashita2014b}.
We calculate the growth rate of a dust grain of radius $a$ as
\begin{equation}
\frac{\diff a}{\diff t} \approx \left( \frac{Z}{Z_\odot} \right) \left( \frac{n_\text{H}}{10^3 \, \text{cm}^{-3}} \right)\left( \frac{T}{10 \, \text{K}} \right)^{1/2} \left( \frac{S_\text{acc}}{0.3} \right) \, \mu\text{m} \, \text{Gyr}^{-1},
\label{EQN:da_dt}
\end{equation}
where $Z_\odot = 0.0127$ is the solar metallicity, $Z$, $n_\text{H}$, and $T$
the local gas metallicity, hydrogen number density, and temperature,
respectively, and $S_\text{acc}$ the accretion sticking efficiency.  As in
equation~(\ref{EQN:kernel_weighting}), we determine $Z$, $n_\text{H}$, and $T$
by interpolating over neighbouring gas cells.

Although the sticking efficiency $S_\text{acc}$ is expected to be a function of
temperature and to vary in different ISM phases \citep[e.g.][]{Zhukovska2016},
the mass resolution available in cosmological simulations is not sufficient to
resolve detailed ISM structure.  Thus, we adopt $S_\text{acc} = 0.3$, as in the
analytic work of \citet{Hirashita2011}.  While this does not capture the
temperature behaviour suggested by some chemisorption and physisorption works
\citep{LeitchDevlin1985, Grassi2011, Chaabouni2012}, it avoids the assumption
of unit sticking efficiency adopted in prior works \citep{Asano2013b,
McKinnon2017, Popping2017} that has been suggested to overdeplete metals
\citep{Zhukovska2016}.  Future work could improve on this assumption when more
explicit ISM models are implemented.

\subsection{Thermal sputtering}\label{SEC:thermal_sputtering}

Dust grains can be eroded through collisions with thermally excited gas.  A
number of works have studied this thermal sputtering process in detail for
various grain materials and compositions \citep{Ostriker1973, Burke1974,
Barlow1978, Draine1979b, Dwek1992, Tielens1994}.  An analytic approximation to
these detailed calculations was provided by equation~14 in \citet{Tsai1995},
with the rate of erosion for a grain of size $a$ given by
\begin{equation}
\frac{\diff a}{\diff t} = -(3.2 \times 10^{-18} \, \text{cm}^{4} \, \text{s}^{-1}) \left( \frac{\rho_\text{g}}{m_\text{p}} \right) \left[ \left( \frac{T_\text{sput}}{T} \right)^{2.5} + 1 \right]^{-1},
\end{equation}
where $\rho_\text{g}$ and $T$ are the gas density and temperature,
respectively, $m_\text{p}$ is the proton mass, and $T_\text{sput} \equiv 2
\times 10^{6} \, \text{K}$.  Thermal sputtering is strongest for $T \gtrsim
10^{6} \, \text{K}$ and can affect the size distribution in hot plasmas like
the intracluster medium \citep{Yahil1973, McGee2010} and in interstellar
SN shocks \citep{Nozawa2006, Bianchi2007, Nath2008, Kozasa2009,
Silvia2010, Silvia2012}.

The sub-resolution ISM model \citep{Springel2003} that we adopt treats
dense, star-forming gas using an effective equation of state.  The star-forming
ISM typically does not resolve hot, $T > 10^{6} \, \text{K}$ gas surrounding
SNe that could thermally sputter dust grains \citep[see Figure 1 in][for an
example gas phase diagram]{Torrey2017b}.  We therefore also require a
sub-resolution scheme that accounts for the sputtering of grains by SNe in a
star-forming ISM, which we introduce in the following section.  Together,
Sections~\ref{SEC:thermal_sputtering} and~\ref{SEC:supernova_destruction}
combine to model grain sputtering outside and inside the star-forming ISM,
respectively.  In the future, more explicit ISM models with better resolution
could attempt to directly capture the multiphase structure of the ISM and avoid
such sub-resolution prescriptions.

\subsection{Supernova destruction}\label{SEC:supernova_destruction}

High-velocity shocks produced by SNe can also destroy dust grains and
shift the grain size distribution to smaller sizes \citep{Nozawa2006,
Bianchi2007, Nozawa2007, Nath2008, Silvia2010}.  Because we do not directly
resolve individual SNe in our galaxy formation model, we account for the
destruction of dust in SN shocks using the sub-resolution ISM model by tying
the dust destruction rate to the local SN rate.

We parallel Section~2.2.3 of \citet{Asano2013b}, which applied the methods
developed in \citet{Yamasawa2011} to determine the influence of SN shocks on
the ISM grain size distribution.  These methods are parameterised in terms of a
function $\xi(a_\text{f}, a_\text{i})$ such that, for our bin
discretisation, $\xi(a^\text{c}_j, a^\text{c}_i) \times (a^\text{e}_{j+1} -
a^\text{e}_{j})$ denotes the fraction of grains starting in bin $i$ that end up
in bin $j$ after one SN shock.  Following the aforementioned works, we use the
$\xi$ values calculated by \citet{Nozawa2006} in detailed modeling of SN
blasts.

Integrating equations~12 and~14 of \citet{Asano2013b} over the width of bin
$j$, we obtain the rate of change of number of grains in bin $j$,
\begin{equation}
\frac{\diff N_j}{\diff t} = \frac{M_\text{swept} \gamma_\text{SN}}{M_\text{ISM}} \left( \sum_{i=0}^{N-1} \xi(a^\text{c}_j, a^\text{c}_i) (a^\text{e}_{j+1} - a^\text{e}_j) N_i(t) - N_j(t) \right),
\label{EQN:dNdt_SN}
\end{equation}
and the rate of change of mass of grains in bin $j$,
\begin{equation}
\begin{split}
&\frac{\diff M_j}{\diff t} = \frac{M_\text{swept} \gamma_\text{SN}}{M_\text{ISM}} \Bigg\{ \sum_{i=0}^{N-1} \left[ \xi(a^\text{c}_j, a^\text{c}_i) N_i(t) \left( \frac{\pi \rho_\text{gr} a^4}{3} \right) \right]^{a=a^\text{e}_{j+1}}_{a=a^\text{e}_j} \\
& \qquad\qquad\qquad\qquad\quad - M_j(t) \Bigg\}.
\end{split}
\label{EQN:dMdt_SN}
\end{equation}
Here, $\gamma_\text{SN} / M_\text{ISM}$ is the ratio of SN rate to mass
in the ISM, and $M_\text{swept}$ is the mass that a SN sweeps up.
Following the fitting function presented in equation~8 of \citet{Yamasawa2011},
we use
\begin{equation}
\frac{M_\text{swept}}{\text{M}_\odot} \equiv 1535 \left(\frac{n}{1 \, \text{cm}^{-3}}\right)^{-0.202} \left(\frac{Z}{Z_\odot} + 0.039 \right)^{-0.289},
\end{equation}
in terms of the local ISM density $n$ and metallicity $Z$.  We calculate the
prefactor $M_\text{swept} \gamma_\text{SN} / M_\text{ISM}$ by kernel-averaging
over neighboring gas cells, where $\gamma_\text{SN}$ and $M_\text{ISM}$ are the
local SN rate and mass of each gas cell.  The local SN rate in a gas cell is
computed using the star formation rate predicted by the sub-resolution ISM model
\citep{Springel2003} and the mass fraction of stars that explode as SNe II for
a chosen initial mass function~(IMF).  After updating the number and mass of
grains in each bin using equations~(\ref{EQN:dNdt_SN}) and~(\ref{EQN:dMdt_SN}),
we then apply equation~(\ref{EQN:M_j_slope_linear}) to determine each bin's
slope and slope limit as before if necessary.

\subsection{Shattering}\label{SEC:shattering}

To this point, we have discussed physical processes that conserve grain number
but not grain mass, with mass either gained from or returned to gas by growing
or shrinking grain radii.  However, it is important to also consider
grain-grain collisional processes like shattering and coagulation that conserve
total grain mass.  In general, such processes could be treated as an
inhomogeneous source term in the grain number continuity equation (see
equation~\ref{EQN:continuity_equation}).  However, it is numerically easier to
separate the treatment of shattering and coagulation from the number-conserving
methods in Section~\ref{SEC:piecewise_linear}.  This enables us to take
advantage of formalisms developed to study particle population dynamics
\citep[e.g.][]{Smoluchowski1916}.

\begin{figure}
\centering
\includegraphics{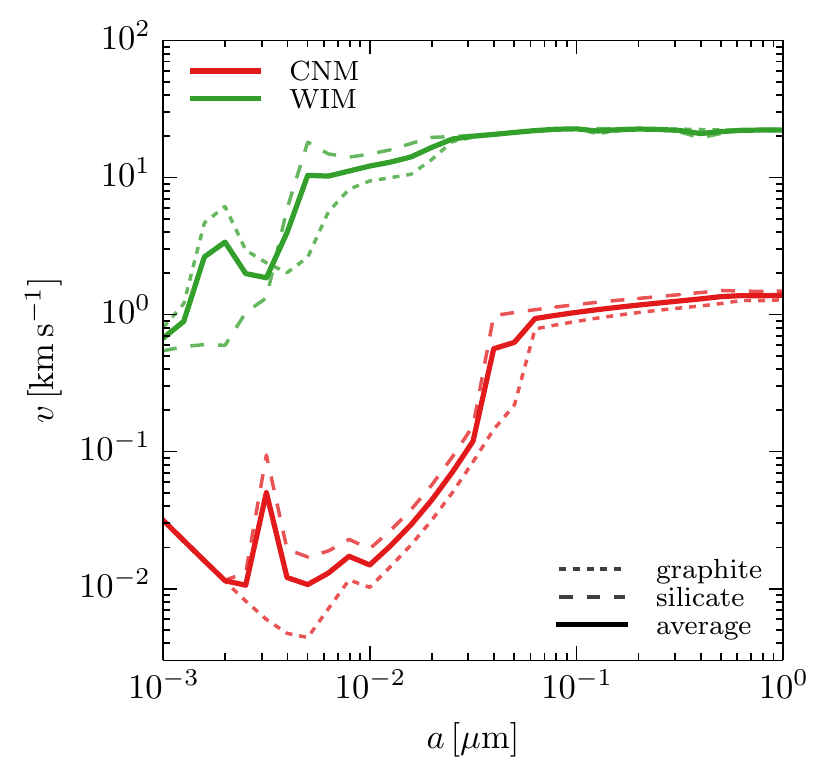}
\caption{Grain velocities for the cold neutral medium (CNM) and warm ionised
medium (WIM) phases of the turbulent ISM predicted by \citet{Yan2004}.  For
each phase, velocity curves are shown for graphite grains (short dashed lines),
silicate grains (long dashed lines), and an average of the two (solid lines).
Bigger grains tend to have larger velocities, and velocities in the WIM exceed
those of the CNM.  Relative velocities between grains of different sizes are
used to compute grain shattering and coagulation rates.}
\label{FIG:grain_velocity_curves}
\end{figure}

Conceptually, shattering causes large grains to fragment and produces many
smaller grains.  Two grains can collisionally shatter when their relative
velocity is above a threshold value.  Suppose grains of size $a_1$ and $a_2$
have speeds $v(a_1)$ and $v(a_2)$, respectively.  In principle, grain speeds
can be influenced by local gas properties like density and temperature.  When
colliding, the grains have relative velocity $v_\text{rel}(a_1, a_2) =
\sqrt{v(a_1)^2 + v(a_2)^2 - 2 v(a_1) v(a_2) \cos \theta}$, where $\cos \theta$
accounts for impact angle.  In this work, we follow \citet{Hirashita2013c} and
stochastically calculate relative velocities between two grains by drawing
$\cos \theta$ values randomly from the interval $[-1, 1]$.  Because of the
limited resolution of our ISM model, we compute grain velocities as a function
of grain size with a sub-resolution scheme, using the small-scale turbulent ISM
models of \citet{Yan2004}.  In particular, \citet{Yan2004} studied the dynamics
of different size grains in a variety of ISM phases, including the cold neutral
medium (CNM) and warm ionised medium (WIM).  Grain velocities as a function of
grain size tabulated for these ISM phases and used in our work are shown in
Figure~\ref{FIG:grain_velocity_curves}.  Appendix~\ref{SEC:appendix_shattering}
details how grain velocities in the CNM and WIM are combined with our equation
of state model to estimate velocities for populations of grains in the ISM.
These velocity curves allow us to calculate the relative velocities
$v_\text{rel}(a_1, a_2)$ that determine shattering rates, which have been
studied in a variety of works \citep{Voelk1980, Markiewicz1991, Cuzzi2003,
Yan2004, Ormel2007, Ormel2009, Hirashita2013c, Paruta2016}.

Because shattering and coagulation are mass-conserving and not
number-conserving processes, it is useful to define a differential mass density
\begin{equation}
\frac{\partial \rho(a, t)}{\partial a} \equiv \left( \frac{m(a)}{V_\text{d}} \right) \frac{\partial n(a, t)}{\partial a},
\label{EQN:differential_mass_density}
\end{equation}
such that $\partial \rho(a, t) / \partial a \times \diff a$ is the mass density
of grains with radii in the interval $[a, a + \diff a]$ at time $t$.  Here,
$V_\text{d} \equiv m_\text{d} / \rho_\text{d}$ is the volume associated to a
dust particle, where $m_\text{d}$ is its known mass and $\rho_\text{d}$ is a
kernel-weighted dust density estimate using neighbouring dust particles.
Because dust particles may vary in mass more than gas cells, when finding dust
neighbors we use a smoothing length enclosing a desired amount of dust mass
instead of a number of neighbors.  Further details on this procedure are
provided in Section~\ref{SEC:dust_ngb_searches}.

Shattering has been studied numerically using piecewise constant
discretisations \citep{ODonnell1997, Hirashita2009} and analytically in the
continuous case \citep{Dubovskii1992, Asano2013b, Mattsson2016}.  We parallel
these implementations in adapting shattering to our piecewise linear
discretisation, noting that in \citet{Hirashita2009} and \citet{Asano2013b},
what we label $\partial \rho(a, t) / \partial a$ they denote $\rho(a, t)$.
Following equation~(23) of \citet{Asano2013b}, shattering causes the mass
density for grains of size $a$ to evolve with the rate
\begin{equation}
\begin{split}
&\frac{\partial}{\partial t} \left[ \frac{\partial \rho(a, t)}{\partial a} \right] = -m(a) \frac{\partial \rho(a, t)}{\partial a} \int_{a_\text{min}}^{a_\text{max}} \alpha(a, a_1) \frac{\partial \rho(a_1, t)}{\partial a_1} \diff a_1 \\
&\qquad \qquad \qquad + \frac{1}{2} \int_{a_\text{min}}^{a_\text{max}} \int_{a_\text{min}}^{a_\text{max}} \bigg[ \alpha(a_1, a_2) \frac{\partial \rho(a_1, t)}{\partial a_1} \frac{\partial \rho(a_2, t)}{\partial a_2} \\
&\qquad \qquad \qquad \qquad \qquad \qquad \qquad \times m_\text{shat}(a, a_1, a_2) \bigg] \diff a_2 \diff a_1,
\end{split}
\label{EQN:continuous_shattering}
\end{equation}
where
\begin{equation}
\alpha(a_1, a_2) \equiv
\begin{dcases}
\frac{\pi (a_1 + a_2)^2 v_\text{rel}(a_1, a_2)}{m(a_1) m(a_2)}, & v_\text{rel}(a_1, a_2) > v_\text{shat}, \\
0, & v_\text{rel}(a_1, a_2) \leq v_\text{shat},
\end{dcases}
\label{EQN:alpha_a1_a2}
\end{equation}
is a function of effective cross-section, grain relative velocity, and grain
masses that only allows collisions when relative velocities are above the
shattering threshold $v_\text{shat}$, and $m_\text{shat}(a, a_1, a_2) \diff a$
is the mass of grains in the size interval $[a, a + \diff a]$ produced through
shattering grains of sizes $a_1$ and $a_2$.  Apart from one test
problem detailed later in this section, in all other applications we calculate
$m_\text{shat}(a, a_1, a_2)$ following Section~2.3 of \citet{Hirashita2009},
which allows grains to fully or partially fragment depending on the sizes of
colliding grains.  In equation~(\ref{EQN:continuous_shattering}), the first
term accounts for the removal of grains of size $a$ in collisions with grains
of size $a_1$, while the second term describes the injection of grains of size
$a$ in collisions with grains of sizes $a_1$ and $a_2$.  Because it is easier
to work with, our definition of $m_\text{shat}$ accounts for mass produced by
both colliding grains and not just one of the grains, as in
\citet{Hirashita2009}.  This necessitates the factor of $1/2$ in the second
term in equation~\ref{EQN:continuous_shattering}.  For $v_\text{shat}$,
\citet{Jones1996} uses $2.7 \, \text{km} \, \text{s}^{-1}$ for silicate grains
and $1.2 \, \text{km} \, \text{s}^{-1}$ for graphite grains.  Because we do not
track detailed grain chemistry, we adopt $v_\text{shat} \approx 2 \, \text{km}
\, \text{s}^{-1}$ for all grain populations.  For simplicity we use an
indicator function to write $\alpha(a_1, a_2) \equiv \pi (a_1 + a_2)^2
v_\text{rel}(a_1, a_2) \mathbbm{1}_{v_\text{rel} > v_\text{shat}}(a_1, a_2) /
(m(a_1) m(a_2))$.  We show in Appendix~\ref{SEC:appendix_shattering} how these
integrals can be discretised given a piecewise linear grain size distribution
and suitable approximations.

After discretising and approximating, the mass evolution for bin $i$ turns into
\begin{equation}
\begin{split}
V_\text{d} \frac{\diff M_i}{\diff t} &= -\pi \sum_{k=0}^{N-1} v_\text{rel}(a^\text{c}_i, a^\text{c}_k) \mathbbm{1}_{v_\text{rel} > v_\text{shat}}(a^\text{c}_i, a^\text{c}_k) \langle m \rangle_i I^{i,k} \\
&\, + \frac{\pi}{2} \sum_{k=0}^{N-1} \sum_{j=0}^{N-1} v_\text{rel}(a^\text{c}_k, a^\text{c}_j) \mathbbm{1}_{v_\text{rel} > v_\text{shat}}(a^\text{c}_k, a^\text{c}_j) m_\text{shat}^{k,j}(i) I^{k,j},
\end{split}
\label{EQN:shattering_sums}
\end{equation}
where $\langle m \rangle_i$ is the average mass of a grain in bin $i$ computed
using only the bin edges and
\begin{equation}
\begin{split}
&I^{k,j}(t) \equiv \int_{a^\text{e}_k}^{a^\text{e}_{k+1}} \int_{a^\text{e}_j}^{a^\text{e}_{j+1}} \bigg[ \left( \frac{N_k(t)}{a^\text{e}_{k+1} - a^\text{e}_{k}} + s_k(t) (a_1 - a^\text{c}_k) \right) \\
& \qquad \times \left( \frac{N_j(t)}{a^\text{e}_{j+1} - a^\text{e}_{j}} + s_j(t) (a_2 - a^\text{c}_j) \right) (a_1 + a_2)^2 \bigg] \diff a_2 \diff a_1
\end{split}
\label{EQN:I_k_j}
\end{equation}
is a polynomial function dependent on the grain size distribution at time $t$.
For brevity, we do not write its analytic form here.  In the limit that
bin slopes go to zero, equation~(\ref{EQN:shattering_sums}) recovers the
piecewise constant update from equation~4 of \citet{Hirashita2009}.
Unlike the number-conserving processes in Section~\ref{SEC:rebinning},
the shattering update in equation~(\ref{EQN:shattering_sums}) requires no
rebinning of grains with radii below $a_\text{min}$ or above $a_\text{max}$.
Instead, we follow the steps in equations~14 through~19 of
\citet{Hirashita2009} to ensure all grains resulting from shattering have radii
in the allowed size range.

Using the grain size distribution at time $t$, for a time-step $\Delta t$ we
compute the change in mass in each bin using the simple first-order update
$M_i(t + \Delta t) = M_i(t) + \mathrm{d} M_i / \mathrm{d} t \times \Delta t
\equiv M_i(t) + \Delta M_i$, where $\Delta M_i$ is the change in dust mass in
bin $i$.  Because of the numerical approximation in
equation~(\ref{EQN:m_shat_kji}), it is possible for the change in dust particle
mass $\Delta m_\text{d} \equiv \sum_{i=0}^{N-1} \Delta M_i$ to deviate slightly
from zero.  In the limit $N \to \infty$, this approximation is exact and does
not introduce any numerical error.  To ensure $\Delta m_\text{d} = 0$ during
the time-step, we use the following rescaling.  When $\Delta m_\text{d} > 0$,
we choose to limit the mass gain in those bins with $\Delta M_i > 0$.  More
precisely, let
\begin{equation}
\Delta m_\text{sub} \equiv \sum_{i | \Delta M_i > 0} \Delta M_i
\end{equation}
be the total change in mass from the subset of bins gaining mass.  We then
subtract $\Delta m_\text{d} \times \Delta M_i / \Delta m_\text{sub}$ from each
bin $i$ with $\Delta M_i > 0$, ensuring the new bins satisfy $\sum_{i=0}^{N-1}
\Delta M_i = 0$.  If instead $\Delta m_\text{d} < 0$, we follow a similar
procedure, this time reducing the magnitude of $\Delta M_i$ of those bins with
$\Delta M_i < 0$.  In the text below, we assume that $\Delta M_i$ values refer
to changes in bin mass after ensuring $\Delta m_\text{d} = 0$.

Because the grain size distribution is parameterised in terms of the number of
grains and slope in each bin, we break the number-slope degeneracy by adding a
heuristic modelling changes in average grain size.  This mirrors the steps
used to handle boundary mass rebinning in Section~\ref{SEC:piecewise_linear}.
Assuming shattered grains have the grain size distribution $\partial n /
\partial a \propto a^{-3.3}$ \citep{Jones1996, Hirashita2009}, a shattered
grain injected into bin $i$ has average size $\langle a \rangle^\text{shat}_i
\equiv 2.3/1.3 \times [(a^\text{e}_{i+1})^{-1.3} - (a^\text{e}_i)^{-1.3}] /
[(a^\text{e}_{i+1})^{-2.3} - (a^\text{e}_i)^{-2.3}]$ and average mass $\langle
m \rangle^\text{shat}_i \equiv 4\pi \rho_\text{gr}/3 \times -2.3/0.7 \times
[(a^\text{e}_{i+1})^{0.7} - (a^\text{e}_i)^{0.7}] / [(a^\text{e}_{i+1})^{-2.3}
- (a^\text{e}_i)^{-2.3}]$.  If shattering injects grains into bin $i$ and
causes it to gain mass ($\Delta M_i \geq 0$), we approximate the new average
grain size as a weighted average of sizes for grains already in the bin and
those added to the bin.  That is, we assume the new average grain size in bin
$i$ is
\begin{equation}
\langle a \rangle_{i}(t + \Delta t) = \frac{N_{i}(t) \times \langle a \rangle_{i}(t) + \Delta N_i \times \langle a \rangle^\text{shat}_i}{N_{i}(t) + \Delta N_i},
\label{EQN:shattering_a_constraint}
\end{equation}
where $\Delta N_i \equiv \Delta M_i / \langle m \rangle^\text{shat}_i$
estimates the number of shattered grains added to bin $i$.  If bin $i$ loses
grain mass ($\Delta M_i < 0$), we assume the leftover grains in bin $i$
maintain their average grain size and set $\langle a \rangle_i(t + \Delta t) =
\langle a \rangle_i(t)$.  Using equations~(\ref{EQN:M_j_slope_linear})
and~(\ref{EQN:avg_a_slope_linear}), we combine the expression for $\langle a
\rangle_i(t + \Delta t)$ and
\begin{equation}
M_{i}(N_{i}(t + \Delta t), s_{i}(t + \Delta t)) = M_{i}(N_{i}(t), s_{i}(t)) + \Delta M_i,
\label{EQN:shattering_m_constraint}
\end{equation}
and simultaneously solve for the new number of grains $N_i(t + \Delta t)$ and
slope $s_i(t + \Delta t)$ in bin $i$.  We slope limit as before if necessary.
This finishes the time-step update due to shattering.

In addition to this piecewise linear discretisation, we also implement a
piecewise constant method.  This follows directly from equation~4 of
\citet{Hirashita2009}, or equivalently from
equation~(\ref{EQN:shattering_sums}) in this work by enforcing that slopes $s_i
\to 0$ and evaluating quantities at bin midpoints.  The development of these
two discretisations for mass-conserving processes parallels our treatment of
number-conserving methods in Section~\ref{SEC:piecewise_linear}.

\begin{figure}
\centering
\includegraphics{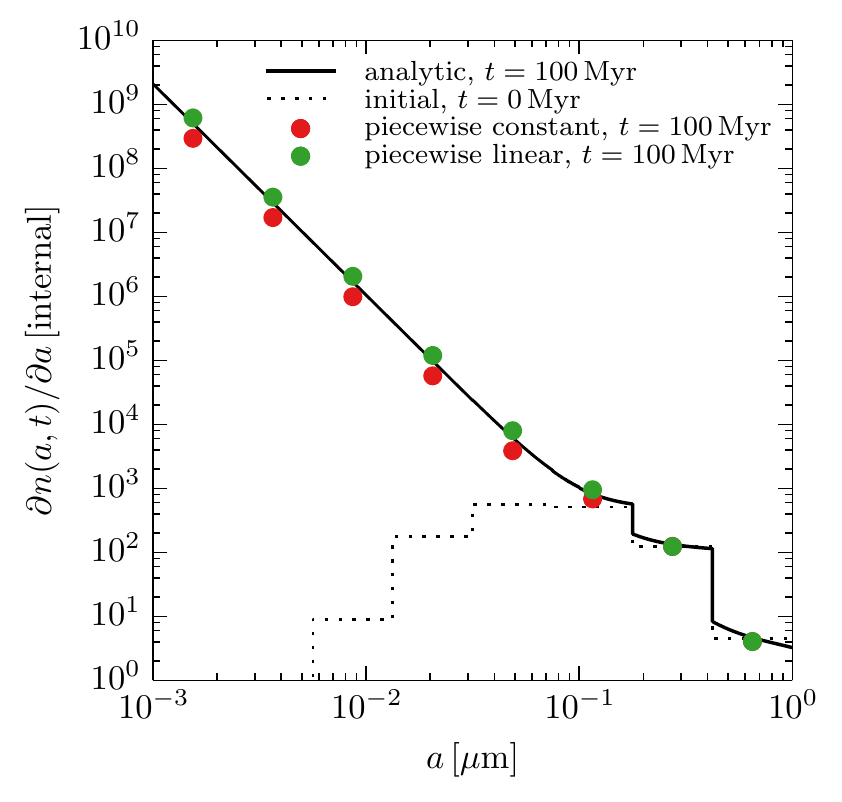}
\caption{Influence of shattering on grain size evolution for an initially
log-normal size distribution (dotted black line), using a simplified set of
grain velocity and mass fragment parameters as detailed in the text.  Coloured
circles show results at $t = 100 \, \text{Myr}$ for the piecewise constant
(red) and piecewise linear (green) discretisations using $N = 8$ bins.  We
compare with the expected solution at $t = 100 \, \text{Myr}$ (solid black
line) computed using a high-precision differential equations solver.
Shattering produces many small grains following the power law $\partial n /
\partial a \sim a^{-3.3}$, although most mass remains in the largest bins.}
\label{FIG:shatteringtest_analytic}
\end{figure}

In the following test problem verifying the numerical implementation of
shattering, we choose to adopt simplified forms of $v_\text{rel}$ and
$m_\text{shat}^{k,j}(i)$ so that the grain size distribution evolves in a more
predictable way.  All other applications -- including the isolated galaxy
simulations presented in Section~\ref{SEC:hernquist_spheres} -- use the grain
velocity and shattering kernel functions detailed in
Figure~\ref{FIG:grain_velocity_curves} and
equation~(\ref{EQN:continuous_shattering}), respectively.  Solely for
this test, we adopt
\begin{equation}
\frac{v_\text{rel}(a_k^\text{c}, a_j^\text{c})}{\text{km} \, \text{s}^{-1}} \equiv
\begin{cases}
3, & a_k^\text{c} > 0.1 \, \mu\text{m} \; \text{and} \; a_j^\text{c} > 0.1 \, \mu\text{m}, \\
0, & \text{else},
\end{cases}
\end{equation}
so that only collisions between large grains cross the shattering threshold.
In this test we do not use the effective relative velocity interpolated
between tabulated CNM and WIM grain velocities, since it introduces more
complicated behaviour.  Additionally, for this test only, we do not
compute $m_\text{shat}^{k,j}(i)$ using Section~2.3 of \citet{Hirashita2009},
which allows for complex size dynamics (e.g.~colliding grains can partially
fragment and leave behind remnants, shatter over a small size range, etc.),
and instead assume that all colliding grains fully fragment and
produce shattered grains in the interval $[a_\text{min}, a_\text{max}]$
according to a size power law with index $-3.3$ \citep{Jones1996}.  Thus, in a
collision between grains in bins $k$ and $j$, the resulting mass entering bin
$i$ is
\begin{equation}
m_\text{shat}^{k,j}(i) = (\langle m \rangle_k + \langle m \rangle_j) \times \left( \frac{(a_{i+1}^\text{e})^{0.7} - (a_i^\text{e})^{0.7}}{a_\text{max}^{0.7} - a_\text{min}^{0.7}} \right),
\end{equation}
where $\langle m \rangle_k$ is the average mass of a grain in bin $k$ computed
for a constant size distribution, as in Section~\ref{SEC:piecewise_linear}.

Paralleling a similar test in Section~2.1 of \citet{Hirashita2010b}, we
initialise one dust particle with a log-normal grain size distribution
\begin{equation}
\frac{\partial n(a, t=0 \, \text{Myr})}{\partial a} = \frac{C}{a} \exp\left( -\frac{\ln^2(a/a_0)}{2 \sigma^2} \right)
\end{equation}
over the interval from $a_\text{min} = 0.001 \, \mu\text{m}$ to $a_\text{max} =
1 \, \mu\text{m}$, where $a_0 = 0.1 \, \mu\text{m}$ and $\sigma = 0.6$.  The
volume has a gas density corresponding to $n_\text{H} \approx 0.4 \,
\text{cm}^{-3}$, and the normalisation constant $C$ is chosen so that the
dust-to-gas ratio is $D = 3.7 \times 10^{-3}$, the average of values used in
\citet{Hirashita2010b}.  We generate piecewise constant initial conditions, so
that they can be used with both discretisations.

Figure~\ref{FIG:shatteringtest_analytic} demonstrates the evolution of the
initially log-normal grain size distribution under the influence of only
shattering.  We compare results at $t = 100 \, \text{Myr}$ using piecewise
constant and piecewise linear discretisations.  Both capture the formation of
small grains following a $\partial n / \partial a \sim a^{-3.3}$ power law,
although the piecewise linear method better reproduces the solution predicted
by a high-accuracy numerical integrator.  Despite shattering forming many small
grains, we caution that most mass remains in large grains: for the piecewise
linear discretisation in this test, the fractions of mass in the smallest and
largest bins are $1 \times 10^{-3}$ and $5 \times 10^{-1}$, respectively.
Although we directly computed these values, one can use $a^4 \times
\partial n(a,t) / \partial a$ as a proxy for the mass size distribution, given
that $\partial n(a, t) / \partial a$ has dimensions of inverse length.  Because
shattering is a collisional process, it will more rapidly transfer mass to
smaller grains in regions of higher dust density.

\subsection{Coagulation}\label{SEC:coagulation}

Although dust grains in high velocity collisions can shatter, grains in low
velocity collisions can stick together and aggregate.  This process of
coagulation shifts the grain size distribution to larger sizes, particularly in
dense regions of the ISM \citep{Chokshi1993, Jones1996, Dominik1997,
Hirashita2009, Mattsson2016}.  The formalism of dust coagulation also shares many
similarities with a wide class of population balance equations
\citep{Smoluchowski1916, Vigil1989, Dubovskii1992, Krivitsky1995, Lee2001,
Filbet2004, Fournier2005}.  A variety of methods have been used to numerically
model dust coagulation, including a piecewise constant grain size
discretisation \citep{Hirashita2009}, a Monte Carlo-based collision evolution
simulator \citep{Ormel2009}, direct numerical integration of the
integro-differential coagulation equation \citep{Asano2013b}, a method of
moments approach that does not explicitly evolve the grain size distribution
\citep{Mattsson2016}, and a finite volume method applied to the conservative
form of the coagulation equation \citep{Paruta2016}.

We explicitly include the effect of coagulation on the grain size distribution
by modifying the piecewise linear formalism developed in
Section~\ref{SEC:shattering} for shattering.  The governing equation for
coagulation is the same as equation~(\ref{EQN:shattering_sums}) for shattering,
except that we replace $m_\text{shat}^{k,j}(i)$ with the kernel
\begin{equation}
m_\text{coag}^{k,j}(i) \equiv
\begin{dcases}
m_k + m_j, & {a^\text{e}_i}^3 \leq \frac{m_k + m_j}{4 \pi \rho_\text{gr}/3} < {a^\text{e}_{i+1}}^3, \\
0, & \text{else}.
\end{dcases}
\label{EQN:coag_kernel}
\end{equation}
That is, when grains in bins $k$ and $j$ coagulate, they form a larger grain
whose mass is the sum of the colliding masses.  We also use velocity indicator
functions of the form
\begin{equation}
\mathbbm{1}_{v_\text{rel} < v_\text{coag}}(a^\text{c}_k, a^\text{c}_j) =
\begin{cases}
1, \quad v_\text{rel}(a^\text{c}_k, a^\text{c}_j) < v_\text{coag}^{k,j}, \\
0, \quad v_\text{rel}(a^\text{c}_k, a^\text{c}_j) \geq v_\text{coag}^{k,j}.
\end{cases}
\end{equation}
This ensures that grains in bins $k$ and $j$ coagulate only when their relative
velocity is below the coagulation threshold velocity, which depends on the
indices $k$ and $j$ and is calculated following equation~8 in
\citet{Hirashita2009}.  Given their high velocities, grains in the
largest size bins do not coagulate \citep{Hirashita2009}.  As a result, the
sizes of grains resulting from coagulation are less than the maximum allowed
value of $a_\text{max}$ and require no rebinning.

We calculate the mass transfer between grain size bins from coagulation
using equation~(\ref{EQN:shattering_sums}) together with the coagulation mass
kernel $m_\text{coag}^{k,j}(i)$.  In order to solve for number of grains,
$N_{i}(t)$, and slope, $s_{i}(t)$, in each bin, we require a second constraint.
For shattering, we utilised a heuristic about a bin's average grain radius,
since the inclusion of shattering is expected to produce new grains following a
roughly $\partial n / \partial a \propto a^{-3.3}$ size distribution.  For
coagulation, we do not have a similar analytic expression for the size
distribution of new grains in a bin.  As a result, we reuse the same form of
equation~(\ref{EQN:shattering_a_constraint}) and solve in bin $i$ for an
estimated average grain size $\langle a \rangle_i(t + \Delta t)$ at the end of
the time-step, where here $\Delta N_i$ denotes the number of grains entering a
bin from coagulation.  Since $\langle a \rangle_i(t)$ and $\langle a
\rangle_i^\text{shat}$ lie within bin $i$, so too will their weighted average
$\langle a \rangle_i(t + \Delta t)$.  We then solve for the number of grains
and slope in each bin by simultaneously solving
equations~(\ref{EQN:shattering_a_constraint})
and~(\ref{EQN:shattering_m_constraint}).

\begin{figure}
\centering
\includegraphics{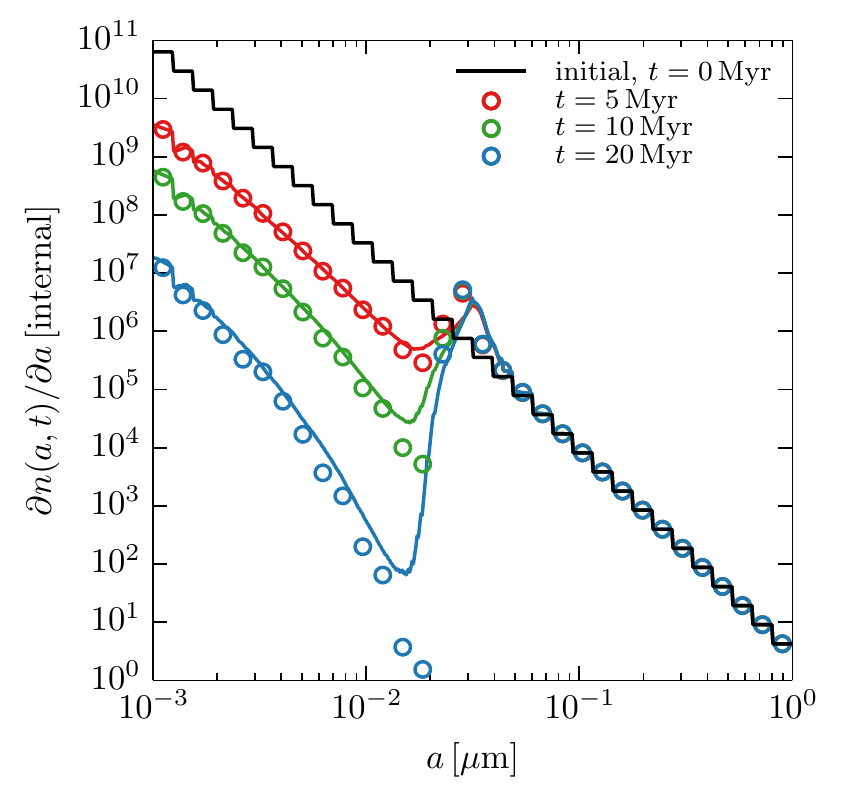}
\caption{Evolution of an initial grain size distribution $\partial n(a, t = 0
\, \text{Myr}) / \partial a \sim a^{-3.5}$ (black line) under the influence of
coagulation.  Grain size distributions are computed at various times (coloured
circles) using the piecewise linear discretisation and $N = 32$ bins.  Coloured
lines show profiles predicted by a numerical differential equations integrator
with many bins.  Coagulation reduces the number of small grains and in this
test produces grains near $a \approx 0.03 \, \mu\text{m}$.}
\label{FIG:coagulationtest_analytic}
\end{figure}

While this procedure does not take into account some physical intuition
for the size distribution of grains within a bin that results from coagulation,
it provides a second constraint that can be used together with the mass in a
bin to solve for the post-coagulation bin state.  As we demonstrate in
Figure~\ref{FIG:coagulationtest_analytic}, even this simplified procedure
allows the grain size distribution to track the effects of coagulation.

Figure~\ref{FIG:coagulationtest_analytic} demonstrates the effect of
coagulation on an initial grain size distribution $\partial n(a, t=0 \,
\text{Myr}) / \partial a \sim a^{-3.5}$.  To avoid unnecessary complexity, in
this test the velocities of grains in individual bins are not calculated by
interpolating the tabulated grain velocities from \citet{Yan2004} but instead
follow the form
\begin{equation}
v(a) = 1.1 \times 10^{3} \, \left( \frac{a}{0.1 \, \mu\text{m}} \right)^{1/2} \; \text{cm} \, \text{s}^{-1}
\end{equation}
from \citet{Hirashita2013c}.  For this test only, we set the threshold velocity
$v_\text{coag}^{k,j} = 1.1 \times 10^3 \, \text{cm} \, \text{s}^{-1}$,
independent of $k$ and $j$.  \citep[That is, we do not use the more complicated
expression in equation~4 of][which depends on the radii of colliding
grains.]{Hirashita2013c}  We note that small grains will coagulate.  For
simplicity, we assume a fixed collision angle $\cos \theta = -1$ when
calculating relative collision velocities.  We adopt a gas density $n_\text{H}
\approx 10^5 \, \text{cm}^{-3}$ and dust-to-gas ratio $D = 0.01$ and integrate
for $20 \, \text{Myr}$ using the piecewise linear discretisation with $N = 32$
bins.  We compare results with those predicted by a numerical integrator
solving coupled ordinary differential equations, starting from the same initial
conditions but using many times more bins.

In this test, coagulation shifts mass from small grains to medium-sized grains,
producing a local peak in the grain size distribution at $a \approx 0.03 \,
\mu\text{m}$.  Because mass is conserved, the number of small grains lost is
greater than the number of medium-sized grains created, and so total grain
number decreases.  Since large grains have velocities exceeding the coagulation
threshold, the grain size distribution for $a \gtrsim 0.05 \, \mu\text{m}$ is
largely unchanged from its initial state.  We do not include shattering in
this test, which would redistribute some of these large grains to smaller
sizes.  The results from the piecewise linear discretisation with $N = 32$
bins capture the qualitative behaviour predicted by the numerical differential
equations integrator.  Although there is some tension near $a \approx 0.02 \,
\mu\text{m}$, where the grain size distribution experiences a sharp increase,
results improve as more bins are added.

Together with Section~\ref{SEC:shattering}, this demonstrates how shattering
and coagulation can shift grains to smaller or larger grain sizes in a
mass-conserving manner.

\subsection{Time-step constraints and sub-cycling for grain size evolution}\label{SEC:gsd_timestep}

We apply a time-step constraint to ensure that changes in a dust particle's
grain size distribution are resolved.  When evolving a grain size distribution
over a time-step $\Delta t$ to account for some grain size process
(e.g.~shattering), we calculate the effective time-scale
\begin{equation}
\tau_\text{GSD} \equiv \min_{i} \frac{m_\text{d}}{\Delta M_i / \Delta t},
\end{equation}
where $m_\text{d}$ is the particle mass, $\Delta M_i$ is the change in mass in
bin $i$, and the minimisation is over all grain size bins.

We then update a grain size distribution with a time-step obeying $\Delta t <
\chi \tau_\text{GSD}$, where $\chi$ is a CFL-like parameter less than
unity.  This restricts the change in mass in a grain size bin to be at most a
fraction $\chi$ of the total particle mass.  We note that even processes that
conserve overall dust particle mass (shattering and coagulation) may transfer
mass between grain size bins and thus impose a time-step constraint.

It is often the case that grain size evolution takes place on shorter
time-scales than those for gravity and drag.  To improve computational
efficiency, we use a sub-cycling procedure that resolves these time-scales for
grain size evolution without subjecting dynamical forces to such short
time-steps.  We introduce a parameter $\lambda \geq 1$ and require the particle
time-step to resolve $\lambda \chi \tau_\text{GSD}$.  This constraint is
combined with the dynamical time-step requirements in
equation~(\ref{EQN:timestep_constraint}) to determine a dust particle's overall
time-step, during which dynamical forces like gravity and drag are applied and
kernel estimates are calculated.  Local grain size distribution updates are
then performed using approximately $\lambda$ time-steps of smaller size $\Delta
t < \chi \tau_\text{GSD}$, using kernel estimates (e.g.~gas density, dust
density, etc.) computed at the start of the larger particle time-steps.  While
these is some flexibility in choosing values for $\chi$ and $\lambda$, in
simulations of isolated galaxies presented in
Section~\ref{SEC:hernquist_spheres} we adopt $\chi = 0.1$ and $\lambda = 2$.

This sub-cycling avoids the need for many tiny updates to a dust particle's
position and velocity from gravity and drag forces when grain size evolution
takes place on time-scales much shorter than these dynamical forces.

\subsection{Dust drag with evolving grain size distributions}

In Section~\ref{SEC:drag_force}, we implemented a dust drag force assuming
grains had one fixed size.  Here, we briefly extend that formulation to account
for drag on dust particles with a grain size distribution.  Since the stopping
time-scale for one grain depends linearly on grain size $a$ (see
equation~\ref{EQN:t_s_full}), let $t_\text{s} \equiv \beta a$, where $\beta$
accounts for all other dependencies.  The magnitude of the drag force on a dust
particle with mass $m_\text{d}$ is given by
\begin{equation}
F_\text{d} = \int_{a_\text{min}}^{a_\text{max}} \left( \frac{\partial n}{\partial a} \right) \left( \frac{4 \pi \rho_\text{gr} a^3}{3} \right) \left( \frac{|\vec{v}_\text{d} - \vec{v}_\text{g}|}{\beta a} \right) \, \diff a,
\end{equation}
recalling that $\partial n / \partial a \times \diff a$ gives the number of
grains with radius in the interval $[a, a + \diff a]$.  We can
alternatively write the drag force as $F_\text{d} = m_\text{d}
|\vec{v}_\text{d} - \vec{v}_\text{g}| / t_\text{s}^\text{eff}$ in terms of an
effective stopping time-scale $t_\text{s}^\text{eff}$.  Equating these two
expressions, applying the piecewise linear grain size discretisation, and
solving for the effective stopping time-scale, we find
\begin{equation}
\begin{split}
t_\text{s}^\text{eff} &= \frac{3 \beta m_\text{d}}{4 \pi \rho_\text{gr}} \left[ \sum_{i=0}^{N-1} \int_{a^\text{e}_i}^{a^\text{e}_{i+1}} \left( \frac{N_i}{a^\text{e}_{i+1} - a^\text{e}_i} + s_i (a - a^\text{c}_i) \right) a^2 \, \diff a \right]^{-1} \\
&= \frac{3 \beta m_\text{d}}{4 \pi \rho_\text{gr}} \left\{ \sum_{i=0}^{N-1} \left[ \frac{N_i a^3 / 3}{a^\text{e}_{i+1} - a^\text{e}_i} + s_i \left(\frac{a^4}{4} - \frac{a^\text{c}_i a^3}{3} \right) \right]^{a^\text{e}_{i+1}}_{a^\text{e}_i} \right\}^{-1}.
\end{split}
\label{EQN:t_s_eff}
\end{equation}
In general, $t_\text{s}^\text{eff}$ is a function of time, as the grain size
distribution (i.e.~$N_i$ and $s_i$ values) will evolve in time.  Going forward,
we use this calculation of effective stopping time-scale when applying drag
kicks to dust particles with a full grain size distribution.

We caution, however, that applying an effective force to an entire dust
particle does not allow grains of different sizes to properly segregate when
moving in one direction.  In the isolated galaxy simulations presented in
Section~\ref{SEC:hernquist_spheres} without feedback, we neglect forces like
radiation pressure or unresolved galactic winds that could drive outflows on
large scales.  However, \citet{Ferrara1991} suggest that radiation pressure can
drive grains more than $100 \, \text{kpc}$ from the galactic centre, with
grains of different sizes and compositions experiencing different strength
forces.  Future simulations including feedback should address the limitation of
effective forces acting on dust particles.

\section{Dust production}\label{SEC:dust_production}

To this point, we have discussed how a dust particle's grain size distribution
evolves in time, but we have not yet specified how the initial grain size
distribution is set.  In practice, dust is injected into the ISM by evolving
stars \citep[e.g.][]{Todini2001, Ferrarotti2006}, and stars of different types
produce dust with different size distributions and chemical compositions.  This
production of solid dust happens simultaneously with the production of
gas-phase metals.

In this section, we first describe a stochastic procedure for forming dust
particles of a certain target mass as star particles evolve.  Then, we describe
the initial grain size distributions assigned to these newly created dust
particles.
There are several competing trends to balance in deciding whether to form many,
lower-mass dust particles or fewer, higher-mass dust particles.  On the one
hand, adopting a low mass threshold for dust particles reduces the
stochasticity of our particle creation scheme and better models continuous dust
injection from stars.  The more dust particles we create, the more finely we
can sample from a star's initial grain size distribution and see grains of
different sizes segregate during drag kicks.  On the other hand, creating many
dust particles can make simulations computationally inefficient.

\subsection{Dust particle creation}

Star formation prescriptions in cosmological simulations often stochastically
convert gas elements into star particles \citep[e.g.][]{Springel2003,
Vogelsberger2013, Hopkins2014, Schaye2015}.  Similarly, stochastic approaches
have been used to model stellar evolution and convert star particles to back
into gas particles in SPH simulations \citep{Torrey2012b}.  We parallel these
methods to stochastically create dust particles.

It is important to draw a distinction between the return of gas-phase metals
from a star to the ISM and the return of dust.  The galaxy formation physics in
\textsc{arepo} handles chemical enrichment of gas-phase metals into the ISM by
spreading the metal mass derived from stellar nucleosynthetic yields over
neighboring gas cells using a kernel-weighting.  However, because dust is not
tracked directly in gas cells but instead as a separate particle type, a
separate procedure is needed for dust than for gas-phase metals.

During a time-step in which a star particle of mass $m_*$ is expected to form
mass $\Delta m_\text{d}$ of dust, a new dust particle of mass $m_\text{d}$ is
created when a number chosen randomly between $0$ and $1$ is less than
\begin{equation}
p_\text{d} = \frac{m_*}{m_\text{d}} \left[ 1 - \exp\left( - \frac{\Delta m_\text{d}}{m_*} \right) \right].
\label{EQN:prob_dust}
\end{equation}
Multiplying equation~(\ref{EQN:prob_dust}) by $m_\text{d}$, this states that
during a time-step the expected dust mass formed equals the change in stellar
mass owing to dust synthesis.  Over the lifetime of a star particle, this
ensures that the correct amount of dust mass is produced in expectation.  We
initialise a dust particle with the same phase space information as the star
particle from which it was spawned.

The choice of desired dust particle mass $m_\text{d}$ affects how often dust
particles are spawned.  A natural parameterisation is $m_\text{d} =
\beta_\text{d} m_{*}^\text{init}$, where $\beta_\text{d}$ is a constant and
$m_{*}^\text{init}$ is the initial mass of the star particle at birth.  We note
that $m_{*}^\text{init}$ will typically be within a factor of a few of the mean
gas cell mass used as a target mass in the (de-)refinement scheme in
\textsc{arepo} \citep{Vogelsberger2012}.  Thus, $\beta_\text{d}$ controls what
fraction of a star's initial mass is converted into dust during each spawn
event.  In Section~\ref{SEC:dust_yields}, we show how $\beta_\text{d}$ impacts
the stochasticity of dust return.

Because chemical enrichment of gas-phase metals into surrounding gas cells does
not involve the creation of new particles, it can be handled during every
time-step in a continuous way.  However, for computational reasons it is
sometimes advantageous to adopt a discrete chemical enrichment scheme that only
periodically performs enrichment updates of accumulated mass in a deterministic
manner. Such discrete enrichment schemes have been used for dust, too.  For
example, the chemical enrichment model in \citet{Bekki2015b} has a star
particle creating dust particles only at three times in its evolution,
corresponding to the typical lifetimes of SNe II, SNe Ia, and asymptotic giant
branch~(AGB) stars.  While this method is deterministic, it introduces
artificial delays in the return of dust to the ISM and does not model
continuous enrichment.  We restrict ourselves to stochastic dust production
schemes in this work.

\subsection{Initial grain size distributions}\label{SEC:initial_gsd}

Once the decision has been made to spawn a dust particle of mass $m_\text{d}$,
we next initialise its grain size distribution.  The form of the grain size
distribution depends on the type of stars evolving off the main sequence
during the time-step.

Hydrodynamical modelling of pulsating AGB stars predicts that newly created SiC
grains obey a log-normal $a^4 \times \partial n / \partial a$ distribution,
with mass concentrated in large grains \citep{Yasuda2012}.  Following
\citet{Asano2013b}, we assume that the initial grain size distribution for all
dust produced by AGB stars takes the form
\begin{equation}
\frac{\partial n}{\partial a} = \frac{C}{a^5} \exp\left( - \frac{\ln^2(a/a_\text{AGB})}{2 \sigma_\text{AGB}^2} \right),
\label{EQN:dnda_agb}
\end{equation}
where $a_\text{AGB} = 0.1 \, \mu\text{m}$, $\sigma_\text{AGB} = 0.47$, and $C$
is a normalisation constant that determines the overall mass of the dust
particle.

Small grains are destroyed in the reverse shocks of SNe due to sputtering
\citep{Bianchi2007, Nozawa2007}, and the resulting mass of dust produced by SNe
is expected to be biased towards large grains \citep{Nozawa2007}.  The initial
grain size distribution used for dust produced by SNe II follows from
Figure~6(b) in \citet{Nozawa2007}, which presents the relative abundance of
dust grains at various discrete sizes for dust formed from a $20 \, \text{M}_\odot$
core-collapse SN in a gas of initial density $n_\text{H} = 1 \,
\text{cm}^{-3}$.  However, the discrete grain size distribution from
\citet{Nozawa2007} is not calculated at exactly the same sizes as the edges of
our grain bins.  To handle this, we calculate the grain size distribution at
each grain bin edge by logarithmically interpolating between neighboring
discrete $\partial n / \partial a$ values calculated in \citet{Nozawa2007}.
From the $\partial n / \partial a$ values at grain bin edges, we can calculate
the number of grains and slope in each bin.  Finally, like for AGB stars, we
scale the initial grain size distribution for dust particles produced by SNe II
by a constant to ensure the total mass in the grain size distribution equals
the particle's mass.

The time-scale for dust grains supplied by AGB stars to be injected into the
ISM is estimated as less than $10^{5} \, \text{yr}$ \citep{Mathews1999}, and
for the purposes of this work we assume no delay in transporting AGB dust into
a dust particle in the surrounding gas.  This is similar to the time-scale over
which dust grains are subjected to reverse shocks in SNe \citep{Bianchi2007}.
Since we employ the same stellar nucleosynthetic yields used by
Illustris \citep{Vogelsberger2013}, AGB stars are assumed to have masses in
the range $1 - 6 \, \text{M}_\odot$, while SNe II have masses in the
range $6 - 100 \, \text{M}_\odot$.

Because the grain size distribution for dust produced by SNe Ia is uncertain,
we assume that dust produced by SNe Ia follows the same size distribution as
that from SNe II.  However, the net amount of dust produced by SNe Ia is
subdominant compared to that from SNe II and AGB stars \citep{Nozawa2011}, and
some works choose to entirely ignore dust production from SNe Ia
\citep[e.g.][]{Asano2013b}.  Because SNe Ia dust yields are so low, as
discussed in Section~\ref{SEC:dust_yields}, our choice for the initial size
distribution of dust from SNe Ia thus does not meaningfully affect results.

\begin{figure}
\centering
\includegraphics{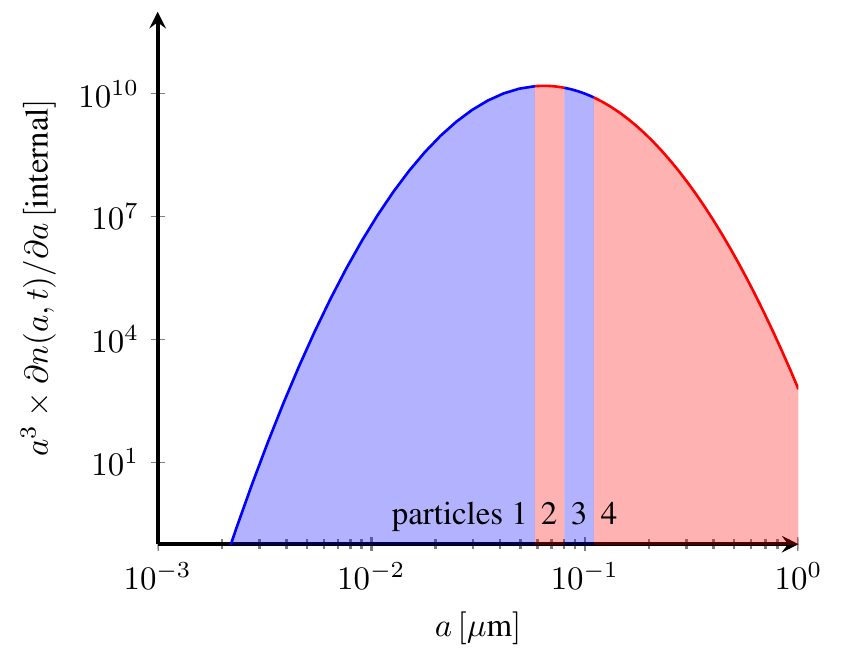}
\caption{Schematic demonstrating the possible creation of multiple dust
particles from an initial grain size distribution.  In this example, we divide
the grain size distribution corresponding to dust from AGB stars (given by
equation~\ref{EQN:dnda_agb}) into $N_\text{d} = 4$ contiguous, equal-mass
segments (shaded regions), each of which is assigned to one new dust particle.
Here, the vertical axis shows the differential mass density $a^3 \times
\partial n / \partial a$.  Alternatively, setting $N_\text{d} = 1$ would create
one dust particle covering the full grain size distribution, an approach used
in Section~\ref{SEC:hernquist_spheres}.}
\label{FIG:production_schematic}
\end{figure}

When deciding to stochastically create total dust mass $m_\text{d}$ with a
corresponding grain size distribution $\partial n / \partial a$, there are a
few possible approaches.  One approach is to break $\partial n / \partial a$
into several contiguous segments and create $N_\text{d}$ dust particles of mass
$m_\text{d} / N_\text{d}$, with each particle's initial grain size distribution
covering only a limited grain size range.  This approach is illustrated in
Figure~\ref{FIG:production_schematic}.  Summing over particles, this procedure
gives the correct initial grain size distribution, and it also allows for
grains of different sizes to stratify under a strong drag force.  However,
splitting the initial grain size distribution in this way increases the number
of dust particles and computational cost.  Additionally, over time the dust
particles' initially narrow size distributions will shift to larger and smaller
sizes as a result of the physical processes detailed in
Section~\ref{SEC:grain_size_evolution}, reducing the advantages of creating
multiple particles.

In the galactic simulations in this paper, we take the simplest approach and
assign the full grain size distribution to one dust particle.  This has the
benefit of treating a large range of grain sizes with just one particle, and
effective drag updates can be applied using equation~(\ref{EQN:t_s_eff}).  This
method has a downside: it does not effectively capture the separation of grains
of different sizes via the drag force.  If constituent grains cover three
orders of magnitude in size and thus have drag accelerations varying by the
same amount, moving the dust particle using an effective drag acceleration
forces its grains to have the same drag acceleration.  However, in galaxy
applications where the drag stopping time-scale is short and dust is
well-coupled to the gas, this is not a serious limitation.

\subsection{Dust elemental yields}\label{SEC:dust_yields}

The probabilities used to stochastically create dust particles are set by
the total dust mass $\Delta m_\text{d}$ produced during a star's time-step.
The total dust mass is the sum of dust masses contributed by individual
chemical elements, and these dust elemental yields are a function of a star's
mass and metallicity.  Dust yields for AGB stars \citep{Zhukovska2008,
Ventura2012, Nanni2013, Schneider2014}, SNe II \citep{Todini2001, Bianchi2007,
Nozawa2007, Nozawa2010, Gall2011c, Temim2013, Gall2014, Marassi2015}, and SNe
Ia \citep{Nozawa2011} have been studied in detail.  Dust formation can also be
characterised in terms of condensation efficiency, the fraction of metals
returned that exist in solid dust grains, with the remainder of metals
occupying the gas phase.  Below, we outline the dust yields that we adopt in
calculating dust mass return from stellar populations.

For AGB stars, we interpolate the results from \citet{Schneider2014}, which
predicts dust yields for stars in the mass range $1 - 8 \, \text{M}_\odot$ and
metallicity range $0.001 \leq Z \leq 0.008$.  These yields are calculated for
four grain types: carbon, silicate, SiC, and iron.  We use these yields to
determine the yields on an element-by-element basis for C, O, Mg, Si, and Fe,
which are tracked in our dust model.  Paralleling \citet{Zhukovska2013}, we
assume that silicate grains are 50 per cent Mg$_2$SiO$_4$, 30 per cent
MgSiO$_3$, and 20 per cent Fe$_2$SiO$_4$ to set the element-by-element dust
mass return and thus condensation efficiencies.

For SNe II, we adopt dust yields from \citet{Nozawa2010}, which presents the
mass of dust formed for each of the elements tracked in our model (C, O, Mg,
Si, and Fe) in the core-collapse of a SN IIb with mass $18 \, \text{M}_\odot$
and metallicity $Z = 0.02$.  We assume that these results hold for
core-collapse SNe of all types, noting that the condensation efficiency of this
SN IIb is similar to that predicted for SNe IIP \citep{Nozawa2003, Nozawa2010}.
Because \citet{Nozawa2010} models only one SN IIb, we assume that the mass of
dust formed from a core-collapse SN scales linearly with progenitor mass.
Future work could explore more detailed models of SN dust condensation
as a function of different progenitor masses \citep[e.g.][]{Bianchi2007,
Nozawa2007}.

For SNe Ia, we assume that the condensation efficiency of individual elements
is the same as for SNe II, noting that dust grains produced by SNe Ia share a
similar elemental distribution as dust grains formed in core-collapse SNe
\citep{Nozawa2011}.  However, because SNe Ia form fewer metals than SNe II in a
stellar population and are not thought to be major contributors of dust
formation \citep{Nozawa2011}, the choice of SNe Ia condensation efficiencies
does not strongly impact our results.

While there may be stochastic deviations from these dust yields as individual
dust particles are spawned, our procedure gives the correct IMF-averaged dust
yields in expectation.  As discussed at the start of
Section~\ref{SEC:grain_size_evolution}, when a dust particle is spawned, we
compute the fraction of its total mass given by individual chemical elements.
These fractions are then updated when the dust particle accretes mass from or
returns mass to the ISM according to the procedure outlined in
Section~\ref{SEC:piecewise_linear}.

\begin{figure}
\centering
\includegraphics{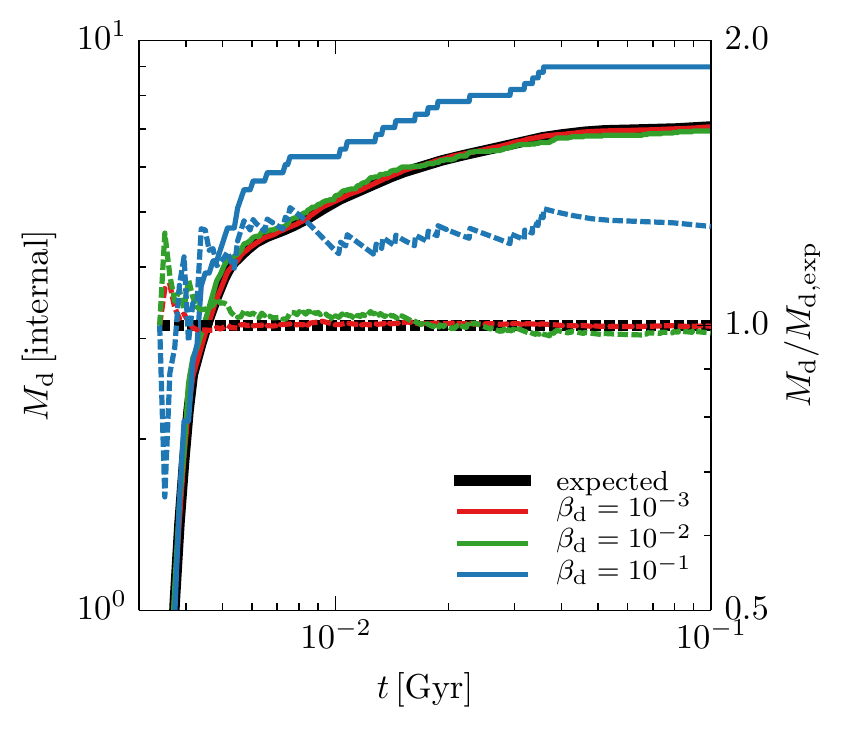}
\caption{Cumulative mass of dust stochastically produced by a group of star
particles born at $t = 0 \, \text{Gyr}$.  Solid coloured lines show dust mass
as a function of time (left axis) for three choices of $\beta_\text{d}$, the
ratio between a spawned dust particle's mass and the initial mass of a star
particle.  The solid black line marks the cumulative amount of dust expected to
form using the dust yields, equivalent to the limit $\beta_\text{d} \to 0$.
Dashed lines show the ratios between the simulated dust mass profiles and the
expected dust mass profile (right axis).  Smaller values of $\beta_\text{d}$
lead to less stochastic behaviour, at the expense of spawning more dust
particles.}
\label{FIG:productiontest_evolution}
\end{figure}

Figure~\ref{FIG:productiontest_evolution} demonstrates the stochastic formation
of dust for a group of 512 star particles, all assumed to be born at $t = 0 \,
\text{Gyr}$ with solar metallicity and subject to a \citet{Chabrier2003}
IMF over the mass range $0.1 - 100 \, \text{M}_\odot$.  For the purposes of
this test, dust particles are not subject to any grain size evolution in the
ISM and thus do not gain or lose mass after creation.  We compare the expected
mass of dust that would be obtained by continually enriching surrounding gas
with the mass of dust obtained via the stochastic spawning of dust particles.
We vary the parameter $\beta_\text{d}$, the ratio between a dust particle's
mass and a star particle's initial mass.  As $\beta_\text{d}$ decreases, the
mass of stochastically spawned dust particles more closely follows the expected
dust mass.  However, this improved accuracy comes at the expense of needing to
spawn more, lower-mass dust particles compared to larger values of
$\beta_\text{d}$.  The optimal value of $\beta_\text{d}$ for a particular
simulation should be determined by balancing the need for accurate dust mass
return with the need for computational efficiency.

\subsection{Dust refinement and de-refinement}\label{SEC:derefinement}

In some circumstances, it may be desirable to constrain the mass of individual
dust particles.  For example, a dust particle that undergoes rapid accretion
may become much more massive than dust particles newly spawned from stars,
while a dust particle in hot gas could see a significant fraction of its mass
thermally sputtered.  Here we outline algorithms that can be used to
reduce the spread in dust particle masses.

Large dust particles can be refined by splitting them in two whenever their
mass exceeds some threshold value $m_\text{d}^\text{max}$.  While the grain
size distribution can be divided between these two new particles in various
ways, it is simplest to divide it equally so that each new particle has half of
the number of grains and slope in every bin.  The two new dust particles are
displaced in opposite directions from the old dust particle's position along a
randomly-chosen axis by a distance of $0.025h$, where $h$ is the smoothing
length enclosing neighboring gas cells computed via equation~(\ref{EQN:N_ngb}).
The new particles keep the same dust velocity so that momentum is conserved.
This procedure has no communication overhead but increases the dust particle
count, adding computational cost.

De-refinement of dust particles works in a similar way.  If the mass of a
dust particle falls below $m_\text{d}^\text{min}$, we search for its nearest
dust particle neighbor with mass above $m_\text{d}^\text{min}$.  A new dust
particle with mass equal to sum of the two particles' masses is placed at the
centre of mass, given a new velocity to conserve momentum, and assigned a grain
size distribution obtained by adding the particles' individual distributions.
In principle, the neighbor lookup could require communication between
processors.

\begin{figure}
\centering
\includegraphics{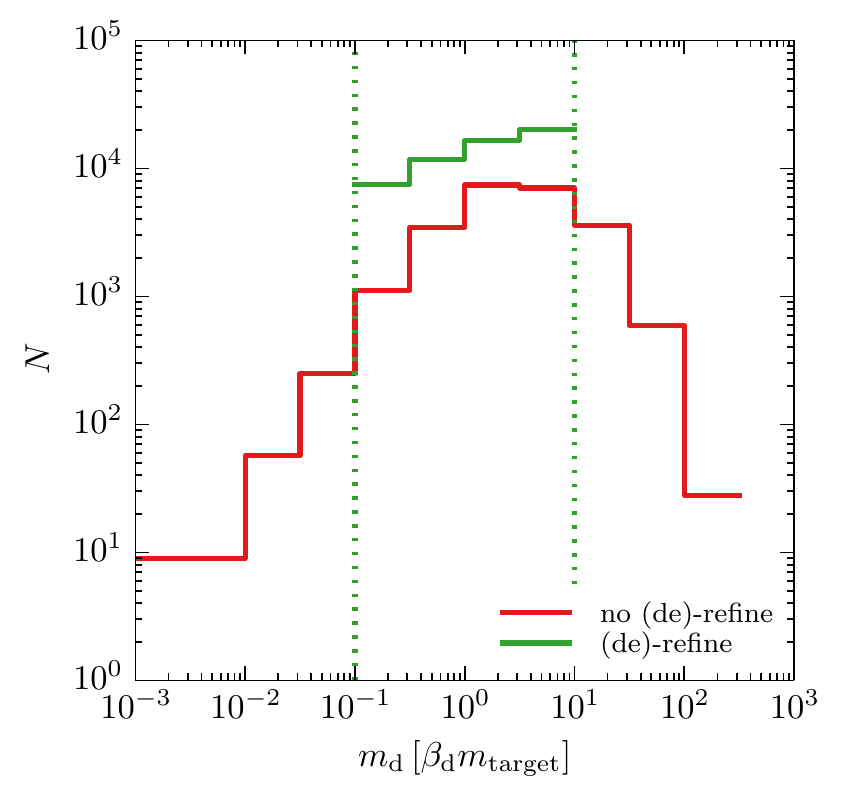}
\caption{Number of dust particles as a function of dust mass at $t = 1.5 \,
\text{Gyr}$ for the isolated disc galaxy presented in
Section~\ref{SEC:hernquist_spheres}.  Distributions are shown for the
medium resolution full physics run with (de)-refinement (green) and a run
without (de)-refinement (red).  Dust masses are shown in units of
$\beta_\text{d} \, m_\text{target}$, where $\beta_\text{d} = 0.1$ and
$m_\text{target}$ is the fixed target gas cell mass.  Vertical dotted lines
show the minimum and maximum dust particle masses allowed by the
(de)-refinement scheme.  Without (de)-refinement, the distribution of dust
particle masses develops tails at low and high mass.}
\label{FIG:example_derefinement}
\end{figure}

We implement these schemes for dust refinement and de-refinement in
\textsc{arepo}, ensuring no significant variation among dust particle masses.
Figure~\ref{FIG:example_derefinement} shows the dust particle mass
distributions that arise at $t = 1.5 \, \text{Gyr}$ in the medium-resolution
isolated galaxy simulations detailed in Section~\ref{SEC:hernquist_spheres}.
We contrast runs with and without (de)-refinement.  Both runs employ
$\beta_\text{d} = 0.1$, meaning dust particles are created with mass one-tenth
of their star particle's initial mass.  Initial star particle masses
are close to $m_\text{target}$, the mean gas cell mass adopted as a target mass
when (de)-refining gas cells \citep{Vogelsberger2012}.  The isolated galaxy run
with dust (de)-refinement limits dust particle masses to be between
$m_\text{d}^\text{min} = 0.01 \, m_\text{target}$ and $m_\text{d}^\text{max} =
m_\text{target}$.  However, in the run without (de)-refinement some dust
particles reach masses more than an order of magnitude beyond these
mass limits.  In this run, the tails in the dust particle mass distribution
become wider with time and are undesirable.

\begin{table*}
\centering
\caption{Description of the grain size physics included in the three dust
models used in isolated galaxy simulations.  Each model adds successively more
physics: the ``production only'' run solely produces dust particles and
includes no grain size evolution, the ``no shattering/coagulation'' run
includes all number-conserving processes, and the ``full physics'' run includes
both number-conserving and mass-conserving processes.}
\begin{tabular}{ll}
\hline
Name & Grain size physics \\
& \\
\hline
\hline
production only & No grain size evolution \\
no shattering/coagulation & Grain growth, thermal sputtering, SN shock-driven destruction \\
full physics & Grain growth, thermal sputtering, SN shock-driven destruction, shattering, coagulation \\
\hline
\end{tabular}
\label{TAB:simulations}
\end{table*}

\subsection{Dust-dust neighbor searches}\label{SEC:dust_ngb_searches}

When dust particles search for neighboring gas cells, smoothing lengths enclose
a weighted number of gas cells (e.g.~see equation~\ref{EQN:N_ngb}).  However,
we also need to perform searches for neighboring dust particles: dust density
estimates are needed for shattering and coagulation.  Because dust particle
masses can vary more strongly than gas cell masses, even with dust
(de)-refinement turned on, we calculate dust-dust smoothing lengths by
enclosing a desired amount of dust mass rather than a desired number of
neighbors.  This avoids circumstances where a dust particle with many low mass
dust neighbors calculates a small smoothing length and estimates a dust density
despite little dust mass enclosed in the kernel.

To be precise, we iteratively solve for dust-dust smoothing lengths by forcing
the kernel to enclose total dust mass in the range $(64 \pm 16) \times
(\beta_\text{d} m_\text{target})$, where $\beta_\text{d} m_\text{target}$ is
the typical mass of dust particles when produced by stars and $\beta_\text{d} =
0.1$ is our fiducial value.  Using this smoothing length, we then perform dust
density estimates with the usual kernel-weighting scheme (see
equation~\ref{EQN:rho_kernel}).

When de-refinement is active, we also require the smoothing length to enclose a
dust neighbor with mass above $m_\text{d}^\text{min}$, the minimum allowable
dust mass.  This way, a dust particle in need of de-refinement can follow the
procedures in Section~\ref{SEC:derefinement} and be de-refined into its
high-mass neighbor.  If necessary, we temporarily allow the kernel's enclosed
dust mass to exceed the upper bound in the previous paragraph in order to find
a high-mass dust neighbor.

\section{Isolated disc galaxy simulations}\label{SEC:hernquist_spheres}

As a first application of our dust model, we simulate the formation and dust
content of an isolated disc galaxy.

\subsection{Initial conditions}

The initial matter distribution consists of slowly rotating gas superimposed on
a collisionless dark matter halo following a~\citet{Hernquist1990} profile.
Initially, the halo has a mass of $10^{12} \, \text{M}_\odot$ with a $10$ per
cent gas fraction.  We set the dimensionless spin parameter to $\lambda = 0.05$
with a concentration $c = 6$.  To start, the number of gas cells and dark
matter particles is $8 \times 10^6$ for each component and $16 \times 10^6$ in
total.  We run this test with cooling and star formation, but without any
feedback processes. The grain size evolution calculations are performed using
the piecewise linear discretisation with $N = 16$ bins covering the size range
from $a_\text{min} = 0.001 \, \mu\text{m}$ to $a_\text{max} = 1 \,
\mu\text{m}$.  Dust particles are stochastically created with mass equal to ten
per cent of a star particle's initial mass (i.e.~$\beta_\text{d} = 0.1$),
refined when the particle mass exceeds the target gas cell mass
(i.e.~$m_\text{d}^\text{max} = m_\text{target}$), and de-refined when the
particle mass is less than one per cent of the target gas cell mass
(i.e.~$m_\text{d}^\text{min} = 0.01 m_\text{target}$).  In this test, we only
create $N_\text{d} = 1$ dust particle per spawn event and do not subdivide the
grain size distribution across multiple particles.  For the grain size
evolution time-step constraint detailed in Section~\ref{SEC:gsd_timestep}, we
adopt $\chi = 0.1$.  We perform grain size evolution updates using the
sub-cycling parameter $\lambda = 2$, meaning that dust particle dynamical
time-steps are only required to resolve twice the grain size evolution
time-step.

We study the evolution of this isolated disc galaxy using three dust models,
each adding progressively more grain size physics as summarised in
Table~\ref{TAB:simulations}.  The first, ``production only,'' creates dust
particles using the stochastic prescription from
Section~\ref{SEC:dust_production} but does not include any grain size evolution
(i.e.~a dust particle's grain size distribution is set upon creation and is
fixed).  The second, ``no shattering/coagulation,'' includes dust production and
also allows grains to undergo number-conserving size evolution processes like
accretion (Section~\ref{SEC:grain_growth}), thermal sputtering
(Section~\ref{SEC:thermal_sputtering}), and SN destruction
(Section~\ref{SEC:supernova_destruction}).  Finally, the ``full physics'' model
adds shattering (Section~\ref{SEC:shattering}) and coagulation
(Section~\ref{SEC:coagulation}).  This latter model thus includes all of the
grain size physics detailed in Section~\ref{SEC:grain_size_evolution}.

In all of these models, dust is dynamically coupled to the gas through
the drag force detailed in Section~\ref{SEC:drag_force}.  However, typical drag
stopping time-scales (e.g.~equation~\ref{EQN:t_s_Myr}) are short compared to
the simulation duration, and dust and gas are not significantly decoupled.  The
results we present below are largely unchanged in the limit where stopping
time-scale $t_\text{s} \to 0$ and drag acts instantaneously to set a dust
particle's velocity equal to the local gas velocity.  However, dust and gas may
be more decoupled in future galaxy simulations including feedback or when
studying smaller portions of the ISM.

Furthermore, to improve the performance of our code, we simplify the integral
in equation~(\ref{EQN:I_k_j}) used during shattering and coagulation to
determine the total cross section for collisions between grains in two
different bins.  We ignore grain size distribution bin slopes when calculating
$I^{k,j}(t)$ (i.e.~we assume that $s_k(t) = s_j(t) = 0$) and instead only use
the numbers of grains $N_k(t)$ and $N_j(t)$ in the bins.  This reduces
considerably the number of floating-point operations needed to evaluate these
integrals, and we have verified that in this isolated galaxy application this
change has no significant effect on our results.

\begin{figure*}
\centering
\includegraphics{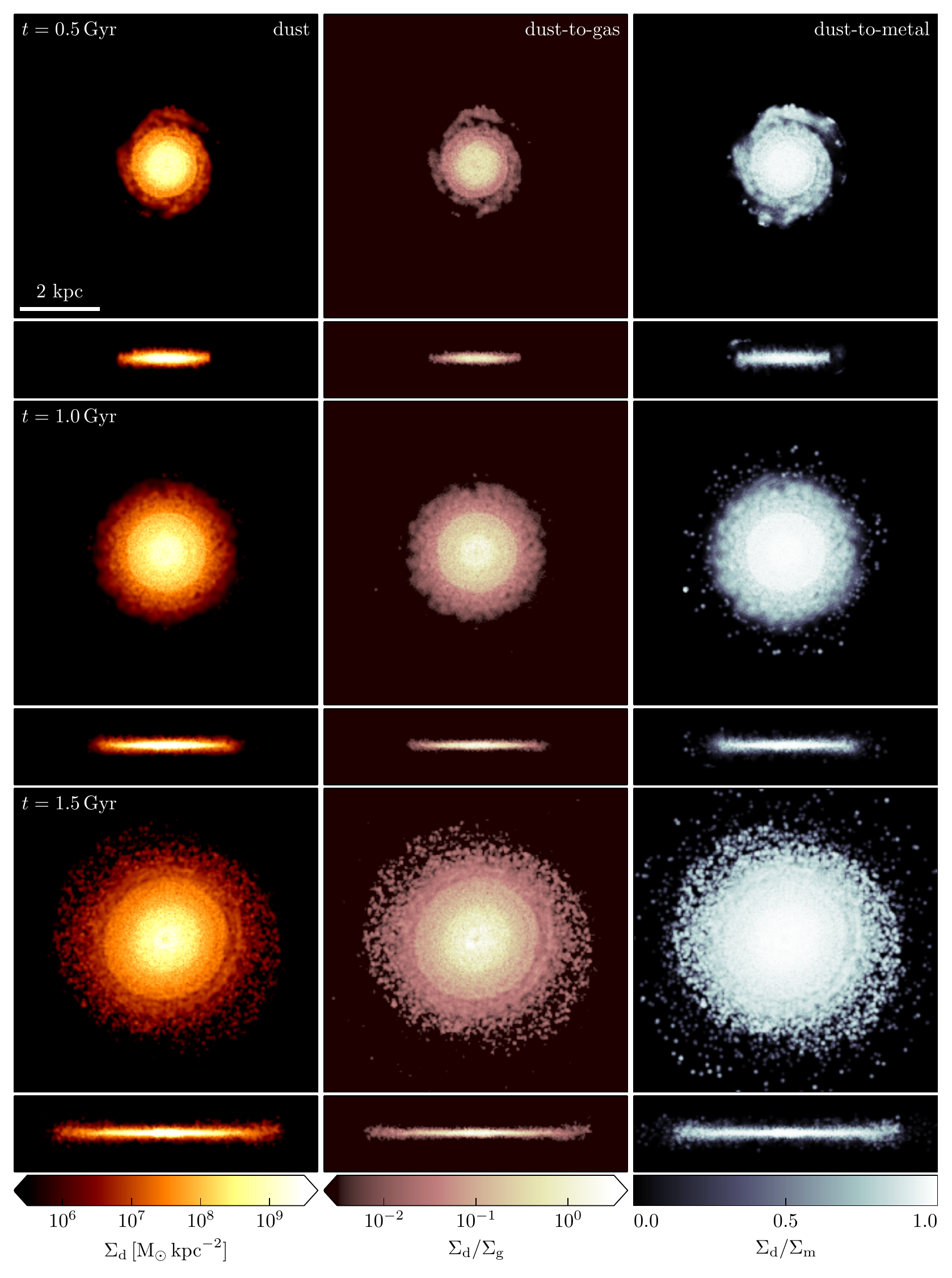}
\caption{Evolution of face-on and edge-on projected dust surface density
(left), dust-to-gas ratio (centre), and dust-to-metal ratio (right) for the
full grain physics run at $t = 0.5$, $1$ and $1.5 \, \text{Gyr}$.  Dust surface
density, dust-to-gas ratio, and dust-to-metal ratio decrease with radius.  This
simulation lacks feedback, overconsuming gas and overproducing dust.}
\label{FIG:hernquisttest_surfacedensity_timeevolution}
\end{figure*}

\begin{figure*}
\centering
\includegraphics{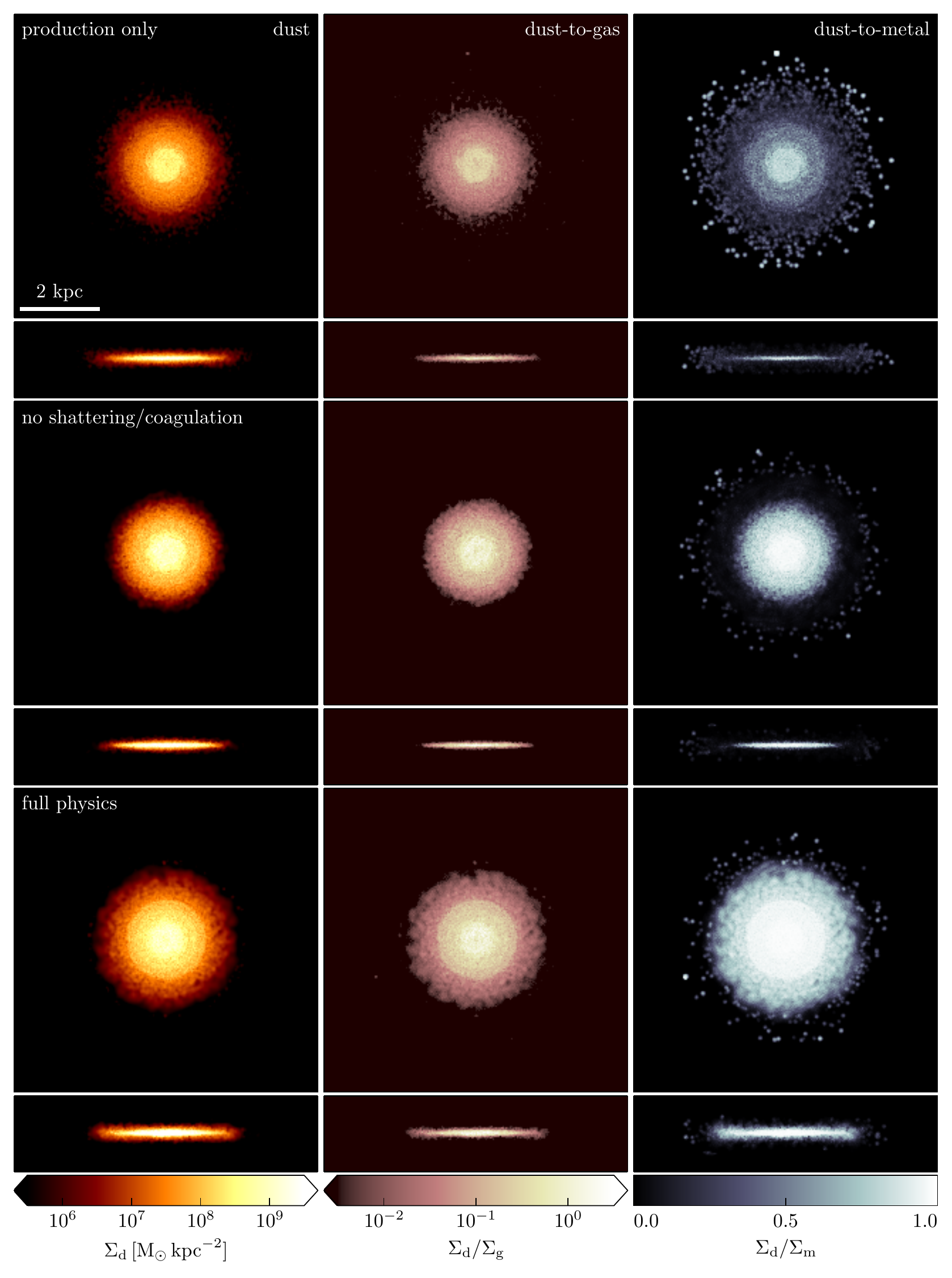}
\caption{Projections of dust surface density (left), dust-to-gas ratio
(centre), and dust-to-metal ratio (right) at $t = 1 \, \text{Gyr}$ for three
grain physics models: the run with only dust production and no grain size
evolution (top), the model with all number-conserving grain processes
(accretion, sputtering, and SN destruction) but lacking shattering and
coagulation (middle), and the full physics model (bottom).}
\label{FIG:hernquisttest_surfacedensity_comparison}
\end{figure*}

\subsection{Predicted dust population}

Figure~\ref{FIG:hernquisttest_surfacedensity_timeevolution} shows the time
evolution of the full physics model, presenting face-on and edge-on projections
of dust surface density, dust-to-gas ratio, and dust-to-metal ratio.  We note
that numerical convergence properties of the dust model across different
resolutions are discussed in Appendix~\ref{SEC:convergence_study}.  The panels
illustrate the formation of a disc galaxy whose dust mass increases with time.
The dust surface density shows a clear radial gradient, with a peak central
value at $t = 1.5 \, \text{Gyr}$ of roughly $10^{9} \, \text{M}_\odot \,
\text{kpc}^{-2}$.  It is important to note that we use no stellar feedback, and
so no winds are driven from the disc that could reduce star formation (and in
turn dust formation) or produce dust outflows.  Thus, our dust surface
densities should not be compared to observations of Milky Way-like systems.
Similarly, dust-to-gas ratios decrease with radius from the galactic centre but
lie above the approximate $10^{-2}$ value associated with the Milky Way
\citep{Draine2007}.

The absence of feedback strongly affects the normalisation in dust-to-gas and
dust-to-metal ratios, since gas is overconsumed and dust is overproduced.  The
dust-to-metal ratio is defined as the ratio of dust mass to total (dust plus
gas-phase) metal mass and is near unity for this run without feedback.
In our current model, dust particles are not removed when nearby gas
cells stochastically convert to star particles.  As a result, star formation in
our model reduces the supply of ISM gas-phase metals but not dust.  However,
the depletion of dust via star formation, known as astration, is not the only
physical process that can reduce ISM dust mass.  In
Appendix~\ref{SEC:astration}, we compare the time-scales for astration and the
destruction of dust in SN shocks, two processes that scale with dust-to-gas
ratio and star-formation rate.  Dust loss via astration is expected to be
subdominant compared with SN destruction.  In our full physics model that
includes SN dust destruction, the dust-to-metal ratio only changes by a few
per-cent when incorporating an estimate of the astration rate.  The dust
content of the production only run is more strongly affected when we include a
model for astration, producing a dust mass and dust-to-metal ratio lower by
about a factor of four at $t = 1 \, \text{Gyr}$.  However, we argue in
Appendix~\ref{SEC:astration} that the production only run with an astration
model is not physically realistic: it includes astration but neglects the SN
dust destruction process that is expected to dominate dust mass loss in the
ISM.  Furthermore, it neglects the ability for dust grains to gain mass and
offset the effects of astration and SN dust destruction.  Although astration is
expected to be subdominant to other dust destruction processes, we plan to
directly model this process in future work.

To assess the impact of grain size evolution,
Figure~\ref{FIG:hernquisttest_surfacedensity_comparison} shows the dust surface
density, dust-to-gas ratio, and dust-to-metal ratio in the isolated disc galaxy
at $t = 1 \, \text{Gyr}$ using the different dust physics models listed in
Table~\ref{TAB:simulations}.  In all three models, dust surface
density, dust-to-gas ratio, and dust-to-metal ratio decrease as a function of
radial distance from the disc centre.  However, the no shattering/coagulation
and full physics runs show higher dust surface density, dust-to-gas ratio, and
dust-to-metal ratio than the production only run, which lacks grain size
evolution.  Overall, the results in
Figure~\ref{FIG:hernquisttest_surfacedensity_comparison} suggest that accretion
increases dust mass more quickly than sputtering and SN destruction decrease
dust mass.  We note that the normalisation in dust-to-metal ratio is strongly
affected by the lack of feedback and overconsumption of gas (and gas-phase
metals) into stars.

\begin{figure}
\centering
\includegraphics{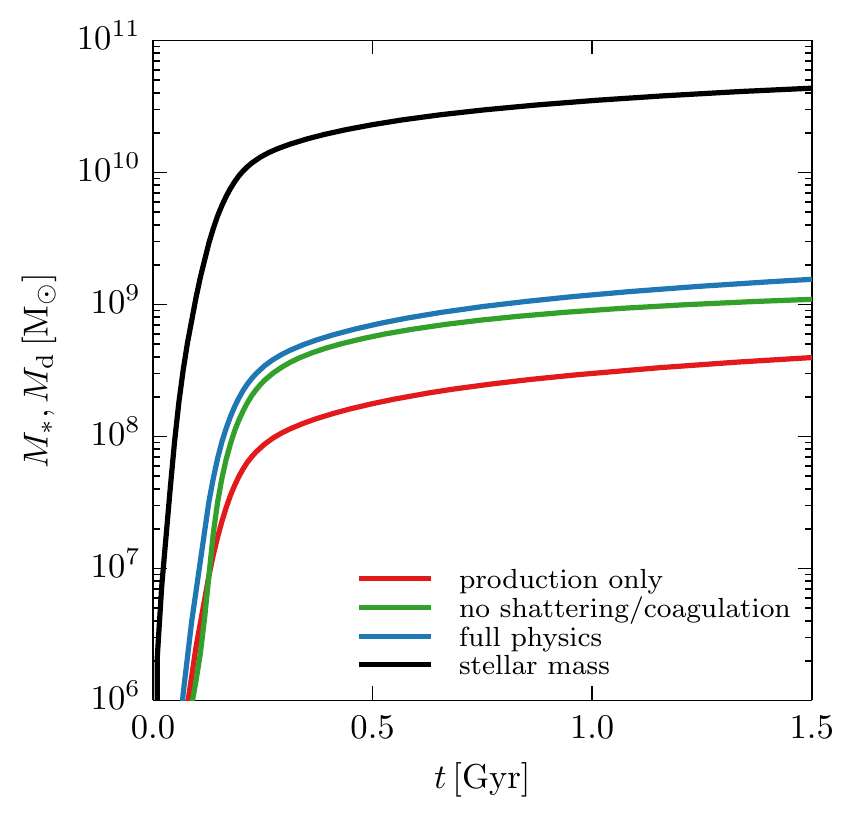}
\caption{Time evolution of the isolated galaxy's dust mass, computed for models
with various dust physics (coloured lines).  Stellar mass evolution is nearly
identical across these dust model variations, and for readability we plot the
stellar mass from only one of these runs (black line).  Dust mass is shown for
the model with dust production but without grain size evolution (red), the
model including solely number-conserving grain processes like accretion,
thermal sputtering, and SN destruction (green), and the full grain physics
model including shattering and coagulation (blue).  The full physics model
produces the largest dust mass.}
\label{FIG:hernquisttest_evolution}
\end{figure}

\begin{figure}
\centering
\includegraphics{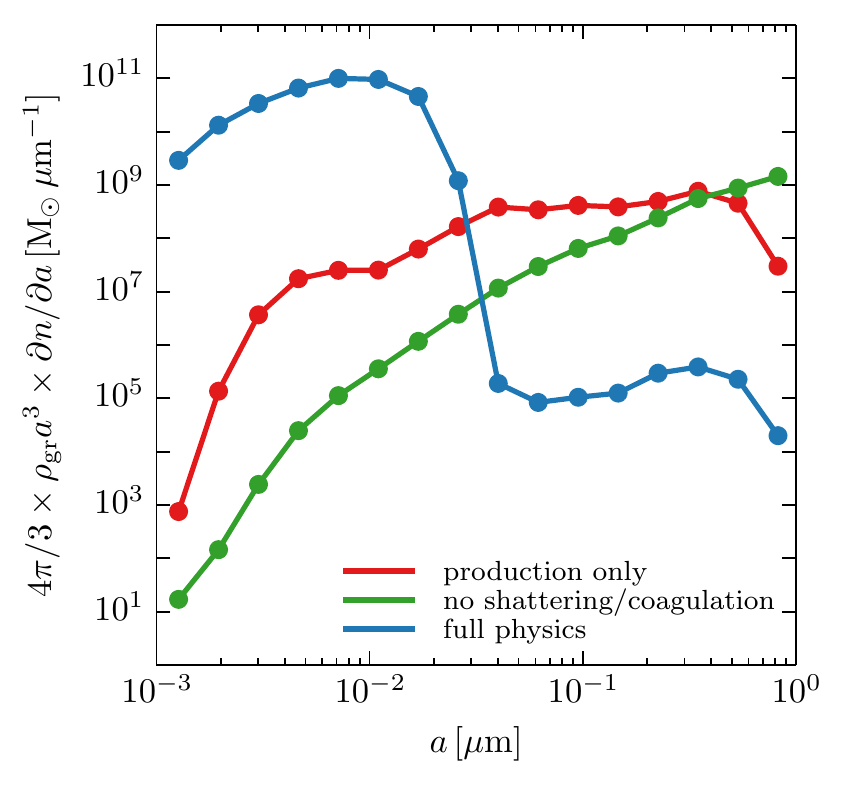}
\caption{Distribution of dust mass among grains of different sizes predicted
for the isolated galaxy at $t = 1 \, \text{Gyr}$ by the same models presented
in Figure~\ref{FIG:hernquisttest_evolution}.  The grain size distribution
($\partial n / \partial a$, with units of inverse length) is summed over all
dust particles in the galaxy, and the vertical axis plots this total grain size
distribution weighted by the masses of grains of different sizes.  Integrating
these profiles gives the total dust masses predicted for the galaxy.  The model
with shattering is qualitatively different, shifting dust mass to smaller grain
sizes, although its total dust mass from
Figure~\ref{FIG:hernquisttest_evolution} is similar to those of the other
models.}
\label{FIG:hernquisttest_gsd}
\end{figure}

Figure~\ref{FIG:hernquisttest_evolution} shows the time evolution of dust and
stellar mass using the three grain size evolution models.  Because the presence
of dust affects gas dynamics and star formation only slightly, we display the
stellar mass evolution for just one model.  Stellar mass increases rapidly at
early times -- reaching roughly $10^{10} \, \text{M}_\odot$ after about $200 \,
\text{Myr}$ -- before slowing.  As suggested by the visuals in
Figure~\ref{FIG:hernquisttest_surfacedensity_comparison}, the three grain size
models show similar qualitative behaviour, characterised by a sharp rise in
dust mass over the first half $\text{Gyr}$.

The production only model with no grain size evolution is easiest to
understand.  In this model, dust particle masses never change, and dust mass
closely traces stellar mass, albeit with a lower normalisation owing to an
effective dust yield for mass return from stars.  The no shattering/coagulation
model with accretion, thermal sputtering, and SN destruction produces about
three times as much dust as in the production only run.  The amount of dust
gained by allowing grains to grow and accrete gas-phase metals thus exceeds the
amount of dust destroyed by thermal sputtering and SN shocks.  This trend is
likely to persist in runs with feedback, given that reduced star formation
rates will lead to reduced SN destruction.

\begin{figure*}
\centering
\includegraphics{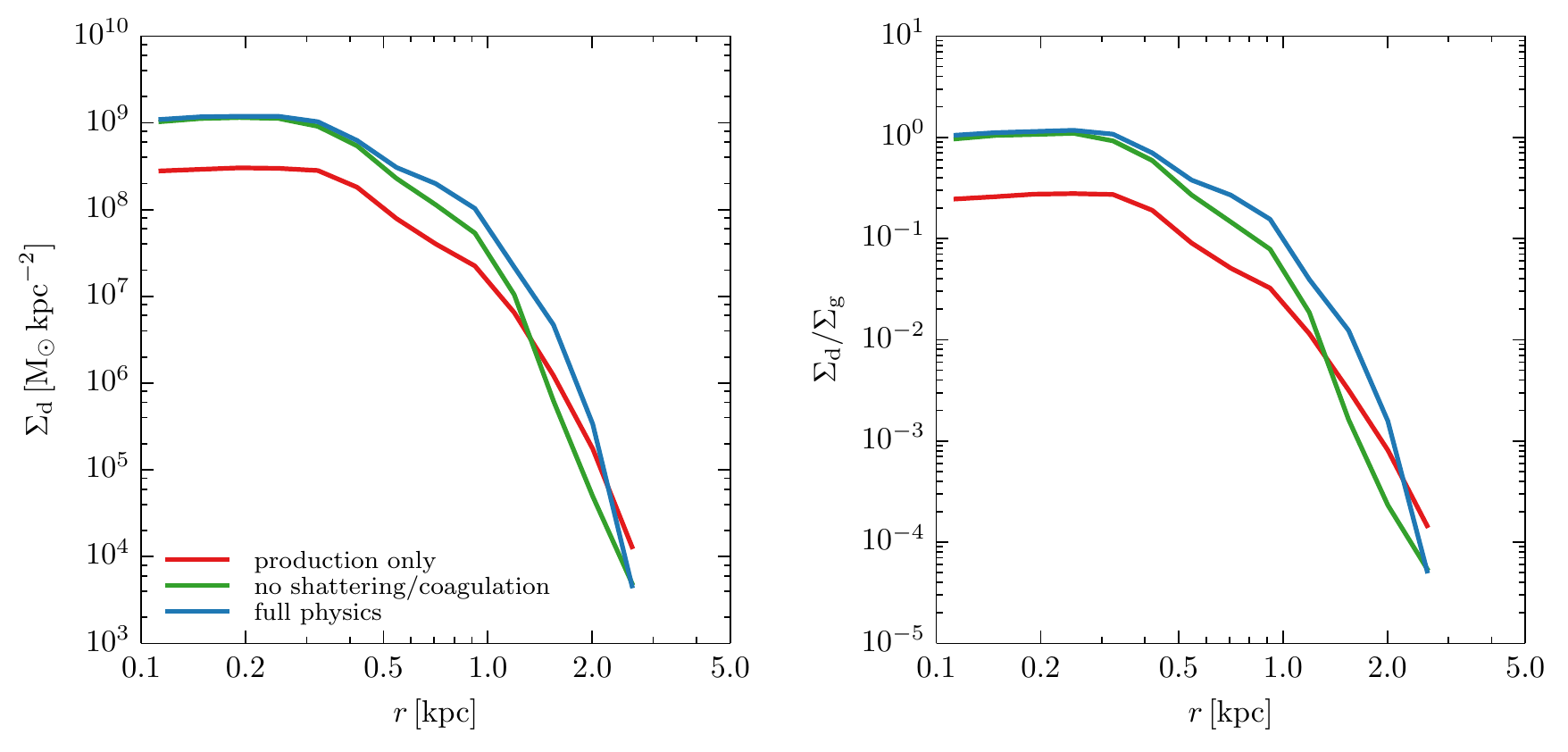}
\caption{Profiles of dust surface density (left) and dust-to-gas ratio (right)
as a function of two-dimensional radial distance from the spin axis at $t = 1
\, \text{Gyr}$ for the runs shown in
Figure~\ref{FIG:hernquisttest_surfacedensity_comparison}.}
\label{FIG:hernquisttest_radial_profiles}
\end{figure*}

The final, full physics model variation increases the dust mass by about 50 per
cent at $t = 1.5 \, \text{Gyr}$ compared to the no shattering/coagulation run.
As we show later in Figure~\ref{FIG:hernquisttest_gsd}, the presence of the
shattering process in the full physics model efficiently shifts grains to
smaller sizes.  Thus, by shattering big grains, this model increases the total
grain surface area.  Since the grain growth mechanism in
Section~\ref{SEC:grain_growth} specifies a form of $\mathrm{d}a / \mathrm{d}t$
dependent on gas quantities but independent of grain size $a$, the radii of
grains of different sizes end up growing at the same rate.  As a result, a
population of small grains gains mass more quickly than a population of large
grains with the same total mass.  Because stars tend to produce large grains
(see Section~\ref{SEC:initial_gsd}), grains in the no shattering/coagulation
model do not gain mass as quickly as in the full physics model with shattering.

While the various models predict similar total dust mass evolution, they differ
in how this mass is distributed among grains.  Although dust particles spawned
by stars have grain size distributions initialised in the same manner, these
models have different components evolving the grain size distribution.
Figure~\ref{FIG:hernquisttest_gsd} shows the total grain size distribution
predicted for the isolated galaxy under these three models at $t = 1 \,
\text{Gyr}$, obtained by summing over all dust particles.  We multiply this
grain size distribution ($\partial n / \partial a$) by the mass of a grain of
size $a$ in order to compare the mass contributed by grains of different sizes.
The two models without shattering and coagulation are most similar: switching
from the production only model lacking grain size evolution to the no
shattering/coagulation model including grain growth shifts grains to larger
sizes, producing an increase in mass contained in grains with size $a \gtrsim
0.4 \, \mu\text{m}$.

The full physics model is qualitatively different from the two previous models.
The inclusion of shattering shifts grains to smaller sizes, strongly enhancing
the amount of mass in grains with $a \lesssim 0.03 \, \mu\text{m}$.  Thus,
although the dust models predict similar total dust masses in
Figure~\ref{FIG:hernquisttest_evolution}, the distribution of this mass into
various size grains significantly differs.  We caution that the absence of
feedback in these runs leads to an overproduction of dust, and inflated dust
densities could shatter grains more rapidly than expected (see
equation~\ref{EQN:continuous_shattering}).  We note that the grain size
at which the distribution starts to rise is set implicitly by our model due to
the shattering velocity scale.

Figures~\ref{FIG:hernquisttest_evolution} and~\ref{FIG:hernquisttest_gsd}
focused on galaxy-integrated dust mass and grain size distribution, but we can
also study predictions of our dust models locally within the galaxy.  Radial
profiles of dust surface density and dust-to-gas ratio for the three different
dust models at $t = 1 \, \text{Gyr}$ are shown in
Figure~\ref{FIG:hernquisttest_radial_profiles}.  The full physics model, which
produced the most amount of dust overall, shows the highest dust surface
density at essentially all radii.  However, all three runs show similar
profiles: a relatively constant surface density near the galactic centre and a
rapid fall off at larger radii.

\begin{figure*}
\centering
\includegraphics{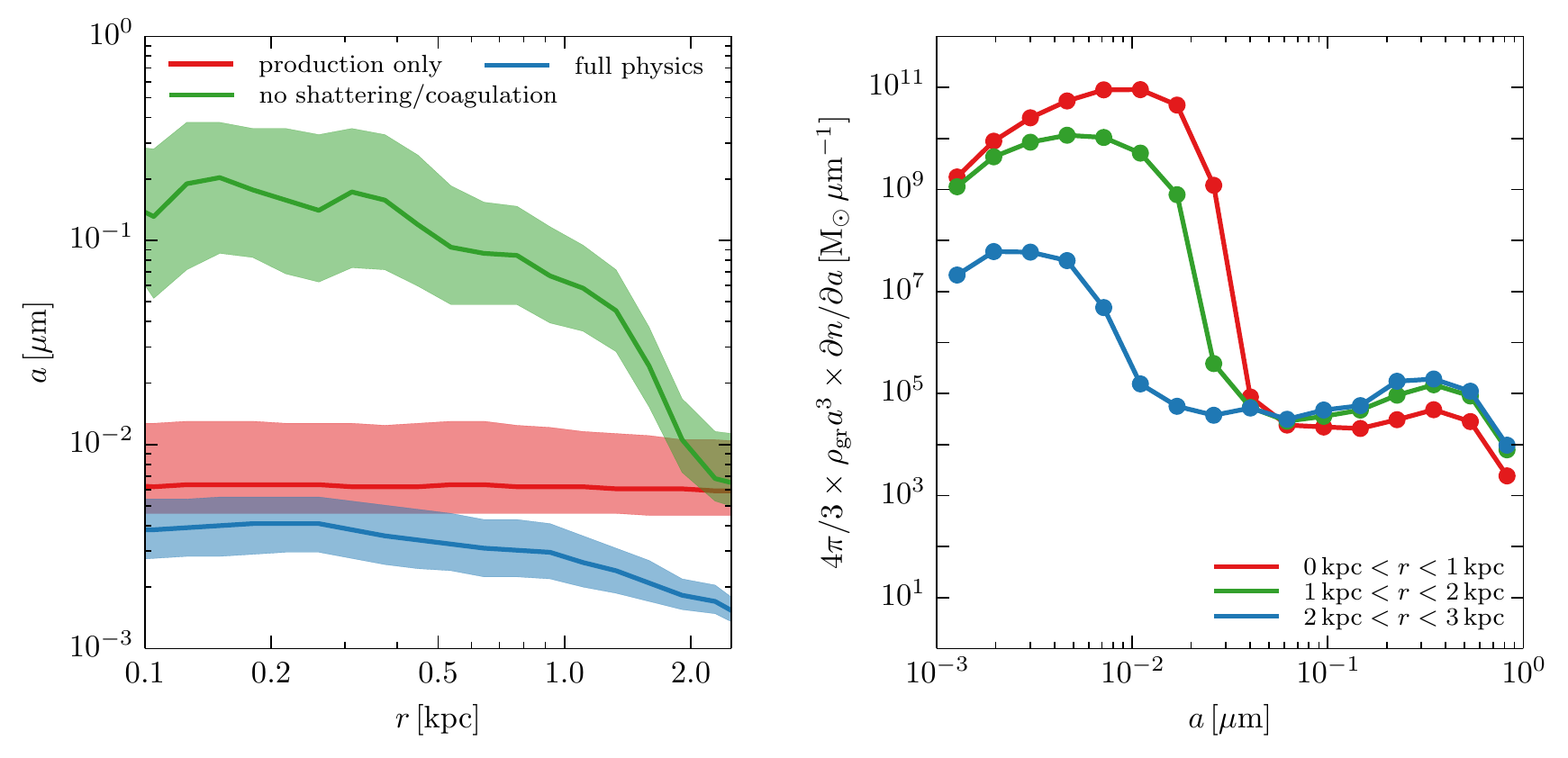}
\caption{Left panel: grain size versus two-dimensional radial distance for the
three dust models at $t = 1 \, \text{Gyr}$.  Solid lines show median grain
size, while shaded regions extend to 32nd and 68th percentiles.  In each radial
bin, these statistics are calculated using the total grain size distribution
obtained by summing over size distributions for constituent dust particles.
Right panel: grain size distribution of dust in different radial slices at $t =
1 \, \text{Gyr}$ in the full physics model.  The prefactor multiplying the
grain size distribution ($\partial n / \partial a$) means that integrating
these profiles yields the total dust mass in each radial slice.}
\label{FIG:hernquisttest_radial_size_slice}
\end{figure*}

In the left panel of Figure~\ref{FIG:hernquisttest_radial_size_slice}, we show
the median grain size as a function of radial distance for the three dust
models at $t = 1 \, \text{Gyr}$.  We assign dust particles to a series of
two-dimensional radial bins, sum the total grain size distribution within each
radial bin, and then use these size distributions to compute median grain sizes
as well as 32nd and 68th percentiles.  In the production only run lacking grain
size evolution, dust particles' grain size distributions are frozen in time.
As a result, the median grain size for this run shows essentially no radial
variation.  The profile is not exactly flat because dust particles created by
AGB stars and SNe II have different initial grain size distributions, and so
not every individual dust particle has the same median grain size.  The runs
with grain size evolution show more variation than the production only run.

Both the no shattering/coagulation and full physics runs show a decline
in median grain size with radius.
Figure~\ref{FIG:hernquisttest_evolution} illustrates that these runs increase
total dust mass above the production only run lacking grain size physics.
For the no shattering/coagulation run, this suggests that accretion,
which increases dust mass, dominates thermal sputtering and SN-based
destruction, which reduce dust mass.  The radial size profile for this no
shattering/coagulation run in Figure~\ref{FIG:hernquisttest_radial_size_slice}
is consistent with a strong accretion mechanism: since accretion is
strongest in regions of high gas density and metallicity (see
equation~\ref{EQN:da_dt}), we expect grain sizes to be highest near the
galactic centre.  While the median grain size in the no shattering/coagulation
run decreases by one order of magnitude out to $2 \, \text{kpc}$, from roughly
$0.1 \, \mu\text{m}$ to $0.01 \, \mu\text{m}$, the full physics model shows a
shallower decline.  Median grain sizes in the full physics model are lower
overall and decrease from about $0.004 \, \mu\text{m}$ to $0.002 \,
\mu\text{m}$ over this same region.  Grain sizes in the full physics run are
subject to a wider variety of processes: for example, both accretion and
coagulation are expected to affect how grains grow \citep{Hirashita2012}.
However, even these simple radial size profiles, which smooth over angular
variations within the disc, suggest that median grain sizes are not
uniform in the galaxy.

For more detail, the right panel of
Figure~\ref{FIG:hernquisttest_radial_size_slice} shows the grain size
distribution in the full physics model at $1 \, \text{Gyr}$ in three, kpc-wide
radial intervals about the galactic centre.  There are two main trends to note.
First, the relative abundance of small grains ($a \lesssim 0.01 \,
\mu\text{m}$) to large grains ($a \gtrsim 0.1 \, \mu\text{m}$) decreases with
radius.  This is intuitive, since the abundance of small grains likely
results from shattering, and shattering is strongest at low radii, where
densities are highest.  Second, among the population of small grains -- which
dominate the overall grain count -- the peak in the mass size distribution in
Figure~\ref{FIG:hernquisttest_radial_size_slice} shifts slightly to smaller
sizes with larger radial distance.  This is similar to the negative slope seen
in the left panel of Figure~\ref{FIG:hernquisttest_radial_size_slice} and is
possibly a consequence of small grains accreting mass from gas more
quickly near the galactic centre, pushing small grain radii somewhat higher.
However, coagulation can also play a role in shifting central grains to
larger sizes \citep[e.g.][]{Hirashita2012}.

We expect the overall cycle between shattering and accretion
to proceed as follows.  Regions of high gas and dust density shatter grains
more quickly (increasing the ratio of small to large grains), but these small
grains then grow in size more quickly (increasing the median size for small
grains and overall dust mass).  This cycle continues itself over time, and so
dust in the central region of the galaxy changes more rapidly than in the
outskirts.

Tracking the grain size distribution locally within the galaxy also allows us
to generate mock extinction curves, with grains of different sizes along a line
of sight contributing different opacities.  We refer the reader to
Appendix~\ref{SEC:extinction_curves} for the full details of how these
extinction curves are constructed.  Here we note that, in addition to dust
particles' grain size distributions, these extinction curves depend on
parameters like the extinction efficiency $Q_\text{ext}(a, \lambda)$, the
dimensionless ratio of extinction cross section to geometric cross section for
grains of size $a$ at wavelength $\lambda$.  The dust mass opacity at
wavelength $\lambda$, $\kappa_\text{ext}(a, \lambda)$, can also be written in
terms of the extinction efficiency via $\kappa_\text{ext}(a, \lambda) = 3
Q_\text{ext}(a, \lambda) / (4 a \rho_\text{gr})$.  In this work, we use
tabulated grain extinction efficiencies from \citet{Draine1984} and
\citet{Laor1993}.  These papers present separate extinction efficiencies for
silicate and graphite grains, $Q_\text{ext}^\text{sil}(a, \lambda)$ and
$Q_\text{ext}^\text{gra}(a, \lambda)$ respectively, which can be converted into
dust mass opacities $\kappa_\text{ext}^\text{sil}$ and
$\kappa_\text{ext}^\text{gra}$.  To calculate extinction curves, we compute an
effective dust mass opacity using the following approximation.  Letting
$f_\text{gra}$ be the fraction of total dust mass in the isolated galaxy
contributed by carbon, the effective dust mass opacity is estimated as
$\kappa_\text{ext}(a, \lambda) = f_\text{gra} \kappa_\text{ext}^\text{gra}(a,
\lambda) + (1 - f_\text{gra}) \kappa_\text{ext}^\text{sil}(a, \lambda)$.  This
dust mass opacity is what enters into extinction curve calculations via
equation~(\ref{EQN:A_lambda}).

\begin{figure*}
\centering
\includegraphics{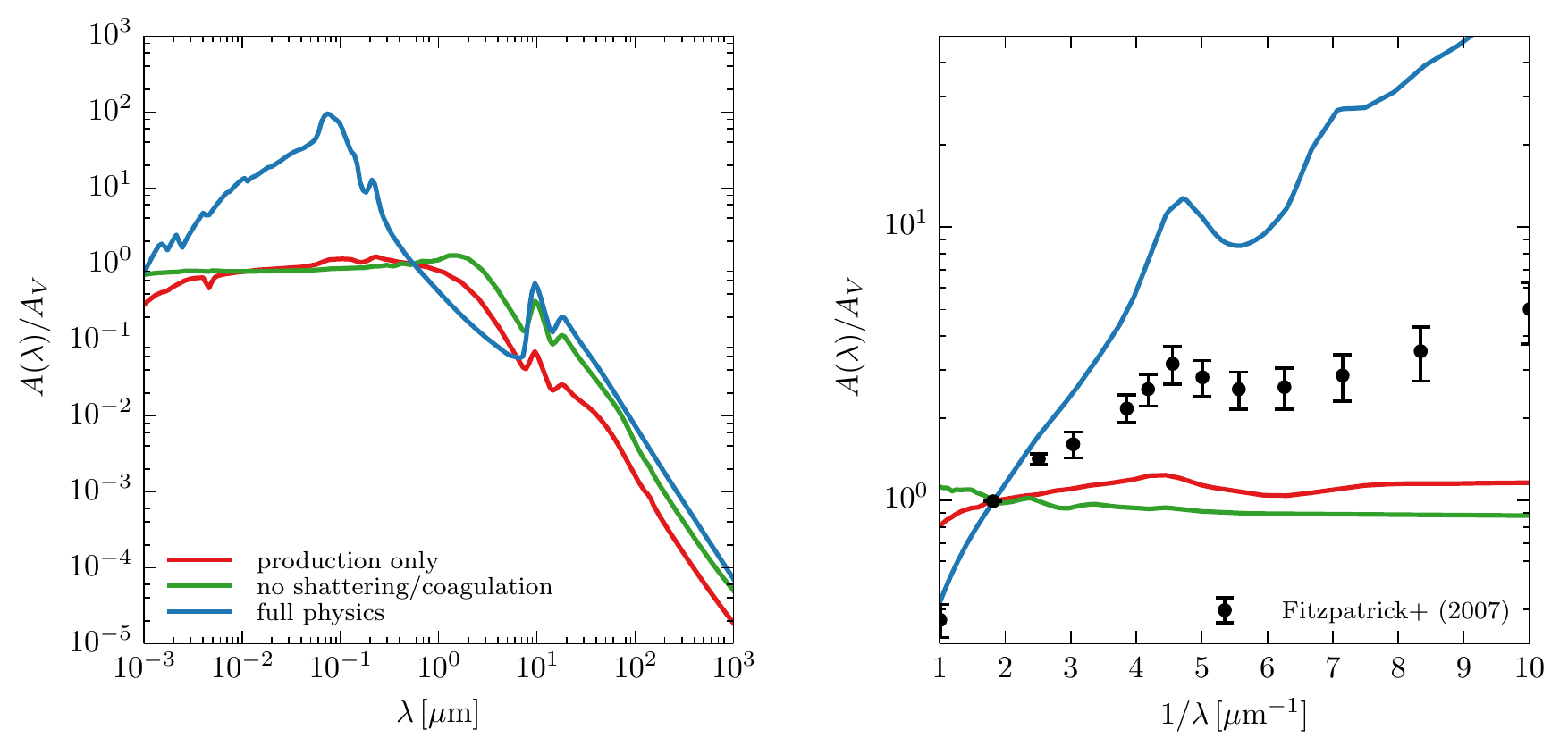}
\caption{Predicted extinction as a function of wavelength at $t = 1 \,
\text{Gyr}$ for the dust models from Figures~\ref{FIG:hernquisttest_evolution}
and~\ref{FIG:hernquisttest_gsd}.  Extinction curves are shown for the full
wavelength range $10^{-3} \, \mu\text{m} \leq \lambda \leq 10^{3} \,
\mu\text{m}$ (left) and for the zoomed in range $0.1 \mu\text{m} \leq \lambda
\leq 1 \mu\text{m}$ (right), with the latter plotted in terms of inverse
wavelength.  Extinction is calculated along a line of sight pointing to the
centre of the galaxy from $5 \, \text{kpc}$ above and $5 \, \text{kpc}$
radially outside and is normalised by extinction in the $V$ band, $A_V$.  The
full dust physics run is the only one to show a $2175 \, \text{\AA}$ bump and a
rise in extinction in the ultraviolet.  Observations of the Galactic extinction
curve are shown in black \citep{Fitzpatrick2007}.}
\label{FIG:hernquisttest_extinction}
\end{figure*}

Figure~\ref{FIG:hernquisttest_extinction} presents a synthetic extinction curve
at $t = 1 \, \text{Gyr}$ for each dust model.  Curves are shown over the full
wavelength range for which there are tabulated extinction efficiencies
($10^{-3} \, \mu\text{m} \leq \lambda \leq 10^{3} \, \mu\text{m}$) and over the
ultraviolet~(UV) and optical region ($0.1 \, \mu\text{m} \leq \lambda \leq 1 \,
\mu\text{m}$, plotted in terms of inverse wavelength as is customary).
Extinction is computed using a line of sight directed towards the galactic
centre from a point $5 \, \text{kpc}$ above and $5 \, \text{kpc}$ radially
outside the disc.  We compare with the observed Galactic extinction
curve from \citet{Fitzpatrick2007}, as compiled by \citet{Hou2017}.

All three dust models predict qualitatively similar extinction curves for
$\lambda \gtrsim 10 \, \mu\text{m}$, wavelengths that are much larger than
typical grain sizes.  In this far-infrared regime, extinction falls off
according to a $A(\lambda) \propto \lambda^{-2}$ power law.  At shorter
wavelengths, the production only and no shattering/coagulation runs lacking
shattering and containing larger grains yield qualitatively different
extinction than the full physics model.  For example, extinction for the
production only and no shattering/coagulation models -- which from
Figure~\ref{FIG:hernquisttest_gsd} predict much of the total dust exists in
large grains -- is nearly flat for $\lambda \lesssim 1 \, \mu\text{m}$.  In
particular, there is essentially no change in extinction from the optical to
the UV, as observed in the Galaxy.

The full physics model, on the other hand, contains many more small grains and
so predicts more features in the extinction curve at small wavelength, like a
$2175 \, \text{\AA}$ bump.  Extinction peaks in the UV near $\lambda \approx
0.1 \, \mu\text{m}$, and the full physics run predicts more than an order of
magnitude more UV extinction than the production only and no
shattering/coagulation runs with large grains.  The full physics model does
show an increase in steepness from optical to UV, albeit one that
rises more steeply than observed in the Galaxy.  This steepness is likely
influenced by the lack of feedback, since the overproduction of dust leads to
high dust densities and thus overly rapid shattering.  An excess of small
grains could exacerbate the rise in UV extinction, motivating future work with
this dust model coupled to feedback methods.

\subsection{Comparison to other models}

We can compare the predictions for our isolated disc galaxy with several other
works that model grain dynamics or size evolution.

For example, recently \citet{Aoyama2017} performed SPH simulations of an
isolated galaxy with total mass $1.3 \times 10^{12} \, \text{M}_\odot$, where
dust evolution calculations take place on each gas particle using a simplified
two-size grain distribution (i.e.~grains are classified as either small,
roughly $0.005 \, \mu\text{m}$, or large, roughly $0.1 \, \mu\text{m}$).
Because there are no separate dust particles, this model implicitly assumes
dust and gas are perfectly coupled and omits a drag force.  At $t = 1 \,
\text{Gyr}$, \citet{Aoyama2017} predict that the mass surface density of small
grains is highest in the galactic centre, a result also suggested by
the right panel of Figure~\ref{FIG:hernquisttest_radial_size_slice}.  In
contrast, the two models disagree about whether most dust mass at $t = 1 \,
\text{Gyr}$ is locked in small or large grains: \citet{Aoyama2017} find that
the total mass ratio of small to large grains is roughly $0.2$, while our
results in Figure~\ref{FIG:hernquisttest_gsd} suggest this ratio is well above
unity.  However, \citet{Aoyama2017} include a model for thermal stellar
feedback, which suppresses the dust mass in their galaxy compared to ours.
This lowers their dust surface densities, in turn slowing the rate of
shattering.  In the future, a fairer comparison of small-to-large grain mass
ratio requires us to couple our dust model to a feedback implementation.

The two-size grain approximation used in \citet{Aoyama2017} has been extended
by \citet{Hou2017} to account for carbonaceous and silicate dust grains.  This
is necessary to predict galactic extinction curves, since carbonaceous and
silicate grains have different extinction cross sections.  Reddening
caused by different grain species can be important not only for extinction in
galaxies but also in the circumgalactic medium \citep{Hirashita2018}.  In
\citet{Hou2017}, the relative abundance of small to large grains increases with
time and grows most rapidly in the central region of the galaxy.  At $t = 1 \,
\text{Gyr}$, the right panel of
Figure~\ref{FIG:hernquisttest_radial_size_slice} also predicts this
small-to-large abundance ratio to peak in the galactic centre.  The presence of
small grains creates more pronounced extinction curve features, like the $2175
\, \text{\AA}$ bump and the UV slope.  The lack of feedback and abundance of
small grains in our runs yields a full physics extinction curve in
Figure~\ref{FIG:hernquisttest_extinction} that rises more quickly from the
optical to the UV than observed.  However, \citet{Hou2017} demonstrate that at
$t = 1 \, \text{Gyr}$ this UV slope is correlated with the small-to-large
abundance ratio: regions in the galaxy with decreased small-to-large abundance
ratio also see decreased the UV slope.  A lower rate of shattering in our model
would produce fewer small grains and a slope in the UV more in line with
observations.

Separately, the two-size grain approximation has also been applied in one-zone
models of galaxy evolution, which lack spatial resolution and instead solve for
galaxy-integrated quantities \citep{Hirashita2015b, Hou2016}.  Other one-zone
models evolve the full grain size distribution, which is more computationally
expensive but tracks the range of sizes grains may have \citep{Hirashita2009,
Asano2013b}.  For example, \citet{Asano2013b} run their one-zone model with and
without shattering, finding that at $t = 1 \, \text{Gyr}$, the inclusion of
shattering increases the abundance of grains with $a \lesssim 0.01 \,
\mu\text{m}$ by orders of magnitude.  The grain size distributions at $t = 0.1$
and $1 \, \text{Gyr}$ are unaffected by the inclusion of coagulation.  Only at
much larger times ($t \approx 10 \, \text{Gyr}$) does coagulation materially
affect the grain size distribution, shifting grain mass to larger sizes.
However, even at late times the effect of shattering is more significant than
that of coagulation.  These predictions from \citet{Asano2013b} parallel our
grain size distribution findings in Figure~\ref{FIG:hernquisttest_gsd}, that at
$t = 1 \, \text{Gyr}$ shattering is more efficient than coagulation and that
the galaxy-wide grain size distribution forms many more small grains than
stars produce.

Another class of model has been developed by \citet{Bekki2015b}, coupling dust
particles to gas in SPH simulations through a drag law but neglecting grain
size evolution.  All dust grains share the same fixed size (roughly $0.1 \,
\mu\text{m}$), limiting the ability to construct extinction curves that capture
the range of grain sizes that exist in the ISM.  However, these dust particles
are coupled to a scheme modelling radiation pressure from stellar sources, and
simulations in \citet{Bekki2015b} predict that radiation pressure can increase
the vertical extent of dust in the disc while reducing radial gradients in the
dust distribution.  As we work to couple our dust dynamics and size evolution
model to more forces like radiation pressure, we will be in position to test
these dynamical predictions and additionally investigate their impact on the
grain size distribution and extinction.

\section{Conclusions}\label{SEC:conclusions}

We have implemented a novel scheme to track the dynamical motion and grain size
evolution of interstellar dust grains in the moving-mesh code \textsc{arepo}.

Simulation dust particles represent ensembles of grains of different sizes and
are characterised by individual grain size distributions that are evolved in
time.  Each grain size distribution is discretised using a piecewise linear
method and updated according to a variety of physical processes.  Processes
like accretion, sputtering, and destruction from supernova shocks
conserve grain number but shift mass between dust and gas phases, while
dust-dust collisional processes like shattering and coagulation conserve total
grain mass but not grain number.  We demonstrate that the piecewise linear
discretisation is second-order accurate in the number of grain size bins.  The
dynamical drag force for each particle is calculated based on its internal
grain size distribution to couple gas and dust motions.

The drag force implementation is based on a second-order semi-implicit
scheme that makes use of analytic properties of the drag force and alleviates
the need for small time-steps when dust and gas are strongly coupled; i.e.~when
the stopping time-scale governing drag is short.  The drag force acting on dust
grains depends on local gas properties, and our methods benefit from the
accurate treatment of hydrodynamics in \textsc{arepo}.  For example, in
simulating gas and dust dynamics in a Sod shock tube test, dust particle
velocities do not suffer from spurious post-shock velocity ringing seen in some
smoothed-particle hydrodynamics methods.

The actual production of dust particles is coupled to the stellar evolution
scheme of our galaxy formation model.  We implement dust mass return during
stellar evolution using a stochastic procedure that probabilistically spawns
dust particles from star particles.  When spawning new dust particles, we adopt
dust elemental yields according to theoretical models of mass return from
AGB stars and SNe.  Similarly, initial grain size distributions for dust
particles are set according to theoretical predictions for grain populations
formed during stellar evolution.  Newly created dust particles are then
subjected to the aforementioned physical processes shaping their grain size
distributions.

Processes like shattering, coagulation, sputtering, and dust growth
can lead to rather significant changes in the number of dust particles and
their masses.  This can lead, for example, to very heavy dust particles or many
low mass particles, which is computationally disadvantageous.  We have
therefore also implemented refinement and de-refinement schemes for dust
particles, to keep the mass distribution of dust particles within predefined
limits.  Furthermore, we have also implemented time-step sub-cycling for the
dust time-steps to avoid too many small dust-dominated time-steps.
While our model currently neglects astration, the consumption of ISM
dust during star formation, we use a time-scale argument to show that SN
destruction of dust is expected to dominate astration as a sink of ISM dust
mass.

To demonstrate the simultaneous application of dust dynamics, grain size
evolution, and dust particle creation, we simulate an isolated disc galaxy with
cooling and star formation but no feedback and study the relative strengths of
various grain size processes.  For example, a model without grain size
evolution and a model with full grain size physics produce galactic dust masses
differing by a factor of four and qualitatively very different grain size
distributions.  The inclusion of shattering is particularly efficient at
shifting large dust grains to smaller sizes.  Using the simulated spatial
distribution of grains, we produce sample extinction curves, with small grains
in the full physics run producing an increase in extinction towards the UV.

Our framework for simulating dust and gas mixtures can readily be extended to
account for other dynamical processes relevant in galaxy formation, like
magnetohydrodynamics, radiation pressure, and thermo-chemical processes.
Ultimately, our model represents a step towards a more comprehensive treatment
of dust dynamics and grain size evolution in galaxy formation.

\section*{Acknowledgements}

Our anonymous referee provided valuable feedback that helped improve
this manuscript. We thank Hiroyuki Hirashita for providing us with the
tabulated form of grain velocities calculated over an extended grain size range
and used in \citet{Hirashita2009}.  We also thank Takaya Nozawa for making
available the tabulated versions of the SN grain size distributions presented
in \citet{Nozawa2007} and the SN dust destruction efficiencies used in
\citet{Asano2013b}.  Finally, we are grateful to Volker Springel for sharing
access to \textsc{arepo}.

The simulations were performed on the joint MIT-Harvard computing cluster
supported by MKI and FAS.  RM acknowledges support from the DOE CSGF under
grant number DE-FG02-97ER25308.  MV acknowledges support through an MIT RSC
award and the support of the Alfred P.~Sloan Foundation.  PT acknowledges
support from NASA through Hubble Fellowship grant HST-HF2-51341.001-A awarded
by STScI, which is operated under contract NAS5-26555.  RK acknowledges support
from NASA through Einstein Postdoctoral Fellowship grant number PF7-180163
awarded by the \textit{Chandra} X-ray Center, which is operated by the
Smithsonian Astrophysical Observatory for NASA under contract NAS8-03060.

\bibliographystyle{mn2e}
\bibliography{../bibliography}

\appendix
\onecolumn

\section{Discretisation of shattering integrals}\label{SEC:appendix_shattering}

This section converts the analytic shattering framework presented in
Section~\ref{SEC:shattering} into one capable of handling a piecewise linear
grain size distribution.  Namely, we show how
equation~(\ref{EQN:shattering_sums}) can be derived from
equation~(\ref{EQN:continuous_shattering}).

To start, multiply equation~(\ref{EQN:continuous_shattering}) by $V_\text{d}^2$,
apply equation~(\ref{EQN:differential_mass_density}) to convert mass densities
into number densities, and rewrite integrals using the partition of
$[a_\text{min}, a_\text{max}]$, so that
\begin{equation}
\begin{split}
V_\text{d} \frac{\diff}{\diff t} \left[ m(a) \frac{\partial n(a, t)}{\partial a} \diff a \right] &= -m(a)^2 \frac{\partial n(a, t)}{\partial a} \diff a \sum_{k=0}^{N-1} \int_{a^\text{e}_k}^{a^\text{e}_{k+1}} \alpha(a, a_1) m(a_1) \frac{\partial n(a_1, t)}{\partial a_1} \diff a_1 \\
&\quad + \frac{1}{2} \sum_{k=0}^{N-1} \sum_{j=0}^{N-1} \diff a \int_{a^\text{e}_k}^{a^\text{e}_{k+1}} \int_{a^\text{e}_j}^{a^\text{e}_{j+1}} \alpha(a_1, a_2) m(a_1) m(a_2) \frac{\partial n(a_1, t)}{\partial a_1} \frac{\partial n(a_2, t)}{\partial a_2} m_\text{shat}(a, a_1, a_2) \diff a_2 \diff a_1.
\end{split}
\end{equation}
Integrating $a$ over the interval $[a^\text{e}_i, a^\text{e}_{i+1}]$ and
substituting in the piecewise linear grain size distribution from
equation~(\ref{EQN:dnda_linear}), we find that the mass of grains in bin $i$,
$M_i$, evolves as
\begin{equation}
\begin{split}
V_\text{d} \frac{\diff M_i}{\diff t} &= \int_{a^\text{e}_i}^{a^\text{e}_{i+1}} -m(a)^2 \left( \frac{N_i(t)}{a^\text{e}_{i+1} - a^\text{e}_{i}} + s_i(t) (a - a^\text{c}_i) \right) \sum_{k=0}^{N-1} \int_{a^\text{e}_k}^{a^\text{e}_{k+1}} \alpha(a, a_1) m(a_1) \left( \frac{N_k(t)}{a^\text{e}_{k+1} - a^\text{e}_{k}} + s_k(t) (a_1 - a^\text{c}_k) \right) \diff a_1 \diff a \\
&\quad + \frac{1}{2} \int_{a^\text{e}_i}^{a^\text{e}_{i+1}} \sum_{k=0}^{N-1} \sum_{j=0}^{N-1} \int_{a^\text{e}_k}^{a^\text{e}_{k+1}} \int_{a^\text{e}_j}^{a^\text{e}_{j+1}} \bigg[ \alpha(a_1, a_2) m(a_1) m(a_2) \left( \frac{N_k(t)}{a^\text{e}_{k+1} - a^\text{e}_{k}} + s_k(t) (a_1 - a^\text{c}_k) \right) \\
&\qquad\qquad\qquad\qquad\qquad\qquad\qquad\qquad \quad \times \left( \frac{N_j(t)}{a^\text{e}_{j+1} - a^\text{e}_{j}} + s_j(t) (a_2 - a^\text{c}_j) \right) m_\text{shat}(a, a_1, a_2) \bigg] \diff a_2 \diff a_1 \diff a.
\end{split}
\end{equation}
Substituting for $\alpha$ using equation~(\ref{EQN:alpha_a1_a2}) and
rearranging, this can be simplified as
\begin{equation}
\begin{split}
V_\text{d} \frac{\diff M_i}{\diff t} &= \sum_{k=0}^{N-1} \int_{a^\text{e}_i}^{a^\text{e}_{i+1}} \int_{a^\text{e}_k}^{a^\text{e}_{k+1}} \bigg[ -\pi (a + a_1)^2 v_\text{rel}(a, a_1) \mathbbm{1}_{v_\text{rel} > v_\text{shat}}(a, a_1) m(a) \left( \frac{N_i(t)}{a^\text{e}_{i+1} - a^\text{e}_{i}} + s_i(t) (a - a^\text{c}_i) \right) \\
&\qquad\qquad\qquad\qquad\qquad \times \left( \frac{N_k(t)}{a^\text{e}_{k+1} - a^\text{e}_{k}} + s_k(t) (a_1 - a^\text{c}_k) \right) \bigg] \diff a_1 \diff a \\
&\quad + \frac{1}{2} \sum_{k=0}^{N-1} \sum_{j=0}^{N-1} \int_{a^\text{e}_i}^{a^\text{e}_{i+1}} \int_{a^\text{e}_k}^{a^\text{e}_{k+1}} \int_{a^\text{e}_j}^{a^\text{e}_{j+1}} \bigg[ \pi (a_1 + a_2)^2 v_\text{rel}(a_1, a_2) \mathbbm{1}_{v_\text{rel} > v_\text{shat}}(a_1, a_2) \left( \frac{N_k(t)}{a^\text{e}_{k+1} - a^\text{e}_{k}} + s_k(t) (a_1 - a^\text{c}_k) \right) \\
&\qquad\qquad\qquad\qquad\qquad\qquad\qquad\qquad \quad \times \left( \frac{N_j(t)}{a^\text{e}_{j+1} - a^\text{e}_{j}} + s_j(t) (a_2 - a^\text{c}_j) \right) m_\text{shat}(a, a_1, a_2) \bigg] \diff a_2 \diff a_1 \diff a.
\end{split}
\label{EQN:dMi_dt_shatter}
\end{equation}
Apart from the terms $v_\text{rel}(a_1, a_2)$ and $m_\text{shat}(a, a_1, a_2)$,
equation~(\ref{EQN:dMi_dt_shatter}) only involves integrals of two-dimensional
polynomials.  We wish to evaluate this integral analytically so that we can
explicitly update the mass in each bin using the piecewise linear grain size
distribution (i.e.~the set of known $N_i(t)$ and $s_i(t)$ values).  We make two
simplifying assumptions.

First, we assume that all grains in the same bin share the same speed.
As a result, the relative velocity between two grains in bins $k$ and $j$ can
be simplified as $v_\text{rel}(a_1, a_2) \approx v_\text{rel}(a^\text{c}_k,
a^\text{c}_j)$, independent of the integration variables $a_1$ and $a_2$.
Following \citet{Hirashita2009} and \citet{Asano2013b}, we adopt relative grain
velocities from \citet{Yan2004}, who calculated grain speeds as a function of
grain size for various ISM phases, assuming a turbulent, magnetised fluid.
Given our intent to use this dust model in cosmological simulations, we
recognise that we will not resolve some of the phases studied by
\citet{Yan2004}, such as the DC1 and DC2 phases with $T = 10 \, \text{K}$ and
$n_\text{H} = 10^4 \, \text{cm}^{-3}$.

The current galaxy formation model in \textsc{arepo} employs the
\citet{Springel2003} multiphase ISM model, which adopts a hybrid mixture of hot
and cold components.  We define the effective relative velocity
$v^\text{eff}_\text{rel}(a_1, a_2) \equiv x
v^\text{CNM}_\text{rel}(a^\text{c}_k, a^\text{c}_j) + (1-x)
v^\text{WIM}_\text{rel}(a^\text{c}_k, a^\text{c}_j)$, where $x$ is a
kernel-smoothed estimate of the cold cloud mass fraction \citep[see Section~3
in][]{Springel2003} in neighbouring gas cells and $v^\text{CNM}_\text{rel}$ and
$v^\text{WIM}_\text{rel}$ are the relative velocities computed for the cold
neutral medium~(CNM) and warm ionised medium~(WIM) phases, respectively, using
\citet{Yan2004}.  As we do not track detailed grain chemistry, the grain
velocities for the CNM and WIM are averaged over the curves calculated in
\citet{Yan2004} for silicate and graphite grains.  This is a minor assumption,
since the silicate and graphite curves are qualitatively similar.  We use this
form of $v^\text{eff}_\text{rel}$ in equation~(\ref{EQN:dMi_dt_shatter}).  A
more realistic ISM model would allow us to probe grain velocities in the
variety of phases studied in \citet{Yan2004}.

Second, we assume that the mass of grains produced with radius $a$ by
shattering grains of sizes $a_1$ and $a_2$ depends only on the bins involved in
the collision.  That is, for grains of sizes $a_1$ and $a_2$ in bins $k$ and
$j$, we adopt $m_\text{shat}(a, a_1, a_2) \approx m_\text{shat}(a,
a^\text{c}_k, a^\text{c}_j)$.  This assumption is reasonable given the physical
uncertainties in grain-grain collisions, and the exact mass distribution of
shattered grains is not expected to strongly affect shattering calculations
\citep{Jones1996, Hirashita2009}.  Performing the second integral in
equation~(\ref{EQN:dMi_dt_shatter}) over $a$, we express the mass of grains
injected into bin $i$ from a collision of grains with sizes $a^\text{c}_k$ and
$a^\text{c}_j$, the midpoints of bins $k$ and $j$, as
\begin{equation}
m_\text{shat}^{k,j}(i) \equiv \int_{a^\text{e}_i}^{a^\text{e}_{i+1}}  m_\text{shat}(a, a^\text{c}_k, a^\text{c}_j) \diff a \approx \int_{a^\text{e}_i}^{a^\text{e}_{i+1}} m_\text{shat}(a, a_1, a_2) \diff a.
\label{EQN:m_shat_kji}
\end{equation}
In practice, we compute $m_\text{shat}^{k,j}(i)$ by following the steps in
Section~2.3 of \citet{Hirashita2009}, which depend on
$v_\text{rel}(a^\text{c}_k, a^\text{c}_j)$.  In dividing shattered grain mass
among different bins, these calculations assume that shattered grains obey the
new size distribution $\partial n / \partial a \propto a^{-3.3}$
\citep{Jones1996}.

Using these steps, we can approximate the integrals in
equation~(\ref{EQN:dMi_dt_shatter}) and bring them to the form presented in
equation~(\ref{EQN:shattering_sums}).

\section{Convergence study of isolated disc galaxies}\label{SEC:convergence_study}

\begin{figure}
\centering
\includegraphics{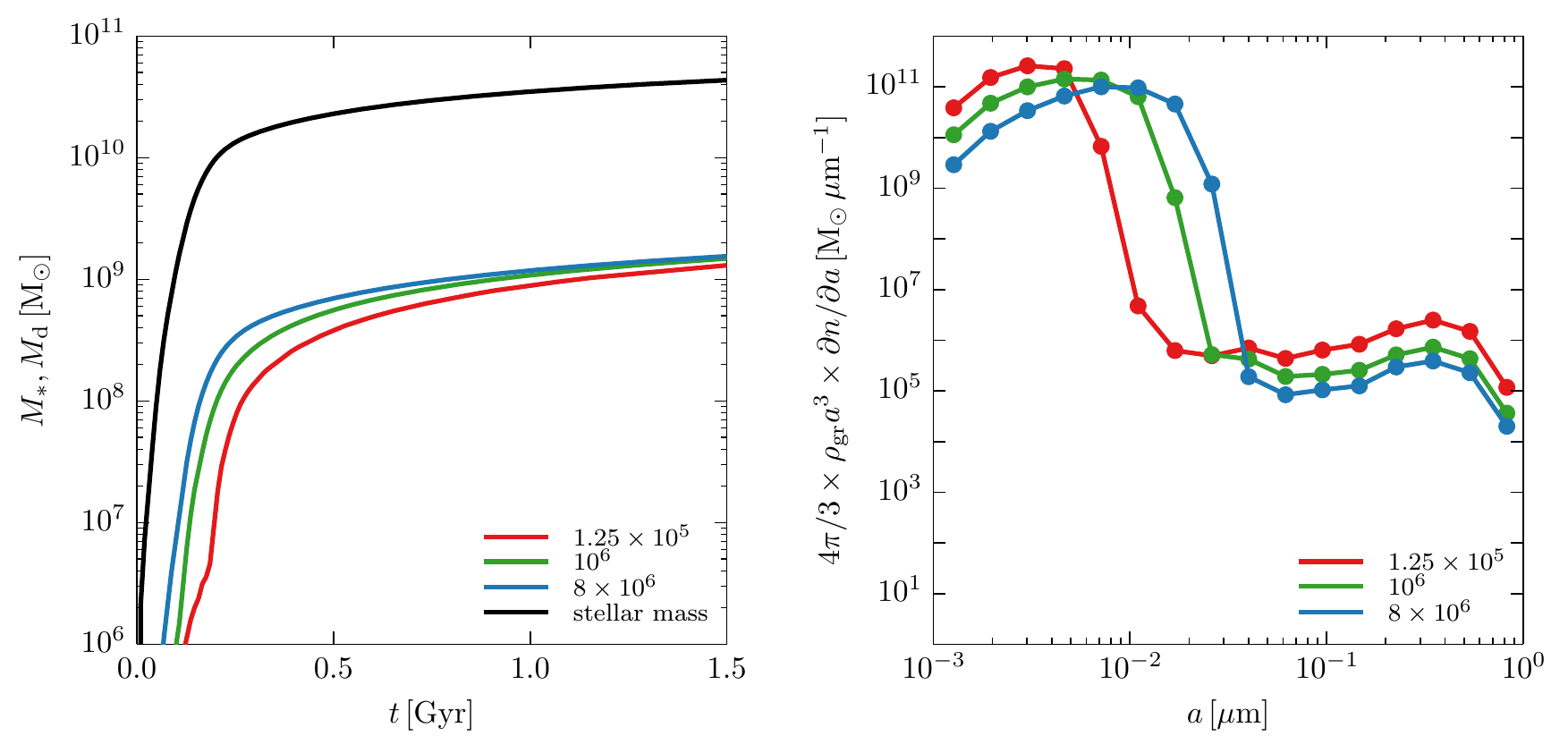}
\caption{Properties of the full physics isolated disc galaxy simulated at three
resolutions, with initial gas cell counts $1.25 \times 10^5$~(red),
$10^6$~(green), and $8 \times 10^6$~(blue).  The left panel shows total dust
mass versus time, while the right panel shows the total grain size distribution
at $1 \, \text{Gyr}$, when the galaxy dust masses differ by about $0.1 \,
\text{dex}$.  For comparison with the dust mass evolution, we also show the
total stellar mass for the high resolution simulation~(black).}
\label{FIG:hernquisttest_convergence}
\end{figure}

We analyse the convergence properties of our full physics dust model using
simulations at three different resolutions.  In these runs, dark matter and gas
each initially have $1.25 \times 10^5$, $10^6$, and $8 \times 10^6$ resolution
elements.  Figure~\ref{FIG:hernquisttest_convergence} shows the isolated
galaxy's total dust mass as a function of time as well as the galaxy-integrated
grain size distribution at $1 \, \text{Gyr}$.  The largest deviation in dust
mass is at early times, with the highest resolution run producing dust more
quickly than the other runs.  However, this trend is largely being driven by
the underlying star formation rate: the star formation rate increases slightly
with resolution at fixed time.  This translates into a small spread in dust
mass in the early stages of the galaxy's formation, before grain size processes
have had much time to act.  Beyond $1 \, \text{Gyr}$, the dust mass profiles
show improved convergence, and, by $1.5 \, \text{Gyr}$, the dust masses differ
by less than $0.1 \, \text{dex}$ across these resolutions.  The grain size
distributions show similar qualitative features, with an abundance of small
grains and a drop off in the mass contained in large grains.  However, the
radius at which the size distribution falls off does vary: the low resolution
simulation predicts this feature at $a \approx 0.01 \, \mu\text{m}$, while the
high resolution run predicts $a \approx 0.03 \, \mu\text{m}$.  The medium
resolution run is more similar to the high resolution one than the low
resolution one, suggesting the profiles are converging, but more simulations
would be needed to fully investigate this.

\section{Comparison of SN destruction and astration time-scales}\label{SEC:astration}

Physically, dust in the ISM is depleted during star formation, a
process known as astration.  In our model used in
Section~\ref{SEC:hernquist_spheres}, star particles form stochastically from
gas cells, which contain gas and gas-phase metals.  Notably, dust is not
treated as a component of gas cells but instead using simulation particles.  As
a result, when gas cells convert to stars, the ISM supply of gas and gas-phase
metals is reduced, but the supply of dust is unchanged.  In this section, we
quantify the expected rate of astration of dust and compare it to the rate at
which dust is depleted through other means (e.g.~SN destruction).

We can estimate the rate at which dust mass is lost due to astration
following equation~3 in \citet{Hjorth2014}, giving $(\mathrm{d} M_\text{d} /
\mathrm{d} t)_\text{astr} = - D \times \mathrm{d} M_{*} / \mathrm{d} t$, where
$M_\text{d}$ denotes dust mass in some region of the ISM, $D$ is the local
dust-to-gas ratio and $\diff M_{*} / \diff t$ is the local star-formation rate.
This rate assumes that when stars form, dust and gas are depleted according to
their relative abundance.

Similarly, the rate of dust destruction in SN shocks is estimated from
equations~2 and~5 in \citet{McKee1989} as
$(\mathrm{d} M_\text{d} / \mathrm{d} t)_\text{dest} = - M_\text{d} / t_\text{SNR}$.
Here, $t_\text{SNR}$ is a time-scale given by the ratio of local ISM
gas mass $M_\text{g}$ to the rate at which gas mass is shocked by SNe, which
depends on the local SN II rate $R_\text{SN}$.  This time-scale is calculated
using $1 / t_\text{SNR} = \epsilon M_\text{cl} f_\text{SN} R_\text{SN} /
M_\text{g}$.
This expression relies on several parameters \citep[with typical values
estimated in Sections~3 and~4 in][]{McKee1989}: $\epsilon \approx 0.4$ denotes
a grain destruction efficiency factor, $f_\text{SN} \approx 0.34$ reduces the
nominal SN rate to account for inefficiencies in correlated SN blasts, and
$M_\text{cl}$ is the mass of gas shocked by a SN.  This latter value is
estimated by \citet{McKee1989} as being in the range $M_\text{cl} \approx 2460
- 6800 \, \text{M}_\odot$.  Since we are interested in whether astration can be
important relative to SN destruction, we will assume $M_\text{cl} \approx
2460 \, \text{M}_\odot$.  This places the SN shock time-scale $t_\text{SNR}$
at the upper end of its expected range and adopts a weak rate of SN dust
destruction.  For our \citet{Chabrier2003} IMF with mass range from $0.1$ to
$100 \, \text{M}_\odot$ and SN II cutoff at $6 \, \text{M}_\odot$, roughly
26.4 per-cent of stellar mass that forms exists as SNe II.  Additionally, the
average SN II mass is calculated as roughly $15.2 \, \text{M}_\odot$.  If we
assume that SNe II immediately die after being formed, then we can estimate
the SN II rate $R_\text{SN}$ from the star-formation rate as
\begin{equation}
R_\text{SN} \approx \left( \frac{0.264}{15.2 \, \text{M}_\odot} \right) \frac{\diff M_{*}}{\diff t}.
\end{equation}
Using all of these values and noting that $M_\text{d} / M_\text{g}$ is
the dust-to-gas ratio $D$, we can write the rate of dust loss due to SN
destruction as a function of star-formation rate and in turn the expected
astration rate via
\begin{equation}
\left(\frac{\diff M_\text{d}}{\diff t}\right)_\text{dest} \approx -5.8 D \frac{\diff M_{*}}{\diff t} = 5.8 \left( \frac{\diff M_\text{d}}{\diff t} \right)_\text{astr}.
\label{EQN:dM_dt_astration}
\end{equation}
That is, the rate of dust loss due to SN destruction is expected to be
roughly five times greater than the rate of dust loss from astration.

To test whether dust mass loss from astration is important, we rerun
the isolated discs presented in Section~\ref{SEC:hernquist_spheres} using the
medium resolution initial conditions and a stronger SN destruction mechanism.
Since both astration and SN dust destruction rates scale with the product of
dust-to-gas ratio and star-formation rate, we can use a larger SN destruction
rate to indirectly model the effects of astration, which is not otherwise
included in our simulations.  To be precise, in these tests we calculate the SN
destruction rate as usual and add an extra dust destruction rate equal to
$\delta$ times the SN destruction rate to model astration.  (In the production
only model that does not include SN destruction of dust, we calculate what the
SN destruction rate would be and use this to estimate the astration rate.)  The
calculations in equation~(\ref{EQN:dM_dt_astration}) suggest $\delta = 1/5.8
\approx 0.17$.  Given that these physical time-scales have some uncertainty and
our desire to assess the maximum impact astration could have, we actually
employ $\delta = 0.25$.  This can be considered an upper bound on the strength
of astration relative to SN destruction.

We acknowledge that this prescription does not perfectly model
astration, since dust destruction via SNe does not directly transfer metal mass
from dust to newly-formed stars but instead star-forming gas.  Additionally,
SNe dust destruction affects grain sizes by shifting them to smaller values.
However, given that the factor by which SN dust destruction is enhanced is only
25 per-cent and not a factor of several or more, this enhancement should not
significantly affect grain size distributions.

Using these tests, we can estimate the impact astration would have on
dust content in our isolated discs.  For the full physics model, the run
without (with) astration predicts a $t = 1 \, \text{Gyr}$ dust mass of $1.1
\times 10^{8} \, \text{M}_\odot$ ($1.1 \times 10^{8} \, \text{M}_\odot$) and
dust-to-metal ratio in the star-forming disc of $0.96$ ($0.93$).  In this
model, astration is subdominant to SN dust destruction in shaping the overall
dust mass and shifts the dust-to-metal ratio down by a few per-cent.  This is
not surprising, given that the astration rate is several times lower than the
SN destruction rate -- and the fact that, overall, dust mass experiences a net
increase in the ISM over time.  The production only run lacking grain size
evolution does experience a stronger effect: without (with) astration, the $t =
1 \, \text{Gyr}$ dust mass and dust-to-metal ratio are $2.8 \times 10^{8} \,
\text{M}_\odot$ ($5.8 \times 10^{7} \, \text{M}_\odot$) and $0.48$ ($0.14$),
respectively.  However, we note that the production only run lacks SN
destruction.  While astration is subdominant to SN dust destruction, if the
latter is not included, then astration can reduce dust masses and dust-to-metal
ratios by roughly a factor of four.  This production only setup should not be
taken as physically plausible: since SN dust destruction dominates astration,
the former should be included in any model accounting for the latter.  In our
full physics model where SN destruction is already present, the addition of
astration affects results less strongly than SN destruction.  Nonetheless, for
completeness we intend to model astration directly in future work.

\section{Generating extinction curves}\label{SEC:extinction_curves}

The optical depth at wavelength $\lambda$ contributed by grains with sizes in
the interval $[a, a + \diff a]$ along a path $\mathcal{P}$ is given by
\begin{equation}
\tau(a, \lambda) \diff a = \int_{\mathcal{P}} \pi a^2 Q_\text{ext}(a, \lambda) n_\text{d}(\vec{r}, a) \diff a \diff s,
\label{EQN:tau_Q}
\end{equation}
where $n_\text{d}(\vec{r}, a) \times \diff a$ is the number density of grains
with sizes in $[a, a + \diff a]$ at position $\vec{r}$, calculated by
interpolating over the grain size distributions of nearby dust particles.  The
extinction efficiency $Q_\text{ext}(a, \lambda) = Q_\text{abs}(a, \lambda) +
Q_\text{sca}(a, \lambda)$ is the ratio of extinction cross section to geometric
cross section, $\pi a^2$, and includes absorption and scattering contributions.
Extinction efficiencies also vary depending on whether grains are assumed to be
silicate or graphite.  We adopt extinction efficiencies for silicate and
graphite grains from \citet{Draine1984} and \citet{Laor1993}, interpolating to
our grain size bins as necessary.

We can rewrite equation~(\ref{EQN:tau_Q}) in terms of
$\kappa_\text{ext}(a, \lambda) = 3 Q_\text{ext}(a, \lambda) / (4 a
\rho_\text{gr})$, the dust mass opacity at wavelength $\lambda$ and grain size
$a$.  This produces
\begin{equation}
\tau(a, \lambda) \diff a = \int_{\mathcal{P}} \frac{4\pi}{3} a^3 \rho_\text{gr} \kappa_\text{ext}(a, \lambda) n_\text{d}(\vec{r}, a) \diff a \diff s.
\end{equation}
The magnitude of the extinction along this line of sight is then obtained by
integrating over the grain size distribution, yielding
\begin{equation}
A(\lambda) = 2.5 \log_{10}(e) \int_{a_\text{min}}^{a_\text{max}} \tau(a, \lambda) \diff a = 2.5 \log_{10}(e) \frac{4\pi}{3} \rho_\text{gr} \int_{a_\text{min}}^{a_\text{max}} a^3 \kappa_\text{ext}(a, \lambda) \int_{\mathcal{P}} n_\text{d}(\vec{r}, a) \diff s \diff a.
\end{equation}
If we break the grain size integral into the sum of integrals over the
$N$ grain size bins and approximate grain sizes with the midpoints of the $N$
bins, we can discretise this as
\begin{equation}
A(\lambda) = 2.5 \log_{10}(e) \frac{4\pi}{3} \rho_\text{gr} \sum_{i = 0}^{N-1} {a_i^\text{c}}^3 \kappa_\text{ext}(a_i^\text{c}, \lambda) (a_{i+1}^\text{e} - a_i^\text{e}) \int_\mathcal{P} n_\text{d}(\vec{r}, a_i^\text{c}) \diff s.
\label{EQN:A_lambda}
\end{equation}

\label{lastpage}

\end{document}